\documentclass[journal,twoside]{asmb}

\begin{document}
\title{Opportunistic Relay Selection over Generalized Fading and Inverse Gamma Composite Fading Mixed Multicast Channels: A Secrecy Tradeoff}

\author[1]{MD. Shakhawat Hossen}
\author[2]{A. S. M. Badrudduza}
\author[3]{S. M. Riazul Islam}
\author[4]{Abu Hanif}
\author[5]{Milton kumar kundu}
\author[6]{Kyung-Sup Kwak}

\affil[1,2,4]{Department of Electronics \& Telecommunication Engineering, Rajshahi University of Engineering \& Technology (RUET), Rajshahi-6204, Bangladesh}
\affil[3]{Department of Computer Science and Engineering, Sejong University, 209 Neungdong-ro, Gwangjin-gu, Seoul 05006, South Korea}
\affil[5]{Department of Electrical \& Computer Engineering, RUET}
\affil[6]{School of Information and Communication Engineering, Inha University, Incheon 22212, South Korea}

\twocolumn[
\begin{@twocolumnfalse}
\maketitle
\begin{abstract}
\section*{Abstract}

The secrecy performance of realistic wireless multicast scenarios can be significantly deteriorated by the simultaneous occurrence of multipath and shadowing. To resolve this security threat, in this work an opportunistic relaying-based dual-hop wireless multicast framework is proposed in which the source dispatches confidential information to a bunch of receivers via intermediate relays under the wiretapping attempts of multiple eavesdroppers. Two scenarios, i.e. non-line of sight (NLOS) and line of sight (LOS) communications along with the multiplicative and LOS shadowing are considered where the first scenario assumes $\eta-\mu$ and $\eta-\mu$/inverse Gamma (IG) composite fading channels and the latter one follows $\kappa-\mu$ and $\kappa-\mu$/IG composite fading channels as the source to relay and relay to receiver's as well as eavesdropper's links, respectively. Secrecy analysis is accomplished by deriving closed-form expressions of three familiar secrecy measures i.e. secure outage probability for multicasting, probability of non-zero secrecy multicast capacity, and ergodic secrecy multicast capacity. We further capitalize on those expressions to observe the effects of all system parameters which are again corroborated via Monte-Carlo simulations. Our observations indicate that a secrecy tradeoff between the number of relays and number of receivers, eavesdroppers, and shadowing parameters can be established to maintain the admissible security level by decreasing the detrimental influences of fading, shadowing, the number of multicast receivers and eavesdroppers.
\end{abstract}

\begin{IEEEkeywords}
\section*{Keywords} 

Composite channel, inverse Gamma shadowing, multiple eavesdroppers, physical layer security, wireless multicasting.

\end{IEEEkeywords}
\end{@twocolumnfalse}
]

\section{Introduction}

Due to the low power requirement and enhanced spatial diversity, relay communication has been established as a proven technique for expanding the coverage of wireless networks. However, such cooperative networks are vulnerable to eavesdropping, where one or more eavesdroppers might overhear transmissions and can thereby possibly pose security threats \cite{lv2019secure,moualeu2018physical}. So maintaining perfect security during data transfer through a wireless medium in general and cooperative communication, in particular,  has become a significant challenge for wireless researchers. To solve this problem, researchers introduced the encryption method where the information was first encrypted using an encryption key and then transmitted to the receiver through the wireless channel. The technique was simple but became troublesome because of the difficulty in encryption key distribution \cite{diamanti2016practical}. On the other hand, physical layer security (PLS) has some tremendous advantages over the encryption method. PLS is more secure as the confidentiality of the transmitted message can be guaranteed by exploiting the physical characteristics of the wireless channel as well as the proper coding and signal processing. Shadowing and multipath fading are two very important characteristics of any wireless channel which co-exist together and affect the channel simultaneously. So to successfully characterize the random fluctuations occurring in the channel, researchers have introduced composite fading models which can efficiently reflect the practical scenarios which go through both fading and shadowing \cite{yoo2017kappa}. Traditionally, lognormal and Gamma distributions were chosen by the researchers for designing the composite fading channels \cite{bhatt2018asep}. But the mathematical intractability of the lognormal distribution made it difficult to use in practical situations. Contrariwise, the inverse Gamma (IG) distribution is mathematically tractable and also exhibits the semi heavy-tailed characteristics of both lognormal and Gamma distribution. As a result, researchers have chosen to design composite fading models for generalized fading channels like $\kappa-\mu$ \cite{yoo2015k} or $\eta-\mu$ \cite{yoo2015eta} fading channels employing the IG distribution. In this paper, we perform a PLS analysis of a relay-based multicast network in the presence of multiple eavesdroppers with the consideration of compound fading channels.

\subsection{Literature Survey}

Yacoub first uncovered the potentials of generalized fading distributions and proposed a set of generalized fading schemes, namely, $\kappa-\mu$, $\eta-\mu$, $\alpha-\mu$, and $\lambda-\mu$ fading models \cite{yacoub2007kappa,yacoub2007alpha, fraidenraich2003spl}. Since then, researchers across the globe have examined various other forms of compound fading channels based on these distributions. Some of the closely related and representative variants of the distributions above are $\alpha-\eta-\mu$, $\alpha-\kappa-\mu$, $\alpha-\eta-\kappa-\mu$, $\alpha-\lambda-\mu-\eta$ fading channels \cite{salahat2014performance, yacoub2016alpha, papazafeiropoulos2009eta}. Researchers have investigated various statistical characterizations, such as probability density function (PDF) and cumulative distribution function (CDF) of the compound fading channels and their variants under different wireless settings, including fading parameter variation \cite{issaid2017fast,9230592,yoo2020effective,pena2014performance}. As evident in their works, one can more precisely analyze the performances of many wireless systems and networks of interest; in terms of channel capacity (CC), outage probability (OP), coverage probability (CP), and bit error rate (BER). For example, the compound fading approaches with or without shadowing have been applied to analyze advanced wireless paradigms, including D2D, amplify and forward, decode and forward, 5G radio access, and optical communication systems \cite{9330523,xu2021performance,gupta2018performance}. As known, PLS can secure information by exploiting the randomness of the wireless channel \cite{nafis2021}. Generalized fading channels and their variants are therefore also being considered in the analysis of PLS of wireless systems and different PLS performance metrics such as secure outage probability (SOP), average secrecy capacity (ASC), and strictly positive secrecy capacity (SPSC) are evaluated \cite{ibrahim2021enhancing, yadav2021comprehensive,a.s.m.physical2021,sumona2021security}.

As mentioned in the early part of this section, IG is mathematically more tractable. As such, researchers are recently paying attention to the IG shadowing-based composite fading models. The effects of multipath and shadowing on the SER performance over $\eta-\mu$ / IG and $\kappa-\mu$ / IG  distribution were analyzed in \cite{sofotasios2018error}; the authors therein derived CFEs for the SER. The expression for the PDF, higher-order moments (HOM), and amount of fading (AF) considering the $\kappa-\mu$ / IG and $\eta-\mu$ / IG composite fading models were derived in \cite{yoo2017kappa}. Here the authors take into account the line of sight (LOS) and non-LOS (NLOS) channel conditions. The performance of wireless communication systems employing  MRC diversity and coherent modulation techniques was investigated over $\eta-\mu$ / IG composite fading channel in \cite{rana2017novel} by deriving expressions for SER. The novel analytical expressions for the envelope and power PDF and OP were derived under $\eta-\mu$/ inverse Gaussian fading channels in \cite{sofotasios2013eta} to show the impacts of multipath and shadowing characteristics. In \cite{yoo2015eta}, the feature of the shadowed fading behavior was inquired over the $\eta-\mu$/ IG composite fading model. Using Kullback-Leibler divergence, the authors claimed that the $\eta-\mu$/ IG compound fading channels demonstrate a better suit than the $\kappa-\mu$/ IG composite fading model. A common approach of characterizing the composite fading distributions based on IG shadowing was introduced by \cite{ramirez2019utility}, where the authors showed the formation of statistical characterization by presenting a composite IG/two-wave with diffuse power fading model. Performance of the efficient capacity (EC) was explored in \cite{ yoo2020effective} considering the different multipath fading, delay constraint, and shadowing conditions. In \cite{sofotasios2018capacity}, EC was investigated over $\eta-\mu$/ IG composite fading channel and the authors exhibited the influence of multipath fading and shadowing severity. ASEP, channel inversion with fixed rate (CIFR), channel capacity under optimal rate adaption (ORA), and truncated CIFR (TIFR) were derived in \cite{pant2019error} for analyzing the CC and error probability of wireless model over IG shadowed fading channel.

\subsection{Motivation}

 {\color{black}It is noteworthy that the aforesaid researches are confined to the analysis of the secrecy performance of wireless networks either in unicast or broadcast scenarios. In many situations, especially in dense networks, however, multicasting is preferred over broadcasting as it delivers the packet to intended recipients only. Hence security issue is more important in a multicast network rather than broadcast one. Moreover, it offers better utilization of bandwidth and better congestion control. On the other hand, whereas relays are usually placed as part of planned network installations or the users experiencing better channel conditions are assigned as relays, the locations of the mobile users are more random and even sparse. Also, the users practically experience shadowing phenomena. Consequently, although the channels between source and relays can be modeled as $\eta-\mu$ and $\kappa-\mu$ distributions, the channels between relays and the end-users might be a better fit with IG shadowing-based fading. On that, a secrecy analysis over $\eta-\mu$ and $\eta-\mu$/ IG, $\kappa-\mu$ and $\kappa-\mu$/ IG dual-hop link for a multicast framework are worth investigating. In this study, we are thus motivated to mathematically address the security of cooperative multicast relay networks over $\eta-\mu$ and $\eta-\mu$/ IG, $\kappa-\mu$ and $\kappa-\mu$/ IG composite fading channels in the presence of a set of eavesdroppers.}

\subsection{Contributions}

The specific contributions of this paper are as follows:

\begin{itemize}

    \item We first individually perform the statistical characterizations of the SNR at each hop and realize the associated PDF expressions. Then the PDF and CDF of the dual-hop end-to-end SNR are derived. In addition to dual-hop channel models under two distinct scenarios, our analysis considers the multicast channel model and eavesdropper channel model by using order statistics.
   
    \item Secondly, we study the secrecy capacity of the relay-based multicast system and derive the closed-form expressions for secure outage probability for multicasting (SOPM), probability of non-zero secrecy multicast capacity (PNSMC), and ergodic secrecy multicast capacity (ESMC). {\color{black}Please note that these expressions are completely novel as no literature has considered the combination of $\eta-\mu$ and $\eta-\mu$/ IG, $\kappa-\mu$ and $\kappa-\mu$/ IG with multiple receivers as well as eavesdroppers. Moreover, these expressions can be used to replicate the results of \cite{gao2016physical, badrudduza2019enhancement,sarkar2009secure, shabab2019enhancement, badrudduza2019performance} just by changing the parameter values as given in \cite[Table 1]{yoo2017kappa}.}
    
    \item Then, we illustrate the results based on the derived expressions of SOPM, PNSMC, and ESMC. The performances of our dual-hop cooperative relay system are in agreement with the secrecy performance of various existing models.
    
    \item Also, we perform Monte-Carlo simulations and conduct a thorough numerical evaluation. Our results, which focus on SOPM, PNSMC, and ESMC, demonstrate the feasibility of adopting PLS in cooperative multicast networks with realistic composite fading. The results show that the secrecy performance enhances with fading and shadowing parameters as well as the average SNR of the main channel. In contrast, secrecy performance degrades with the target secrecy rate, number of receivers, eavesdroppers, and average SNR of eavesdropper channel.
    
\end{itemize}

\subsection{Organization}

The rest of the paper is embodied as follows. The system model is shown in section II. The channel models for both LOS and NLOS scenarios are presented in Sections III-VI. The expressions of the performance metrics are derived in Sections VII, VIII, and IX. Section X illustrates the numerical results and finally, section XI draws the ending remarks of this research.

\section{System Model} 

\begin{figure*}[!ht]
\vspace{0mm}
    \centerline{\includegraphics[width=0.9\textwidth,angle =0]{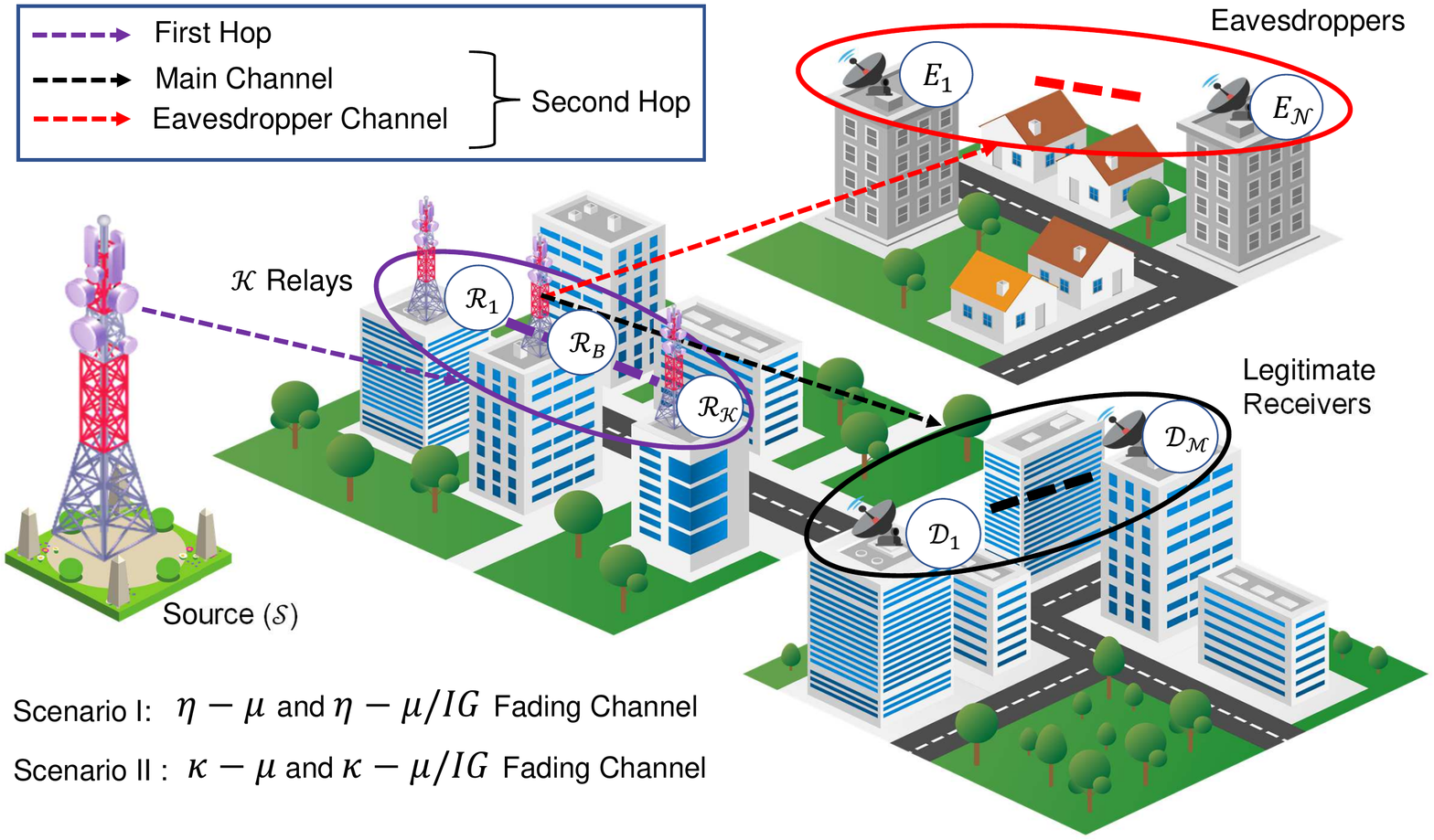}}
        \vspace{0mm }
    \caption{
  Proposed multicast scenario with opportunistic relaying.
    }
    \label{fig:1}
\end{figure*}
The proposed system network is depicted in Fig \ref{fig:1}, where a single antenna source, $\mathcal{S}$ dispatches secret messages to a group of $\mathcal{M}$ legitimate receivers via $\mathcal{K}$ relays. A bunch of $\mathcal{N}$ eavesdroppers are also present in that network which is presumed to be passive and attempting to decode the private information. But we are intended to defend the eavesdroppers from decoding the transmitted message. Here every relay is furnished with a single antenna, contrariwise each receiver and each eavesdropper is involved with an antenna, respectively. The distances between the source and the receivers as well as the eavesdroppers are considered very large and hence no direct link exists from source to receivers and from source to eavesdroppers. Hence the total communication occurs only through the relays. Here the channels between $\mathcal{S}$ and $\mathcal{M}$ are notified as main channels and the channels between $\mathcal{S}$ to $\mathcal{N}$ are indicated as eavesdropper channels. Entire communication occurs at two different phases. In the first phase, the source transmits information to the relays. Then each relay contends to be the best relay and receivers receive the signal from only the best relay which is considered as the second phase. 
We consider two scenarios depending on the LOS and NLOS communication links. In scenario I, we assume all the links of the first and second hops are NLOS. The channels between source to relays i.e $\mathcal{S}\rightarrow \mathcal{K}$ links are assumed to undergo $\eta-\mu$ fading. On the other hand relays to desire receivers i.e $\mathcal{K}\rightarrow \mathcal{M}$  links undergo $\eta-\mu$ / IG fading. When all the communication links are LOS i.e. scenario II, we assume the $\mathcal{S}\rightarrow \mathcal{K}$ links undergo $\kappa-\mu$ fading whereas the $\mathcal{K}\rightarrow \mathcal{M}$ and $\mathcal{K}\rightarrow \mathcal{N}$ links undergo $\kappa-\mu$ / IG fading.

The channel coefficients between $\mathcal{S}$ and $p$\textit{th} $(p=1,2,3,\cdots, \mathcal{K})$ link is denoted by $\mathbf{a}_{p,j} \in \mathbb{C}^{1\times 1}$, $\mathcal{K}$ and $m$\textit{th} $(m=1,2,3,\cdots, \mathcal{M})$ link is notified by $\mathbf{b}_{m,j} \in \mathbb{C}^{1\times 1}$ , $\mathcal{K}$ and $t$\textit{th} $(t=1,2,3,\cdots, \mathcal{N})$ link is denoted by $\mathbf{c}_{t,j} \in \mathbb{C}^{1\times 1}$, respectively. Here $j\in[1,2]$ indicates the first and second scenarios, respectively. Hence the corresponding SNRs of $\mathcal{S}\rightarrow \mathcal{K}$, $\mathcal{K}\rightarrow \mathcal{M}$, and $\mathcal{K}\rightarrow \mathcal{N}$ links are respectively, given by $\Upsilon_{sp,j}=\frac{P_{s,j}}{N_{p,j}}\|a_{p,j}\|^{2}$, $\Upsilon_{pm,j}=\frac{P_{k,j}}{N_{m,j}}\|b_{m,j}\|^{2}$, and $\Upsilon_{pt,j}=\frac{P_{k,j}}{N_{t,j}}\|c_{t,j}\|^{2}$. Here $P_{s,j}$, and $P_{k,j}$ are the transmit powers from $\mathcal{S}$, and $\mathcal{K}$. The additive white Gaussian noise terms at $\mathcal{K}$, $\mathcal{M}$ and $\mathcal{N}$ are denoted by $N_{p,j}$, $N_{m,j}$, and $N_{t,j}$, respectively.

\section{Channel Model (Scenario I)}

\subsection{PDFs of SNRs for $\mathcal{S}\rightarrow \mathcal{K}$, $\mathcal{K}\rightarrow \mathcal{M}$ and $\mathcal{K}\rightarrow \mathcal{N}$ Links}

The PDF of SNR for $\mathcal{S}\rightarrow \mathcal{K}$ link denoted as $\Upsilon_{sp,1}$ is given by  \cite[eq.~2]{peppas2013performance}
\begin{align}\label{eqn:sp1}
f_{sp,1}(\Upsilon)&=\alpha_{1}\Upsilon^{\mu_{p1}-0.5}e^{-\beta_{s}\Upsilon}I_{\mu_{p1}-0.5}(\varepsilon_{1}\Upsilon),
\end{align}   
where
$\alpha_{1}=\frac{2\sqrt\pi\mu_{p1}^{\mu_{p1}+0.5}h_{1}^{\mu_{p1}}}{\Gamma(\mu_{p1})H_{1}^{\mu_{p1}-0.5}\phi_{p1}^{\mu_{p1}+0.5}}$,
$h_{1}=\frac{1}{1-\eta_{p}^2}$, $H_{1}=\frac{\eta_{p}}{1-\eta_{p}^2}$, $\varepsilon_{1}=\frac{2\mu_{p1}H_{1}}{\phi_{p1}}$,
$\beta_{s}=\frac{2\mu_{p1}h_{1}}{\phi_{p1}}$, $\mu_{p1}>$ 0 symbolizes the number of multipath clusters of the first hop, $\eta_{p}$ which is bounded in $-1<\eta_{p}<1$ demonstrates the correlation coefficient between the in-phase and quadrature scattered waves in each multi-path cluster of the first hop \cite{almaeeni2016error}, and $\phi_{p1}$ is the average SNR of $\mathcal{S}\rightarrow \mathcal{K}$ link and $I_{\nu}(.)$ is the modified Bessel function  which is defined in \cite[eq~8.440]{GR:07:Book}. Simplifying \eqref{eqn:sp1} using  \cite[eq~8.445]{GR:07:Book}, the PDF of $\Upsilon_{sp,1}$ can be expressed as
\begin{align}\label{eqn:msp1}
f_{sp,1}(\Upsilon)=\sum _{n_{1}=0}^{\infty}\alpha_{2}\Upsilon^{2\mu_{p1}-1+2n_{1}}e^{-\beta_{s}\Upsilon},
\end{align} 
where $\alpha_{2}=\frac{a_{1}}{n_{1}!\Gamma(\mu_{p1}+0.5+n_{1})}(\frac{\varepsilon_{1}}{2})^{\mu_{p1}-0.5+2n_{1}}$. Now the PDF of  $\Upsilon_{pm,1}$ is given by \cite[eq.~ 13]{yoo2017kappa}
\begin{align}\label{eqn:mp1}
\nonumber
f_{pm,1}(\Upsilon)&=\frac{2^{2 \mu _{m}}\mu_{m1}^{2\mu_{m1}}}{\left(m_{m1}\phi_{m1}+2h_{2}\mu_{m1}\Upsilon\right){}^{m_{m1}+2\mu_{m1}}}
\\
\nonumber
&\times \frac{\left(m_{m1}\phi_{m1}\right){}^{m_{m1}}h_{2}^{\mu _{m}}\Upsilon^{2\mu _{m}-1}}{B\left(m_{m1},2\mu_{m1}\right)} \\
&\times \,_2F_{1}\left(\epsilon_{1},\epsilon_{2};\epsilon_{3};\left(\frac{2H_{2}\mu_{m1}\Upsilon}{m_{m1}\phi_{m1}+2h_{2}\mu_{m1} \Upsilon}\right)^2\right),
\end{align} 
where
$h_{2}=\frac{1}{1-\eta_{m}^2}$, $H_{2}=\frac{\eta_{m}}{1-\eta_{m}^2}$, $\epsilon_{1}=\frac{1}{2}\left(m_{m1}+2\mu_{m1}\right)$, $\epsilon_{2}=\frac{1}{2}\left(m_{m1}+2\mu_{m1}+1\right)$, $\epsilon_{3}=\frac{1}{2}\left(2\mu_{m1}+1\right)$, $\mu_{m1}>$ 0 is related to the number of multipath clusters of the $\mathcal{K}\rightarrow \mathcal{M}$ link and $\eta_{m}$ is concerned with the scattered wave power ratio between the in-phase and quadrature components of each multipath cluster of the $\mathcal{K}\rightarrow \mathcal{M}$ link, B($\cdot$, $\cdot$) denotes the Beta function as defined in \cite[eq~ 8.384.1]{GR:07:Book}, and $_2F_1$($\cdot$, $\cdot$; $\cdot$; $\cdot$) notifies the Gaussian hyper-geometric function as defined in  \cite[eq~ 9.111]{GR:07:Book}. Using \cite[eq~ 9.14.1]{GR:07:Book} and modifying \eqref{eqn:mp1} accordingly, the PDF of $\Upsilon_{pm,1}$ can be expressed as
\begin{align}\label{eqn:mmp1}
f_{pm,1}(\Upsilon)&=\sum _{r_{1}=0}^{\infty}\frac{\lambda _{2}\Upsilon^{2\mu_{m1}+2r_{1}-1}}{\left(1+\beta_{m}\Upsilon\right){}^{m_{m1}+2\mu_{m1}+2r _{1}}},
\end{align}
where
$\beta_{m}=\frac{2h_{2}\mu_{m1}}{m_{m1}\phi_{m1}}$, $\phi_{m1}$ is the average SNR of $\mathcal{K}$ to   $\mathcal{M}$ link, $\lambda_{2}=\frac{\lambda_{1}\left(\epsilon_{1}\right)_{r_{1}}\left(\epsilon_{2}\right)_{r_{1}}\left(2H_{2}\mu_{m1}\right){}^{2r_{1}}}{r_{1}! \left(\epsilon_{3}\right)_{r_{1}}\left(m_{m1}\phi_{m1}\right){}^{m_{m1}+2\mu_{m1}+2r_{1}}}$, and $\lambda_{1}=\frac{2^{2\mu_{m1}} \mu_{m1}^{2\mu_{m1}}\left(m_{m1}\phi_{m1}\right){}^{m_{m1}}h_{2}^{\mu_{m1}}}{B\left(m_{m1},2\mu_{m1}\right)}$,
Similar to \eqref{eqn:mmp1}, the PDF of $\Upsilon_{pt,1}$ can be expressed as 
\begin{align}\label{eqn:ep1}
f_{pt,1}(\Upsilon)&=\sum_{\nu_{1}=0}^{\infty}\frac{\xi_{2}\Upsilon^{2\mu_{t1}+2\upsilon_{1}-1}}{\left(1+\beta_{e}\Upsilon\right){}^{2\mu_{t1}+m_{t1}+2 \upsilon_{1}}},
\end{align}
where
$\beta_{e}=\frac{2h_{3}\mu_{t1}}{m_{t1}\phi_{t1}}$, $\phi_{t1}$ is the average SNR of  $\mathcal{K}\rightarrow\mathcal{N}$ link, $\mu_{t1}>$ 0 symbolizes the number of multipath clusters of the $\mathcal{K}\rightarrow \mathcal{N}$ link and $\eta_{t}$ is concerned with the scattered wave power ratio between the in-phase and quadrature components of each multipath cluster of the $\mathcal{K}\rightarrow \mathcal{N}$ link, 
$\xi_{2}=\frac{\xi_{1}\left(\kappa_{1}\right)_{\upsilon_{1}}\left(\kappa_{2}\right)_{\upsilon_{1}} \left(2H_{3}\mu_{t1}\right){}^{2\upsilon_{1}}}{\upsilon_{1}!\left(\kappa_{3}\right)_{\upsilon_{1}} \left(m_{t1}\phi_{t1}\right){}^{2\mu_{t1}+m_{t1}+2\upsilon_{1}}}$, $\xi_{1}=\frac{2^{2\mu_{t1}}\mu_{t1}^{2\mu_{t1}}h_{3}^{\mu_{t1}}\left(m_{t1}\phi_{t1}\right){}^{m_{t1}}}{B\left(m_{t1} ,2\mu_{t1}\right)}$, $h_{3}=\frac{1}{1-\eta_{t}^2}$, $H_{3}=\frac{\eta_{t}}{1-\eta_{t}^2}$, $\kappa _{1}=\frac{1}{2}\left(2\mu_{t1}+m_{t1}\right)$, $\kappa_{2}=\frac{1}{2}\left(2\mu_{t1}+m_{t1}+1\right)$,  
and $\kappa_{3}=\frac{1}{2} \left(2\mu_{t1}+1\right)$.

\subsection{PDFs of Dual-hop SNRs}

Denoting the SNR of $\mathcal{S}\rightarrow \mathcal{K}\rightarrow\mathcal{M}$ link by $\Upsilon_{sm,1}$, its PDF can be obtained as
\begin{align}\label{eqn:theorem 1}
\nonumber
f_{sm,1}(\Upsilon)&=\sum _{r_{2}=0}^{\infty}\sum _{r_{1}=0}^{\infty}\sum_{n_{1}=0}^{\infty}\Lambda_{2}\biggl[  \Upsilon^{-n_{2}+2\mu_{p1}+2n_{1}}e^{-\beta_{s}\Upsilon}
\\
&+\Upsilon^{-n_{2}}\Gamma(2\mu_{p1}+2n_{1},\beta_{s}\Upsilon)\biggl],
\end{align}
where
 $n_{2}=m_{m1}+r_{2}+1$.

\quad

\textit{Proof}: See Appendix \ref{dual1}.

\quad


Similarly, by denoting the SNR of $\mathcal{S}\rightarrow \mathcal{K}\rightarrow \mathcal{N}$ link by $\Upsilon_{st,1}$, its PDF can be given by
\begin{align}\label{eqn:theorem 2}
\nonumber
f_{st,1}(\Upsilon)&=\sum_{\nu_{2}=0}^{\infty}\sum_{\nu_{1}=0}^{\infty}\sum _{n_{1}=0}^{\infty}\chi_{2} \biggl[\Upsilon^{-n_{3}+2\mu_{p1}+2n_{1}}e^{-\beta_{s}\Upsilon}
\\
&+\Upsilon^{-n_{3}}\Gamma(2\mu_{p1}+2n_{1},\beta_{s}\Upsilon)\biggl],
\end{align} 
where
$n_{3}=m_{t1}+\nu_{2}+1$. 

\quad

\textit{Proof}: See Appendix \ref{dual2}.


\subsection{PDFs of SNRs for the Best Relay}

Let,$\Upsilon_{bm,1}$ denotes the SNR between the best relay, and $m$\textit{th} receiver which is defined as
\begin{align}\label{eqn:brelay1}
\Upsilon_{bm,1}=arg_{p\varepsilon\varsigma}^{max}\min(\Upsilon_{sp,1},\Upsilon_{pm,1}),
\end{align}
where
$\varsigma$ = (1,2,\ldots,$\mathcal{K}$) is the relay set. 
The PDF of $\Upsilon_{bm,1}$ can be obtained as the following.

\begin{align}\label{eqn:theorem 3}
\nonumber
&f_{bm,1}(\Upsilon)=\sum _{r_{2}=0}^{\infty}\sum _{r_{1}=0}^{\infty}\sum_{n_{1}=0}^{\infty}\Lambda_{2}\mathcal{K}\biggl[\Upsilon^{-n_{2}+2\mu_{p1}+2n_{1}}e^{-\beta_{s}\Upsilon}
\\
\nonumber 
&+\Upsilon^{-n_{2}}\Gamma(2\mu_{p1}+2n_{1},\beta_{s}\Upsilon)\biggl]
\\
&\times\biggl[1-\sum _{r_{2}=0}^{\infty}\sum_{r_{1}=0}^{\infty}\sum _{n_{1}=0}^{\infty}\Lambda_{2}\Upsilon^{-m_{m1}-r_{2}}\Gamma(2\mu_{p1}+2n_{1},\beta_{s}\Upsilon)\biggl]^{\mathcal{K}-1}. 
\end{align}

\quad

\textit{Proof}: See Appendix \ref{best1}.

\quad

Similar to \eqref{eqn:brelay1}, notifying the SNR between the best relay and the $t$\textit{th} eavesdropper by $\Upsilon_{bt,1}$, we have
\begin{align}\label{eqn:brelay2}
\Upsilon_{bt,1}=arg_{p\varepsilon\varsigma}^{max}\min(\Upsilon_{sp,1},\Upsilon_{pt,1}).
\end{align}
The PDF $\Upsilon_{bt,1}$ can be obtained as the following.
\begin{align}\label{eqn:theorem 4}
\nonumber
&f_{bt,1}(\Upsilon)=\sum_{\nu_{2}=0}^{\infty}\sum_{\nu_{1}=0}^{\infty}\sum _{n_{1}=0}^{\infty}\chi_{2}\mathcal{K} \biggl[\Upsilon^{-n_{3}+2\mu_{p1}+2n_{1}}e^{-\beta_{s}\Upsilon} 
\\
\nonumber 
&+\Upsilon^{-n_{3}}\Gamma(2\mu_{p1}+2n_{1},\beta_{s}\Upsilon)\biggl]
\\
&\times\biggl[1-\sum_{\nu_{2}=0}^{\infty}\sum_{\nu_{1}=0}^{\infty}\sum _{n_{1}=0}^{\infty}\chi_{2}\Upsilon^{-m_{t1}-\nu_{2}}\Gamma(2\mu_{p1}+2n_{1},\beta_{s}\Upsilon)\biggl]^{\mathcal{K}-1}.
\end{align}

\quad

\textit{Proof}: See Appendix \ref{best2}.


\section{Channel Model (Scenario II)}

\subsection{PDFs of SNRs for $\mathcal{S}\rightarrow \mathcal{K}$, $\mathcal{K}\rightarrow \mathcal{M}$ and $\mathcal{K}\rightarrow \mathcal{N}$ Links }

The PDF of $\Upsilon_{sp,2}$  is given by  \cite[Eq.~(2)]{bhargav2016secrecy}      
\begin{align}\label{eqn:sp2} 
\nonumber 
f_{sp,2}(\Upsilon)&=\frac{\mu_{p2}(1+\kappa_{p})^{\frac{\mu_{p2}+1}{2}}\Upsilon^{\frac{\mu_{p2}-1}{2}}e^{\frac{-\mu_{p2}(1+\kappa_{p})\Upsilon}{\phi_{p2}}}}{\kappa_{p}^{\frac{\mu_{p2}-1}{2}}\phi_{p2}^{\frac{\mu_{p2}+1}{2}}e^{\mu_{p2}\kappa_{p}}}
\\
&\times I_{\mu_{p2}-1}\left(2\mu_{p2}\sqrt{\frac{\kappa_{p}(1+\kappa_{p})\Upsilon}{\phi_{p2}}}\right),
\end{align}
where $\kappa_{p} > 0$ is the ratio of the total power of the dominant components to the total power of the scattered waves, $\mu_{p2} > 0 $ is related to the number of multipath clusters that is given by $\mu_{p2}=\frac{\mathbb{E}^{2}(\Upsilon_{sp,2})(1+2\kappa_{p})}{\mathbb{V}(\Upsilon_{sp,2})(1+\kappa_{p})^{2}}$ \cite{bhargav2016secrecy}, where $\mathbb{E}(\cdot)$ and $\mathbb{V}(\cdot)$ indicates the expectation and variance operators, apart from that $\phi_{p2}=\mathbb{E}(\Upsilon_{sp,2})$, is the average signal-to-noise-ratio (SNR) of $\mathcal{S}\rightarrow \mathcal{K}$ link  and $I_{n}(\cdot)$ is the modified Bessel function of the first kind and order $n$. 
Simplifying \eqref{eqn:sp2} by using \cite[eq.~ 8.445]{GR:07:Book}, the PDF of $\Upsilon_{sp,2}$ can be evaluated as 
\begin{align}\label{eqn:msp2} 
f_{sp,2}(\Upsilon)&=\sum_{\rho_{1}=0}^{\infty}\delta_{2}\Upsilon^{\mu_{p2}-1+\rho_{1}}e^{\frac{-\mu_{p2}(1+\kappa_{p})\Upsilon}{\phi_{p2}}},
\end{align}
where $\delta_{2}=\delta_{1}\frac{\mu_{p2}^{\mu_{p2}-1+2\rho_{1}}(\frac{\kappa_{p}(1+\kappa_{p})}{\phi_{p2}})^{\frac{\mu_{p2}-1+2\rho_{1}}{2}}}{\rho_{1}!\Gamma(\mu_{p2}+\rho_{1})}$, $\delta_{1}=\frac{\mu_{p2}(1+\kappa_{p})^{\frac{\mu_{p2}+1}{2}}}{\kappa_{p}^{\frac{\mu_{p2}-1}{2}}\phi_{p2}^{\frac{\mu_{p2}+1}{2}}e^{\mu_{p2}\kappa_{p}}}$, and $\Gamma(\cdot)$ denotes the Gamma function \cite{GR:07:Book}. 
Now the PDF of $\Upsilon_{pm,2}$ is given by \cite[Eq.~(4)]{yoo2017kappa}
\begin{align}\label{eqn:mPDF1}
\nonumber
&f_{pm,2}(\Upsilon)=\frac{e^{-\mu_{m2}\kappa_{m}}\mu_{m2}^{\mu_{m2}}}{[\mu_{m2}(\kappa_{m}+1)\Upsilon+m_{m2}\phi_{m2}]^{m_{m2}+\mu_{m2}}}
\\
\nonumber
&\times \frac{(\kappa_{m}+1)^{\mu_{m2}}(m_{m2}\phi_{m2})^{m_{m2}}\Upsilon^{\mu_{m2}-1}}{B(m_{m2},\mu_{m2})}
\\
&\times\, {}_{1}F_{1}\left(m_{m2}+\mu_{m2};\mu_{m2};\frac{\mu_{m2}^{2}\kappa_{m}(\kappa_{m}+1)\Upsilon}{\mu_{m2}(\kappa_{m}+1)\Upsilon+m_{m2}\phi_{m2}}\right),
\end{align}
where $\phi_{m2}$ is the average SNR of $\mathcal{K}\rightarrow \mathcal{M}$ links, B($\cdot$, $\cdot$) and ${}_{1}F_{1}$($\cdot$; $\cdot$; $\cdot$) denotes the Beta and the Confluent Hypergeometric functions as defined in \cite[eq.~ 9.210.1]{GR:07:Book}, and $(j_{1})_{p_{1}}=\frac{\Gamma(j_{1}+p_{1})}{\Gamma(j_{1})}$ denotes the Pochhammer symbol. Using \cite[eq.~ 9.14.1]{GR:07:Book}, \eqref{eqn:mPDF1} can be simplified as 
\begin{align}\label{eqn:mmp2} 
f_{pm,2}(\Upsilon)=\sum _{\rho _2=0}^\infty  \frac{\delta_{5}\Upsilon^{\rho_{2}+\mu_{m2}-1}}{(1+\delta_{4}\Upsilon)^{m_{m2}+\mu_{m2}+\rho_{2}}}, 
\end{align}
where $\delta _5=\delta _3\frac{\left(\kappa _m \left(\kappa _m+1\right)\mu_{m2}^2\right){}^{\rho _2} \left(m_{m2}+\mu_{m2}\right)_{\rho _2}}{\rho _2! \left(\mu_{m2}\right)_{\rho _2}}$, 
$\delta_{4}= \frac{\kappa _m \mu_{m2}+\mu_{m2}}{m_{m2}\Phi_{m2}}$, and 
$\delta_{3}= \frac{\mu_{m2}^{\mu_{m2}} \left(m_{m2}\Phi_{m2}\right){}^m_{m2} e^{\kappa _m \left(-\mu_{m2}\right)} \left(\kappa _m+1\right){}^{\mu_{m2}}}{B\left(m_{m2},\mu_{m2}\right) \left(m_{m2} \Phi_{m2}\right){}^{m_{m2}+\mu_{m2}+\rho _2}}$.
Similarly, the PDF of $\Upsilon_{pt,2}$ for $\mathcal{K}\rightarrow \mathcal{N}$ link can be illustrated as
\begin{align}\label{eqn:ep2} 
f_{pt,2}(\Upsilon)=\sum _{\sigma _2=0}^\infty \frac{\beta_{5}\Upsilon^{\sigma_{2}+\mu_{t2}-1}}{(1+\beta_{4}\Upsilon)^{m_{t2}+\mu_{t2}+\sigma_{2}}},
\end{align}
where $\beta_{3}=\beta_{1}\frac{\left(\kappa_{t} \left(\kappa_{t}+1\right) \mu_{t2}^2\right){}^{\sigma _2} \left(m_{t2}+\mu_{t2}\right)_{\sigma _2}}{\sigma _2! \left(\mu_{t2}\right)_{\sigma _2}}$, 
$\beta_{2}= \frac{\kappa_{t} \mu_{t2}+\mu_{t2}}{m_{t2}\Phi_{t2}}$, and $\beta_{1}= \frac{\mu_{t2}^{\mu_{t2}} \left(m_{t2}\Phi_{t2}\right){}^{m_{t2}} e^{\kappa_{t} \left(-\mu_{t2}\right)} \left(\kappa_{t}+1\right){}^{\mu_{t2}}}{B\left(m_{t2},\mu_{t2}\right) \left(m_{t2}\Phi_{t2}\right){}^{m_{t2}+\mu_{t2}+\sigma _2}}$.

\subsection{PDF of Dual-Hop SNRs}
Denoting $\Upsilon_{sm,2}$ as the  SNR of $\mathcal{S}\rightarrow \mathcal{K}\rightarrow\mathcal{M}$ link, the PDF of
$\Upsilon_{sm,2}$ can be obtained as the following.

\begin{align}\label{eqn:theorem 5}
\nonumber 
f_{sm,2}&(\Upsilon)=\sum_{\rho _3=0}^\infty\sum_{\rho _2=0}^\infty\sum_{\rho_{1}=0}^{\infty}\delta_{7}\biggl[\Upsilon^{-\rho_{3}-m_{m2}-1+\mu_{p2}+\rho_{1}}
\\
&\times e^{-\Upsilon\delta_{b_{1}}}+\Upsilon^{-1-\rho_{3}-m_{m2}}\Gamma[\mu_{p2}+\rho_{1},\Upsilon\delta_{b_{1}}]\biggl].
\end{align}

\quad

\textit{Proof}: See Appendix \ref{dual.s2.5}.

\quad

Similarly, denoting $\Upsilon_{st,2}$ as the  SNR of $\mathcal{S}\rightarrow \mathcal{K}\rightarrow \mathcal{N}$ link, the PDF of
$\Upsilon_{st,2}$ can be obtained as the following.

\begin{align}\label{eqn:theorem 6}
\nonumber 
f_{st,2}(\Upsilon)=&\sum_{\sigma _3=0}^\infty\sum_{\sigma _2=0}^\infty\sum_{\rho_{1}=0}^{\infty}\beta_{7}\biggl[e^{-\Upsilon\delta_{b_{1}}}\Upsilon^{-\sigma_{3}-m_{t2}-1+\mu_{p2}+\sigma_{1}}
\\
&+\Upsilon^{-1-\sigma_{3}-m_{t2}}\Gamma[\mu_{p2}+\rho_{1},\Upsilon\delta_{b_{1}}]\biggl].
\end{align}

\quad

\textit{Proof}: See Appendix \ref{dual.s2.6}.

\subsection{PDFs of SNRs for best relay}

Let, $\Upsilon_{bm,2}$ denotes the SNR between the best relay and $m$\textit{th} receiver which is defined as
\begin{align}\label{eqn:brelay3}
    \Upsilon_{bm,2}=arg_{p\in\varsigma}^{max}min(\Upsilon_{sp,2},\Upsilon_{pm,2}).
\end{align}
The PDF of $\Upsilon_{bm,2}$ can be obtained as the following.

\begin{align}\label{eqn:theorem 7}
\nonumber
&f_{bm,2}(\Upsilon)=\sum_{\rho _3=0}^\infty\sum_{\rho _2=0}^\infty\sum_{\rho_{1}=0}^{\infty}\delta_{7}\mathcal{K}\biggl[\Upsilon^{-\rho_{3}-m_{m2}-1+\mu_{p2}+\rho_{1}}
\\
\nonumber 
&\times e^{-\Upsilon\delta_{b_{1}}}+\Upsilon^{-1-\rho_{3}-m_{m2}}\Gamma[\mu_{p2}+\rho_{1},\Upsilon\delta_{b_{1}}]\biggl]
\\
&\times\biggl[1-\sum_{\rho _3=0}^\infty\sum _{\rho _2=0}^\infty\sum_{\rho_{1}=0}^{\infty}\delta_{7}\Upsilon^{-m_{m2}-\rho_{3}}
\Gamma(\mu_{p2}+\rho_{1},\delta_{b_{1}}\Upsilon)\biggl]^{\mathcal{K}-1}.
\end{align}

\textit{Proof}: See Appendix \ref{dual5}.

\quad

Similar to \eqref{eqn:brelay3}, the SNR between best relay and $t\textit{th}$ eavesdropper is defined by $\Upsilon_{bt,2}$, we have
\begin{align}\label{eqn:brelay4}
    \Upsilon_{bt,2}=arg_{p\in\varsigma}^{max}min(\Upsilon_{sp,2},\Upsilon_{pt,2}).
\end{align}
The PDF of $\Upsilon_{bt,2}$ can be obtained as 
\begin{align}\label{eqn:theorem 8}
\nonumber
&f_{bt,2}(\Upsilon)=\sum_{\sigma _3=0}^\infty\sum_{\sigma _2=0}^\infty\sum_{\rho_{1}=0}^{\infty}\beta_{7}\mathcal{K}\biggl[\Upsilon^{-\sigma_{3}-m_{t2}-1+\mu_{p2}+\sigma_{1}}
\\
\nonumber 
&\times e^{-\Upsilon\delta_{b_{1}}}+\Upsilon^{-1-\sigma_{3}-m_{t2}}\Gamma[\mu_{p2}+\rho_{1},\Upsilon\delta_{b_{1}}]\biggl]
\\
&\times \biggl[1-\sum_{\sigma _3=0}^\infty\sum_{\sigma _2=0}^\infty\sum_{\rho_{1}=0}^{\infty}\beta_{7}\Upsilon^{-m_{t2}-\sigma_{3}}\Gamma(\mu_{p2}+\rho_{1},\delta_{b_{1}}\Upsilon)\biggl]^{\mathcal{K}-1}.
\end{align}
\textit{Proof}: See Appendix \ref{dual6}.

\section{Multicast Channel Model}
We consider multiple receivers in the proposed multicast system and the secrecy analysis is carried out by means of assuming the worst case of the multicast channels i.e. minimum SNR among $\mathcal{M}$ instantaneous SNRs at the legitimate receivers. The significance of this assumption is that if the security can be enhanced for the worst case, then we can clearly declare that the total system is also secure for any other possible cases of the multicast channels. Let $\sigma_{min,j}= min_{1\leqslant m\leqslant\mathcal{M}}\Upsilon_{bm,j}$ denotes the minimum SNR of the multicast channels. Since $\Upsilon_{b1,j}$,  $\Upsilon_{b2,j}$, $\Upsilon_{b3,j}$, \ldots, $\Upsilon_{b\mathcal{M},j}$ are independent, hence the PDF of $\sigma_{min,j}$ is defined using order statistics as \cite[Eq.~12]{badrudduza2019performance}
\begin{align}
\nonumber 
f_{\sigma _{min,j}}(\Upsilon_{bm,j})=\mathcal{M}f_{bm,j}(\Upsilon)[1-F_{bm,j}(\Upsilon)]^{\mathcal{M}-1}.
\end{align}
The following subsections include derivations of the expressions of $f_{\sigma _{min,1}}(\Upsilon)$ and $f_{\sigma _{min,2}}(\Upsilon)$ for the two considered scenarios.

\subsection{Scenario I}
The final expression of $f_{\sigma_{min,1}}(\Upsilon)$ can be derived as,
\label{tmin1}

\begin{align}\label{eqn:theorem 9}
\nonumber 
&f_{\sigma_{min,1}}(\Upsilon)=\sum_{\psi_{r_{6}}}\sum_{r_{6}=0}^{\mathcal{K}+\mathcal{K}r_{5}-1}\sum_{r_{5}=0}^{\mathcal{M}-1}\sum_{r_{2}=0}^{\infty}\sum _{r_{1}=0}^{\infty}\sum _{n_{1}=0}^{\infty}
\\
&\times\biggl(\alpha_{3}\Upsilon^{\alpha_{5}}+\sum_{r_{7}=0}^{2\mu_{p1}+2n_{1}-1}\alpha_{4}\Upsilon^{\alpha_{6}}\biggl)e^{-\beta_{4}\Upsilon},
\end{align}
where $\alpha_{3}=\Lambda_{2}\Lambda_{6}\Lambda_{9}\mathcal{M}\mathcal{K}$, $\alpha_{4}=\Lambda_{2}\Lambda_{6}\Lambda_{7}\Lambda_{9}\mathcal{M}\mathcal{K}$, $\alpha_{5}=-n_{2}+2\mu_{p1}+2n_{1}+\varphi_{\psi_{r_{6}}}$, $\alpha_{6}=-n_{2}+r_{7}+\varphi_{\psi_{r_{6}}}$, and $\beta_{4}=\beta_{s}+\phi_{\psi_{r_{6}}}$.

\quad

\textit{Proof}: See Appendix \ref{min1}.

\subsection{Scenario II}
Similarly, the final expression of $f_{\sigma _{min,2}}(\Upsilon_{bm,2})$ can be derived as
\label{tmin2}

\begin{align}\label{eqn:theorem 10}
\nonumber 
&f_{\sigma_{min,2}}(\Upsilon)=\sum_{\psi_{\rho_{7}}}\sum_{\rho_{7}=0}^{\mathcal{K}+\mathcal{K}\rho_{6}-1}\sum_{\rho_{6}=0}^{\mathcal{M}-1}\sum_{\rho _3=0}^\infty\sum_{\rho_2=0}^\infty\sum_{\rho_{1}=0}^{\infty}
\\
&\times\biggl(\varpi_{1}\Upsilon^{\varpi_{4}}+\sum_{\rho_{8}=0}^{\mu_{p2}+\rho_{1}-1}\varpi_{2}\Upsilon^{\varpi_{5}}\biggl)e^{-\Upsilon\varpi_{3}},
\end{align}
where
$\delta_{14}=\sum_{\psi_{\rho_{7}}}^{}\binom{\rho_{7}}{g_{0,0,0,0},\ldots,g_{\rho_{9},\rho_{3}\rho_{2},\rho_{1}},\ldots,n_{\mu_{p2}+\rho_{1}-1,\infty,\infty,\infty}}
\\
\chi_{\psi_{\rho_{7}}}$, $\varpi_{1}=\mathcal{M}\mathcal{K}\delta_{7}\delta_{10}\delta_{14}$,
$\varpi_{2}=\mathcal{M}\mathcal{K}\delta_{7}\delta_{10}\delta_{12}\delta_{14}$,
$\varpi_{3}=\delta_{b_{1}}+\Theta_{\psi_{\rho_{7}}}$,
$\varpi_{4}=\eta_{\psi_{\rho_{7}}}+\delta_{11}+\mu_{p2}+\rho_{1}$, and
$\varpi_{5}=\eta_{\psi_{\rho_{7}}}+\delta_{11}+\rho_{8}$.

\quad

\textit{Proof}: See Appendix \ref{min2}.

\section{Eavesdropper Channel Model}
Since multiple eavesdroppers exist in the proposed system, the worst secure condition in that particular case is obtained by maximizing the impacts of the eavesdroppers i.e. assuming the maximum SNR among $\mathcal{N}$ instantaneous SNRs at the eavesdropper terminals. Let $\sigma_{max,j}=max_{1\leqslant t\leqslant\mathcal{N}}\Upsilon_{bt,j}$ denotes the maximum SNR of the eavesdropper channels. Again, since $\Upsilon_{b1,j}$,  $\Upsilon_{b2,j}$,  $\Upsilon_{b3,j}$,\ldots,$\Upsilon_{b\mathcal{N},j}$ are independent, hence the PDF of $\sigma_{max,j}$ is defined as \cite[Eq.~15]{badrudduza2019performance} 
\begin{align}
\nonumber 
f_{\sigma _{max,j}}(\Upsilon)=\mathcal{N}f_{bt,j}(\Upsilon)[F_{bt,j}(\Upsilon)]^{\mathcal{N}-1}.
\end{align}

The following subsections illustrate the formation of the expressions for $f_{\sigma _{max,1}}(\Upsilon)$ and $f_{\sigma _{max,2}}(\Upsilon)$ for the both scenarios (I and II).

\subsection{Scenario I}
The final expression of $f_{\sigma_{max,1}}(\Upsilon)$ can be derived as,
\label{tmax1}
\begin{align}\label{eqn:theorem 11}
\nonumber 
&f_{\sigma_{max,1}}(\Upsilon)=\sum_{\psi_{\nu_{6}}}\sum_{\nu_{6}=0}^{\mathcal{K}\mathcal{N}-1}\sum_{\nu_{2}=0}^{\infty}\sum_{\nu_{1}=0}^{\infty}\sum _{n_{1}=0}^{\infty}
\\
&\times\biggl(\alpha_{10}\Upsilon^{\alpha_{12}}+\sum_{\nu_{7}=0}^{2\mu_{p1}+2n_{1}-1}\alpha_{11}\Upsilon^{\alpha_{13}}\biggl)e^{-\beta_{5}\Upsilon},
\end{align} 
where $\chi_{9}=\sum_{\psi_{\nu_{6}}}\binom{\nu_{6}}{q_{0,0,0,0,\ldots,}q_{\nu_{8},\nu_{2},\nu_{1},n_{1}.\ldots,}n_{2\mu_{p1}+2n_{1}-1,\infty,\infty,\infty}}
\\
\Xi_{\psi_{\nu_{6}}}$,$\alpha_{10}=\chi_{2}\chi_{6}\chi_{9}\mathcal{N}\mathcal{K}$, $\alpha_{11}=\chi_{2}\chi_{6}\chi_{7}\chi_{9}\mathcal{N}\mathcal{K}$, $\alpha_{12}=-n_{3}+2\mu_{p1}+2n_{1}+\varphi_{\psi_{\nu_{6}}}$, $\alpha_{13}=-n_{3}+\nu_{7}+\varphi_{\psi_{\nu_{6}}}$, and $\beta_{5}=\beta_{s}+\phi_{\psi_{\nu_{6}}}$.

\quad

\textit{Proof}: See Appendix \ref{max1}.

\subsection{Scenario II}
Similarly, the final expression of $f_{\sigma _{max,2}}(\Upsilon_{bt,2})$ can be derived as,
\label{tmax2}
\begin{align}\label{eqn:theorem 12}
\nonumber 
&f_{\sigma _{max,2}}(\Upsilon)=\sum_{\psi_{\sigma_{10}}}\sum_{\sigma_{10}=0}^{\mathcal{K}\mathcal{N}-1}\sum_{\sigma _3=0}^\infty\sum_{\sigma _2=0}^\infty\sum_{\rho_{1}=0}^{\infty}
\\
&\times\biggl(\omega_{1}\Upsilon^{\omega_{4}}+\sum_{\sigma_{11}=0}^{\mu_{p2}+\rho_{1}-1}\omega_{2}\Upsilon^{\omega_{5}}\biggl)e^{-\Upsilon\omega_{3}},
\end{align}
where
$\beta_{18}=\sum_{\psi_{\sigma_{10}}}\binom{\sigma_{10}}{f_{0,0,0,0},\ldots,f_{\sigma_{12},\sigma_{3},\sigma_{2},\rho_{1}},\ldots,n_{\mu+\rho_{1}-1,\infty,\infty,\infty}}
\\
\chi_{\psi_{\sigma_{10}}}$, $\omega_{1}=\mathcal{N}\mathcal{K}\beta_{7}\beta_{15}\beta_{18}$, $\omega_{2}=\mathcal{N}\mathcal{K}\beta_{7}\beta_{15}\beta_{16}\beta_{18}$, $\omega_{3}=\delta_{b_{1}}+\Theta_{\psi_{\sigma_{10}}}$,  
$\omega_{4}=\eta_{\psi_{\sigma_{10}}}+\beta_{10}+\mu_{p1}+\rho_{1}$, and $\omega_{5}=\eta_{\psi_{\sigma_{10}}}+\beta_{10}+\sigma_{11}$. 

\quad

\textit{Proof}: See Appendix \ref{max2}.

\setcounter{equation}{28}
\begin{figure*}[!t]
\begin{align}\label{eqn:sopmeta 3}
\nonumber
&P_{out}(\sigma_{s1})=1-\sum_{\psi_{r_{6}}}\sum_{r_{6}=0}^{\mathcal{K}+\mathcal{K}r_{5}-1}\sum_{r_{5}=0}^{\mathcal{M}-1}\sum_{r_{2}=0}^{\infty}\sum _{r_{1}=0}^{\infty}\sum _{n_{1}=0}^{\infty}\sum_{\psi_{\nu_{6}}}\sum_{\nu_{6}=0}^{\mathcal{K}\mathcal{N}-1}\sum_{\nu_{2}=0}^{\infty}\sum_{\nu_{1}=0}^{\infty}
\biggl[\sum_{r_{9}=0}^{\alpha_{5}}\sum_{r_{11}=0}^{r_{9}}\biggl(\frac{\Lambda_{10}\alpha_{10}(r_{11}+\alpha_{12})!}{(\beta_{5}+\beta_{4}q_{\zeta})^{r_{11}+\alpha_{12}+1}}+\sum_{\nu_{7}=0}^{2\mu_{p1}+2n_{1}-1} 
 \\
 &\frac{\Lambda_{10}\alpha_{11}(r_{11}+\alpha_{13})!}{(\beta_{5}+\beta_{4}q_{\zeta})^{r_{11}+\alpha_{13}+1}}\biggl)-\sum_{r_{10}=0}^{\alpha_{6}}\sum_{r_{12}=0}^{r_{10}}\sum_{r_{7}=0}^{2\mu_{p1}+2n_{1}-1}\biggl(\frac{\Lambda_{11}\alpha_{10}(r_{12}+\alpha_{12})!}{(\beta_{5}+\beta_{4}q_{\zeta})^{r_{12}+\alpha_{12}+1}}+\sum_{\nu_{7}=0}^{2\mu_{p1}+2n_{1}-1}\frac{\Lambda_{11}\alpha_{11}(r_{12}+\alpha_{13})!}{(\beta_{5}+\beta_{4}q_{\zeta})^{r_{12}+\alpha_{13}+1}}\biggl)\biggl].
\end{align}
\hrulefill
\end{figure*}
\setcounter{equation}{31}
\begin{figure*}[!t]
\begin{align}\label{eqn:sopmkappa 3}
\nonumber
&P_{out}(\sigma_{s2})=1-\sum_{\psi_{\rho_{7}}}\sum_{\rho_{7}=0}^{\mathcal{K}+\mathcal{K}\rho_{6}-1}\sum_{\rho_{6}=0}^{\mathcal{M}-1}\sum_{\rho _3=0}^\infty\sum_{\rho_2=0}^\infty\sum_{\rho_{1}=0}^{\infty}\sum_{\psi_{\sigma_{10}}}\sum_{\sigma_{10}=0}^{\mathcal{K}\mathcal{N}-1}\sum_{\sigma _3=0}^\infty\sum_{\sigma _2=0}^\infty\biggl[\sum_{\sigma_{13}=0}^{\varpi_{4}}\sum_{\sigma_{15}=0}^{\sigma_{13}}\biggl(\frac{\delta_{19}\omega_{1}(\sigma_{15}+\omega_{4})!}{(\varpi_{3}+\omega_{3}q_{\theta})^{\sigma_{15}+\omega_{4}+1}}
\\ 
&+\sum_{\sigma_{11}=0}^{\mu_{p2}+\rho_{1}-1}\frac{\delta_{19}\omega_{2}(\sigma_{15}+\omega_{5})!}{(\varpi_{3}+\omega_{3}q_{\theta})^{\sigma_{15}+\omega_{5}+1}}\biggl)-\sum_{\sigma_{16}=0}^{\varpi_{5}}\sum_{\sigma_{16}=0}^{\sigma_{14}}\sum_{\rho_{8}=0}^{\mu_{p2}+\rho_{1}-1}\biggl(\frac{\delta_{20}\omega_{1}(\sigma_{16}+\omega_{4})!}{(\varpi_{3}+\omega_{3}q_{\theta})^{\sigma_{16}+\omega_{4}+1}}+\sum_{\sigma_{11}=0}^{\mu_{p2}+\rho_{1}-1}\frac{\delta_{20}\omega_{2}(\sigma_{16}+\omega_{5})!}{(\varpi_{3}+\omega_{3}q_{\theta})^{\sigma_{16}+\omega_{5}+1}}\biggl)\biggl].
\end{align} 
\hrulefill
\end{figure*}

\section{Secure Outage Analysis}

Denoting $C_{s,j}$ as the instantaneous secrecy multicast rate \cite{wyner1975wire} and $\sigma_{sj}$ as the target secrecy rate, the SOPM is defined as \cite[Eq.~23]{badrudduza2019performance}
\begin{align}
\nonumber
&P_{out}(\sigma_{sj})=\Pr(C_{s,j}<\sigma_{sj})=1-\int_{0}^{\infty}\int_{\zeta}^{\infty}f_{\sigma_{min,j}}(\Upsilon_{bm,j})
\\
\nonumber
&\times f_{\sigma _{max,j}}(\Upsilon_{bt,j})d\Upsilon_{bm,j}d\Upsilon_{bt,j},
\end{align}
where
$\zeta=2^{\sigma_{sj}}(1+\Upsilon_{bt,j})-1$ and $\sigma_{sj}>0$. The definition indicates that, the secure transmission is effective only if $C_{s,j}>\sigma_{sj}$. Here $f_{\sigma_{min,j}}(\Upsilon_{bm,j})$ is obtained from  Eqs. \eqref{eqn:theorem 9}, and \eqref{eqn:theorem 10} and similarly, $f_{\sigma _{max,j}}(\Upsilon_{bt,j})$ is represented by Eqs. \eqref{eqn:theorem 11} and \eqref{eqn:theorem 12}, respectively.

\subsection{Scenario I}

In the case of scenario-I, SOPM is defined as 
\setcounter{eqnback}{\value{equation}} \setcounter{equation}{26}
\begin{align}\label{eqn:sopmeta 1}
\nonumber
&P_{out}(\sigma_{s1})=\Pr(C_{s,1}>\sigma_{s1})=1-
\\
&\int_{0}^{\infty}\int_{\zeta}^{\infty}f_{\sigma_{min,1}}(\Upsilon_{bm,1})f_{\sigma _{max,1}}(\Upsilon_{bt,1})
 d\Upsilon_{bm,1}d\Upsilon_{bt,1},
\end{align}
where
$\zeta=2^{\sigma_{s1}}(1+\Upsilon_{bt,1})-1$ and $\sigma_{s1}>0$.
Substituting \eqref{eqn:theorem 9}, and \eqref{eqn:theorem 11} into \eqref{eqn:sopmeta 1}, we have
\begin{align}\label{eqn:sopmeta 2}
\nonumber
&P_{out}(\sigma_{s1})=1-\sum_{\psi_{r_{6}}}\sum_{r_{6}=0}^{\mathcal{K}+\mathcal{K}r_{5}-1}\sum_{r_{5}=0}^{\mathcal{M}-1}\sum_{r_{2}=0}^{\infty}\sum _{r_{1}=0}^{\infty}\sum _{n_{1}=0}^{\infty}\int_{0}^{\infty}\int_{\zeta}^{\infty}
\\
\nonumber 
&\times \biggl(\alpha_{3}\Upsilon_{bm,1}^{\alpha_{5}}+\sum_{r_{7}=0}^{2\mu_{p1}+2n_{1}-1}\alpha_{4}\Upsilon_{bm,1}^{\alpha_{6}}\biggl)e^{-\beta_{4}\Upsilon_{bm,1}}
\sum_{\psi_{\nu_{6}}}\sum_{\nu_{6}=0}^{\mathcal{K}\mathcal{N}-1}
\\\nonumber 
&\times \sum_{\nu_{2}=0}^{\infty}\sum_{\nu_{1}=0}^{\infty} \sum _{n_{1}=0}^{\infty}\biggl(\alpha_{10}\Upsilon_{bt,1}^{\alpha_{12}}+\sum_{\nu_{7}=0}^{2\mu_{p1}+2n_{1}-1}\alpha_{11}\Upsilon_{bt,1}^{\alpha_{13}}\biggl)
\\
&\times e^{-\beta_{5}\Upsilon_{bt,1}}d\Upsilon_{bm,1}d\Upsilon_{bt,1}.
\end{align}

Now, executing integration using \cite[eq~ 3.351.2]{GR:07:Book}, the closed-form expression of SOPM is obtained in \ref{eqn:sopmeta 3}, where
$p_{\zeta}=2^{\sigma_{s1}}-1$, $q_{\zeta}=2^{\sigma_{s1}}$,
$\Lambda_{10}=\frac{\alpha_{3}\alpha_{5}!\binom{r_{9}}{r_{11}}p_{\zeta}^{r_{9}-r_{11}}q_{\zeta}^{r_{11}}}{r_{9}!(\beta_{4})^{\alpha_{5}-r_{9}+1}e^{\beta_{4}p_{\zeta}}}$, and
$\Lambda_{11}=\frac{\alpha_{4}\alpha_{6}!\binom{r_{10}}{r_{12}}p_{\zeta}^{r_{10}-r_{12}}q_{\zeta}^{r_{12}}}{r_{10}!(\beta_{4})^{\alpha_{6}-r_{10}+1}e^{\beta_{4}p_{\zeta}}}$.

\subsection{Scenario II}

The SOPM in the case of  scenario-II is defined as
\setcounter{eqnback}{\value{equation}} \setcounter{equation}{29}
\begin{align}\label{eqn:sopmkappa 1}
\nonumber
&P_{out}(\sigma_{s2})=\Pr(C_{s,2}>\sigma_{s2})=1-
\\
&\int_{0}^{\infty}\int_{\varrho}^{\infty}f_{\sigma_{min,2}}(\Upsilon_{bm,2})f_{\sigma_{max,2}}(\Upsilon_{bt,2})d(\Upsilon_{bm,2}) d(\Upsilon_{bt,2}),
\end{align}
where
$\varrho=2^{\sigma_{s2}}(1+\Upsilon_{bt,2})-1$ and $\sigma_{s2}>0$.
Substituting the values of  \eqref{eqn:theorem 10} and \eqref{eqn:theorem 12}
into \eqref{eqn:sopmkappa 1}, we have
\begin{align}\label{eqn:sopmkappa 2}
\nonumber
&P_{out}(\sigma_{s2})=1-\sum_{\psi_{\rho_{7}}}\sum_{\rho_{7}=0}^{\mathcal{K}+\mathcal{K}\rho_{6}-1}\sum_{\rho_{6}=0}^{\mathcal{M}-1}\sum_{\rho_3=0}^\infty\sum_{\rho_2=0}^\infty\sum_{\rho_{1}=0}^{\infty}\int_{0}^{\infty}\int_{\varrho}^{\infty}
\\
\nonumber 
&\times e^{-\Upsilon_{bm,2}\varpi_{3}}(\varpi_{1}\Upsilon_{bm,2}^{\varpi_{4}}+\varpi_{2}\Upsilon_{bm,2}^{\varpi_{5}})\sum_{\psi_{\sigma_{10}}}\sum_{\sigma_{10}=0}^{\mathcal{K}\mathcal{N}-1}\sum_{\sigma _3=0}^\infty\sum_{\sigma _2=0}^\infty
\\
&\times \sum_{\rho_{1}=0}^{\infty} e^{-\Upsilon_{bt,2}\omega_{3}}(\omega_{1}\Upsilon_{bt,2}^{\omega_{4}}+\omega_{2}\Upsilon_{bt,2}^{\omega_{5}})d(\Upsilon_{bm,2})d(\Upsilon_{bt,2}).
\end{align}
Now, performing integration on \eqref{eqn:sopmkappa 2} using \cite[eq~ 3.351.2]{GR:07:Book},
the closed-form expression of SOPM is expressed in \ref{eqn:sopmkappa 3}, where
$p_{\theta}=2^{\sigma_{s2}}-1$, 
$q_{\theta}=2^{\sigma_{s2}}$, 
$\delta_{19}=\frac{\varpi_{1}\varpi_{4}!\binom{\sigma_{13}}{\sigma_{15}}p_{\theta}^{\sigma_{13}-\sigma_{15}}q_{\theta}^{\sigma_{15}}}{\sigma_{13}!(\varpi_{3})^{\varpi_{4}-\sigma_{13}+1}e^{\varpi_{3}p_{\theta}}} $, and $\delta_{20}=\frac{\varpi_{2}\varpi_{5}!\binom{\sigma_{14}}{\sigma_{16}}p_{\theta}^{\sigma_{14}-\sigma_{16}}q_{\theta}^{\sigma_{16}}}{\sigma_{14}!(\varpi_{3})^{\varpi_{5}-\sigma_{14}+1}e^{\varpi_{3}p_{\theta}}}$.

\setcounter{equation}{33}
\begin{figure*}[!b]
\hrulefill
\begin{align}\label{eqn:pnsmceta 2}
\nonumber
&\Pr(C_{s,1}>0)=\sum_{\psi_{r_{6}}}\sum_{r_{6}=0}^{\mathcal{K}+\mathcal{K}r_{5}-1}\sum_{r_{5}=0}^{\mathcal{M}-1}\sum_{r_{2}=0}^{\infty}\sum _{r_{1}=0}^{\infty}\sum _{n_{1}=0}^{\infty}\sum_{\psi_{\nu_{6}}}\sum_{\nu_{6}=0}^{\mathcal{K}\mathcal{N}-1}\sum_{\nu_{2}=0}^{\infty}\sum_{\nu_{1}=0}^{\infty}\biggl[\biggl(\chi_{10}+\sum_{\nu_{7}=0}^{2\mu_{p1}+2n_{1}-1}\chi_{12}\biggl)\alpha_{3}\alpha_{5}!\beta_{4}^{-(\alpha_{5}+1)}
\\
\nonumber
&+\sum_{r_{7}=0}^{2\mu_{p1}+2n_{1}-1}\biggl(\chi_{10}+\sum_{\nu_{7}=0}^{2\mu_{p1}+2n_{1}-1}\chi_{12}\biggl)\frac{\alpha_{4}\alpha_{6}!}{\beta_{4}^{\alpha_{6}+1}}-\sum_{\nu_{9}=0}^{\alpha_{12}}\biggl(\frac{\chi_{11}\alpha_{3}(\alpha_{5}+\nu_{9})!}{(\beta_{4}+\beta_{5})^{\alpha_{5}+\nu_{9}+1}}+\sum_{r_{7}=0}^{2\mu_{p1}+2n_{1}-1}\frac{\chi_{11}\alpha_{4}(\alpha_{6}+\nu_{9})!}{(\beta_{4}+\beta_{5})^{\alpha_{6}+\nu_{9}+1}}\biggl)
\\
&-\sum_{\nu_{10}=0}^{\alpha_{13}}\sum_{\nu_{7}=0}^{2\mu_{p1}+2n_{1}-1}\biggl(\frac{\chi_{13}\alpha_{3}(\alpha_{5}+\nu_{10})!}{(\beta_{4}+\beta_{5})^{\alpha_{5}+\nu_{10}+1}}+\sum_{r_{7}=0}^{2\mu_{p1}+2n_{1}-1}\frac{\chi_{13}\alpha_{4}(\alpha_{6}+\nu_{10})!}{(\beta_{4}+\beta_{5})^{\alpha_{6}+\nu_{10}+1}}\biggl)\biggl].
\end{align}
\end{figure*}
\setcounter{equation}{35}
\begin{figure*}[!b]
\hrulefill
\begin{align}\label{eqn:pnsmckappa 2}
\nonumber
&\Pr(C_{s,2}>0)=\sum_{\psi_{\rho_{7}}}\sum_{\rho_{7}=0}^{\mathcal{K}+\mathcal{K}\rho_{6}-1}\sum_{\rho_{6}=0}^{\mathcal{M}-1}\sum_{\rho_3=0}^\infty\sum_{\rho_2=0}^\infty\sum_{\rho_{1}=0}^{\infty}\sum_{\psi_{\sigma_{10}}}\sum_{\sigma_{10}=0}^{\mathcal{K}\mathcal{N}-1}\sum_{\sigma_3=0}^\infty\sum_{\sigma_2=0}^\infty\biggl[\biggl(\delta_{15}+\sum_{\sigma_{11}=0}^{\mu_{p2}+\rho_{1}-1}\delta_{17}\biggl)\varpi_{1}\varpi_{4}!\varpi_{3}^{-(\varpi_{4}+1)}
\\
\nonumber
& +\sum_{\rho_{8}=0}^{\mu_{p2}+\rho_{1}-1}\biggl(\delta_{15}+\sum_{\sigma_{11}=0}^{\mu_{p2}+\rho_{1}-1}\delta_{17}\biggl)\frac{\varpi_{2}\varpi_{5}!}{\varpi_{3}^{\varpi_{5}+1}}-\sum_{\sigma_{13}=0}^{\omega_{4}}\biggl(\frac{\delta_{16}\varpi_{1}(\sigma_{13}+\varpi_{4})!}{(\omega_{3}+\varpi_{3})^{\sigma_{13}+\varpi_{4}+1}}+\sum_{\rho_{8}=0}^{\mu_{p2}+\rho_{1}-1}\frac{\delta_{16}\varpi_{2}(\sigma_{13}+\varpi_{5})!}{(\omega_{3}+\varpi_{3})^{\sigma_{13}+\varpi_{5}+1}}\biggl)
\\
&-\sum_{\sigma_{14}=0}^{\omega_{5}}\sum_{\sigma_{11}=0}^{\mu_{p2}+\rho_{1}-1}\biggl(\frac{\delta_{18}\varpi_{1}(\sigma_{14}+\varpi_{4})!}{(\omega_{3}+\varpi_{3})^{\sigma_{14}+\varpi_{4}+1}}+\sum_{\rho_{8}=0}^{\mu_{p2}+\rho_{1}-1}\frac{\delta_{18}\varpi_{2}(\sigma_{14}+\varpi_{5})!}{(\omega_{3}+\varpi_{3})^{\sigma_{14}+\varpi_{5}+1}}\biggl)\biggl].
\end{align}
\end{figure*}

\section{Non-zero Secrecy Capacity Analysis} 

The PNSMC is defined as \cite[Eq.~18]{badrudduza2019performance}
\begin{align}
\nonumber 
&P_{r}(C_{s,j}>0)=
\\\nonumber
&\int_{0}^{\infty}\int_{0}^{\Upsilon_{bm,j}}f_{\sigma_{min,j}}(\Upsilon_{bm,j})f_{\sigma _{max,j}}(\Upsilon_{bt,j})d\Upsilon_{bt,j}d\Upsilon_{bm,j}.
\end{align}
For a secure communication, the secrecy multicast capacity must be a positive quantity, otherwise the secrecy of the transmitted information can not be guaranteed. Using the expressions of $f_{\sigma _{min,j}}(\Upsilon_{bm,j})$, and $f_{\sigma _{max,j}}(\Upsilon_{bt,j})$ from \eqref{eqn:theorem 9}, \eqref{eqn:theorem 10}, \eqref{eqn:theorem 11}, and \eqref{eqn:theorem 12}, we derive the expressions of PNSC as the following: 

\subsection{Scenario I}
The PNSMC in the case of Scenario-I is defined as 
\setcounter{eqnback}{\value{equation}} \setcounter{equation}{32}
\begin{align}\label{eqn:pnsmceta 1}
\nonumber
&\Pr(C_{s,1}>0)=
\\
& \int_{0}^{\infty}\int_{0}^{\Upsilon_{bm,1}}f_{\sigma_{min,1}}(\Upsilon_{bm,1})f_{\sigma _{max,1}}(\Upsilon_{bt,1})s d\Upsilon_{bt,1}d\Upsilon_{bm,1}.
\end{align} 
Substituting \eqref{eqn:theorem 9} and \eqref{eqn:theorem 11} into \eqref{eqn:pnsmceta 1} and performing integration using \cite[eqs~(3.351.1, 3.351.3)]{GR:07:Book}, the closed-form expression of PNSMC is obtained in \eqref{eqn:pnsmceta 2}, where
$\chi_{10}=\frac{\alpha_{10}\alpha_{12}!}{\beta_{5}^{\alpha_{12}+1}}$, $\chi_{11}=\frac{\alpha_{10}\alpha_{12}!}{\nu_{9}!(\beta_{5})^{\alpha_{12}-\nu_{9}+1}}$,
$\chi_{12}=\frac{\alpha_{11}\alpha_{13}!}{\beta_{5}^{\alpha_{15}+1}}$, and $\chi_{13}=\frac{\alpha_{11}\alpha_{13}!}{\nu_{10}!(\beta_{5})^{\alpha_{13}-\nu_{10}+1}}$.

\subsection{ For Scenario II:}
In the case of Scenario-II mixed, the PNSMC is defined as
\setcounter{eqnback}{\value{equation}} \setcounter{equation}{34}
\begin{align}\label{eqn:pnsmckappa 1}
\nonumber
&\Pr(C_{s,2}>0)=
\\
&\int_{0}^{\infty}\int_{0}^{\Upsilon_{bm,2}}f_{\sigma_{min,2}}(\Upsilon_{bm,2})f_{\sigma_{max,2}}(\Upsilon_{bt,2}) d\Upsilon_{bt,2}d\Upsilon_{bm,2}.
\end{align}
Substituting the value of  \eqref{eqn:theorem 10} and \eqref{eqn:theorem 12}
into \eqref{eqn:pnsmckappa 1} and then performing integration using \cite[eqs.~ 3.351.2, and 3.351.3]{GR:07:Book}, we get the final expression of PNSMC in \eqref{eqn:pnsmckappa 2}, where $\delta_{15}=\frac{\omega_{1}\omega_{4}!}{\omega_{3}^{\omega_{4}+1}}$,
$\delta_{16}=\frac{\omega_{1}\omega_{4}!}{\sigma_{13}!\omega_{3}^{\omega_{4}-\sigma_{13}+1}}$,
$\delta_{17}=\frac{\omega_{2}\omega_{5}!}{\omega_{3}^{\omega_{5}+1}}$, and
$\delta_{18}=\frac{\omega_{2}\omega_{5}!}{\sigma_{14}!\omega_{3}^{\omega_{5}-\sigma_{14}+1}}$.

\setcounter{equation}{37}
\begin{figure*}[!t]
\begin{align}\label{eqn:esmceta 2}
\nonumber
&\langle C_{s,1}\rangle=\sum_{\psi_{r_{6}}}\sum_{r_{6}=0}^{\mathcal{K}+\mathcal{K}r_{5}-1}\sum_{r_{5}=0}^{\mathcal{M}-1}\sum_{r_{2}=0}^{\infty}\sum _{r_{1}=0}^{\infty}\sum _{n_{1}=0}^{\infty}\sum_{\psi_{\nu_{6}}}\sum_{\nu_{6}=0}^{\mathcal{K}\mathcal{N}-1}\sum_{\nu_{2}=0}^{\infty}\sum_{\nu_{1}=0}^{\infty}\biggl[\sum_{r_{13}=0}^{\alpha_{5}}\frac{\alpha_{3}\alpha_{5}!\beta_{4}^{-\alpha_{5}-1}}{(\alpha_{5}-r_{13})!}\biggl[\frac{(-1)^{\alpha_{5}-r_{13}-1}\mathbf{Ei}[-\beta_{4}]}{e^{-\beta_{4}}\beta_{4}^{-\alpha_{5}+r_{13}}}
\\   
\nonumber 
&+\sum_{r_{14}=1}^{\alpha_{5}-r_{13}}\frac{(r_{14}-1)!}{(-\beta_{4})^{-\alpha_{5}+r_{13}+r_{14}}}\biggl]+\sum_{r_{15}=0}^{\alpha_{6}}\sum_{r_{7}=0}^{2\mu_{p1}+2n_{1}-1}\frac{\alpha_{4}\alpha_{6}!\beta_{4}^{-\alpha_{6}-1}}{(\alpha_{6}-r_{15})!}\biggl[\frac{(-1)^{\alpha_{6}-r_{15}-1}\mathbf{Ei}[-\beta_{4}]}{e^{-\beta_{4}}\beta_{4}^{-\alpha_{6}+r_{15}}}+\sum_{r_{16}=1}^{\alpha_{6}-r_{15}}\frac{(r_{16}-1)!}{(-\beta_{4})^{-\alpha_{6}+r_{15}+r_{16}}}\biggl]
\\
\nonumber
&-\sum_{\nu_{11}=0}^{\alpha_{12}}\frac{\alpha_{10}\alpha_{12}!\beta_{5}^{-\alpha_{12}-1}}{(\alpha_{12}-\nu_{11})!}\biggl[\frac{(-1)^{\alpha_{12}-\nu_{11}-1}\mathbf{Ei}[-\beta_{5}]}{e^{-\beta_{5}}\beta_{5}^{-\alpha_{12}+\nu_{11}}}+
\sum_{\nu_{12}=1}^{\alpha_{12}-\nu_{11}}\frac{(\nu_{12}-1)!}{(-\beta_{5})^{-\alpha_{12}+\nu_{11}+\nu_{12}}}\biggl]
\\
&-\sum_{\nu_{13}=0}^{\alpha_{13}}\sum_{\nu_{7}=0}^{2\mu_{p1}+2n_{1}-1}\frac{\alpha_{11}\alpha_{13}!\beta_{5}^{-\alpha_{13}-1}}{(\alpha_{13}-\nu_{13})!}\biggl[\frac{(-1)^{\alpha_{13}-\nu_{13}-1}\mathbf{Ei}[-\beta_{5}]}{e^{-\beta_{5}}\beta_{5}^{-\alpha_{13}+\nu_{13}}}+\sum_{\nu_{12}=1}^{\alpha_{13}-\nu_{13}}\frac{(\nu_{14}-1)!}{(-\beta_{5})^{-\alpha_{13}+\nu_{13}+\nu_{14}}}\biggl]\biggl].
\end{align}
\hrulefill
\end{figure*}
\setcounter{equation}{39}
\begin{figure*}[!t]
\begin{align}\label{eqn:esmckappa 2}
\nonumber  
&\langle C_{s,2} \rangle  =\sum_{\psi_{\rho_{7}}}\sum_{\rho_{7}=0}^{\mathcal{K}+\mathcal{K}\rho_{6}-1}\sum_{\rho_{6}=0}^{\mathcal{M}-1}\sum_{\rho_3=0}^\infty\sum_{\rho_2=0}^\infty\sum_{\rho_{1}=0}^{\infty}\sum_{\psi_{\sigma_{10}}}\sum_{\sigma_{10}=0}^{\mathcal{K}\mathcal{N}-1}\sum_{\sigma_3=0}^\infty\sum_{\sigma_2=0}^\infty\biggl[\sum_{\sigma_{15}=0}^{\varpi_{4}}\frac{\varpi_{1}\varpi_{4}!}{\varpi_{3}^{\varpi_{4}+1}(\varpi_{4}-\sigma_{15})!}\biggl[\frac{(-1)^{\varpi_{4}-\sigma_{15}-1}\mathbf{Ei}(-\varpi_{3})}{e^{-\varpi_{3}}(\frac{1}{\varpi_{3}})^{\varpi_{4}-\sigma_{15}}}
\\
\nonumber 
&+\sum_{\sigma_{16}=1}^{\varpi_{4}-\sigma_{15}}\frac{(\sigma_{16}-1)!}{(\frac{-1}{\varpi_{3}})^{\varpi_{4}-\sigma_{15}-\sigma_{16}}}\biggl]+\sum_{\sigma_{17}=0}^{\varpi_{5}}\sum_{\rho_{8}=0}^{\mu_{p2}+\rho_{1}-1}\frac{\varpi_{2}\varpi_{5}!\varpi_{3}^{-(\varpi_{5}+1)}}{(\varpi_{5}-\sigma_{17})!}\biggl[\frac{(-1)^{\varpi_{5}-\sigma_{17}-1}\mathbf{Ei}(-\varpi_{3})}{e^{-\varpi_{3}}(\frac{1}{\varpi_{3}})^{\varpi_{5}-\sigma_{17}}}+\sum_{\sigma_{18}=1}^{\varpi_{5}-\sigma_{17}}\frac{(\sigma_{18}-1)!}{(\frac{-1}{\varpi_{3}})^{\varpi_{5}-\sigma_{17}-\sigma_{18}}}\biggl]
\\
\nonumber  
&-\sum_{\sigma_{19}=0}^{\omega_{4}}\frac{\omega_{1}\omega_{4}!}{\omega_{3}^{\omega_{4}+1}(\omega_{4}-\sigma_{19})!}\biggl[\frac{(-1)^{\omega_{4}-\sigma_{19}-1}\mathbf{Ei}(-\omega_{3})}{e^{-\omega_{3}}(\frac{1}{\omega_{3}})^{\omega_{4}-\sigma_{19}}}+\sum_{\sigma_{20}=1}^{\omega_{4}-\sigma_{19}}\frac{(\sigma_{20}-1)!}{(\frac{-1}{\omega_{3}})^{\omega_{4}-\sigma_{19}-\sigma_{20}}}\biggl]
\\
&-\sum_{\sigma_{21}=0}^{\omega_{5}}\sum_{\sigma_{11}=0}^{\mu_{p2}+\rho_{1}-1}\frac{\omega_{2}\omega_{5}!}{\omega_{3}^{\omega_{5}+1}(\omega_{5}-\sigma_{21})!}\biggl[\frac{(-1)^{\omega_{5}-\sigma_{21}-1}\mathbf{Ei}(-\omega_{3})}{e^{-\omega_{3}}(\frac{1}{\omega_{3}})^{\omega_{5}-\sigma_{21}}}+\sum_{\sigma_{22}=1}^{\omega_{5}-\sigma_{21}}\frac{(\sigma_{22}-1)!}{(\frac{-1}{\omega_{3}})^{\omega_{5}-\sigma_{21}-\sigma_{22}}}\biggl]\biggl].  
\end{align}
\hrulefill
\end{figure*}

\section{Ergodic Secrecy Multicast Capacity Analysis}
The ESMC is defined as the average value of the instantaneous secrecy capacity. It can be mathematically expressed as    \cite[Eq.~21]{badrudduza2019performance}
\begin{align}
\nonumber
\langle C_{s,j}\rangle&=\int_{0}^{\infty}log_{2}(1+\Upsilon_{bm,j})f_{\sigma_{min,j}}(\Upsilon_{bm,j})d\Upsilon_{bm,j}
\\
\nonumber&-\int_{0}^{\infty}log_{2}(1+\Upsilon_{bt,j})f_{\sigma_{max,j}}(\Upsilon_{bt,j}))d\Upsilon_{bt,j}.
\end{align}
Utilizing the expressions of $f_{\sigma_{min,j}}(\Upsilon_{bm,j})$, and 
\\
$f_{\sigma_{max,j}}(\Upsilon_{bm,j})$ from \eqref{eqn:theorem 9}, \eqref{eqn:theorem 10}, \eqref{eqn:theorem 11}, and \eqref{eqn:theorem 12}, we derive the expressions for ESMC in closed-form in the following sub-sections.

\subsection{Scenario I}
The ESMC for the case of Scenario-I is defined as 
\setcounter{eqnback}{\value{equation}} \setcounter{equation}{36}
\begin{align}\label{eqn:esmceta 1}
\nonumber
\langle C_{s,1}\rangle&=\int_{0}^{\infty}log_{2}(1+\Upsilon_{bm,1})f_{\sigma_{min,1}}(\Upsilon_{bm,1})d\Upsilon_{bm,1}
\\&-\int_{0}^{\infty}log_{2}(1+\Upsilon_{bt,1})f_{\sigma_{max,1}}(\Upsilon_{bt,1}))d\Upsilon_{bt,1}.
\end{align}
Substituting \eqref{eqn:theorem 9} and \eqref{eqn:theorem 11} into \eqref{eqn:esmceta 1}, and performing integration by making use of \cite[eq~ 4.222.8]{GR:07:Book}, we get the closed-form expression of ESMC is shown in \ref{eqn:esmceta 2}, where $\mathbf{Ei}[.]$ denotes the exponential integral function. 

\subsection{For Scenario II}
In the case of  Scenario-II, the ESMC is defined as 
\setcounter{eqnback}{\value{equation}} \setcounter{equation}{38}
\begin{align}\label{eqn:esmckappa 1}
\nonumber
\langle C_{s,2}\rangle&=\int_{0}^{\infty}log_{2}(1+\Upsilon_{bm,2})f_{\sigma_{min,2}}(\Upsilon_{bm,2})d\Upsilon_{bm,2}
\\
&-\int_{0}^{\infty}log_{2}(1+\Upsilon_{bt,2})f_{\sigma_{max,2}}(\Upsilon_{bt,2})d\Upsilon_{bt,2}.
\end{align}
Substituting \eqref{eqn:theorem 10} and \eqref{eqn:theorem 12} into \eqref{eqn:esmckappa 1},
and then performing integration utilizing \cite[eq~ 4.222.8]{GR:07:Book}, the closed-form expression of ESMC is obtained as given in \ref{eqn:esmckappa 2}.

\section{NUMERICAL RESULTS}

\begin{figure*}[!ht]
\vspace{0mm}
    \centerline{\includegraphics[width=1\textwidth,angle =0]{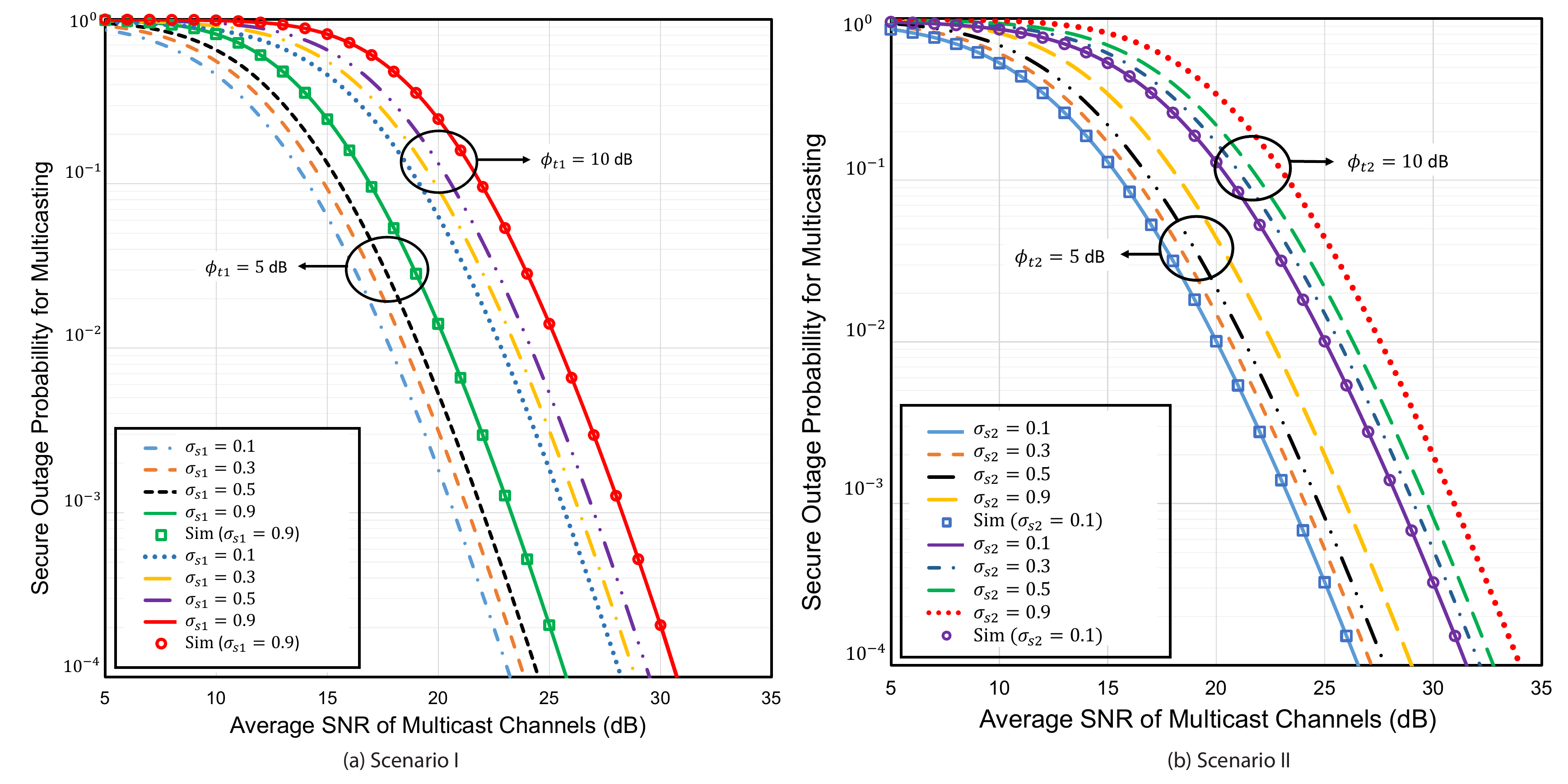}}
        \vspace{0mm }
    \caption{
  The SOPM versus average SNR of  multicast channels varying $\sigma_{s1}$ and $\sigma_{s2}$ for selected values of $\phi_{t1}$ and $\phi_{t2}$ with $\mathcal{M}=\mathcal{N}=$ 2, $\mathcal{K}=4$, (a) $m_{m1}=m_{t1}=5$, $\mu_{p1}=\mu_{m1}=\mu_{t1}=0.5$ and $\eta_{p}=\eta_{m}=\eta_{t}=0.5$, and (b) $m_{m2}=m_{t2}=5$, $\mu_{p2}=\mu_{m2}=\mu_{t2}=0.5$, $\kappa_{p}=\kappa_{m}=\kappa_{t}=0.5$.
    }
    \label{fig:2}
\end{figure*}
\begin{figure*}[!ht]
\vspace{0mm}
    \centerline{\includegraphics[width=1\textwidth,angle =0]{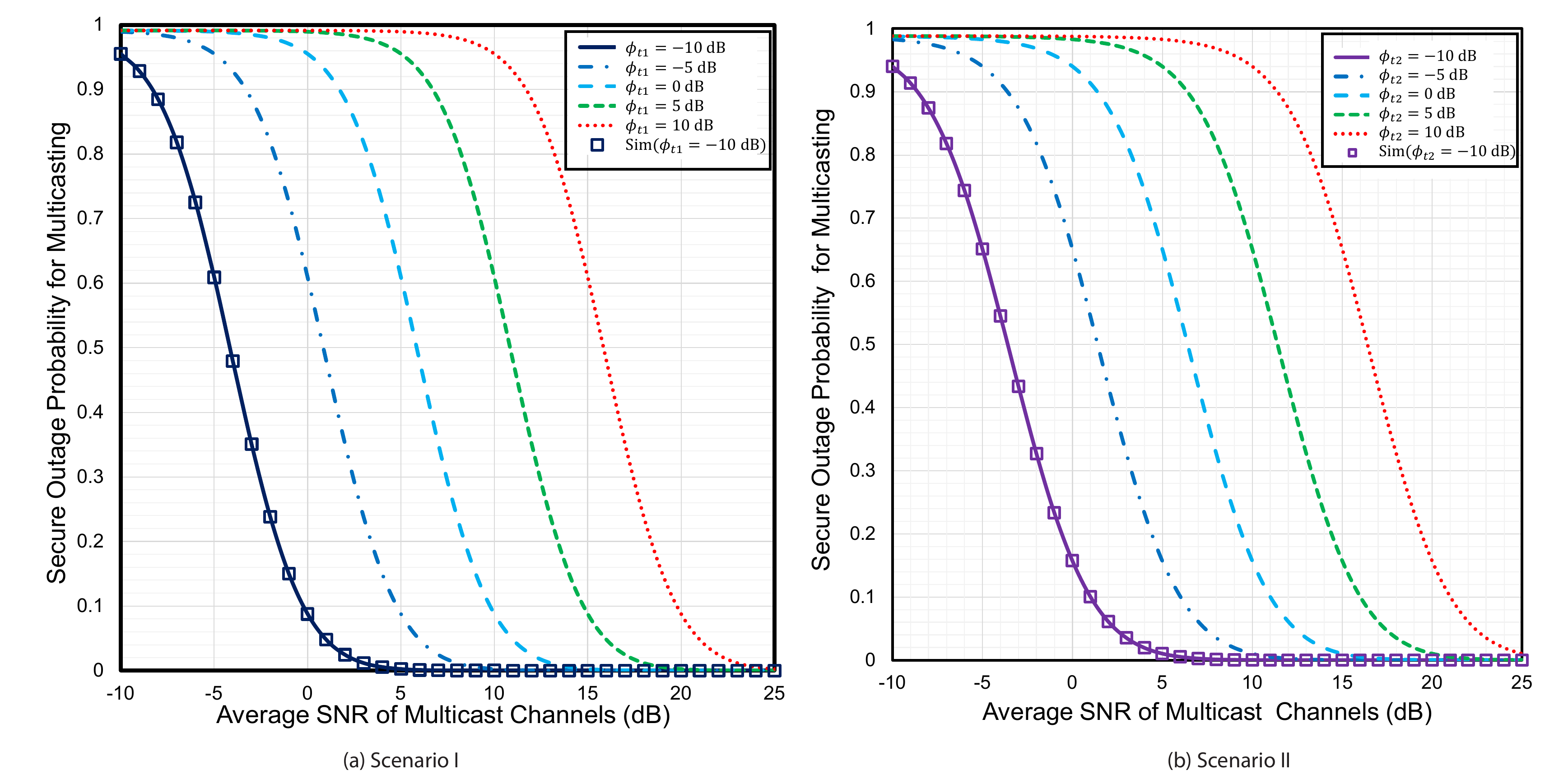}}
        \vspace{0mm }
    \caption{
  The SOPM versus average SNR of case-I and case-II multicast channels varying $\phi_{t1}$ and  $\phi_{t2}$ with $\mathcal{M}=\mathcal{N}$=2, $\mathcal{K}=$ 2, (a) $m_{m1}=m_{t1}=5$, $\mu_{p1}=\mu_{m1}=\mu_{t1}=2$, $\eta_{p}=\eta_{m}=\eta_{t}=$ 0.5 and $\sigma_{s1}=0.5$ bits/s/Hz, and (b) $m_{m2}=m_{t2}=5$, $ \mu_{p2}=\mu_{m2}=\mu_{t2}=2$, $\kappa_{p}=\kappa_{m}=\kappa_{t}=0.5$, $\sigma_{s2}=0.5$ bits/s/Hz.}
    \label{fig:3}
\end{figure*}

In this section, we graphically represent the numerical results corresponding to the derived analytical expressions for the SOPM, PNSMC, and ESMC and observe how the system parameters, e.g., fading severity, shadowing, the number of relays, receivers, and eavesdroppers, etc. influence the multicast system's secrecy characteristics. Analytical results are obtained using Mathematica software and for simulation purposes, we averaged $10^6$ random samples of each channel. {\color{black}Note that the infinite series in \eqref{eqn:sopmeta 3}, \eqref{eqn:sopmkappa 3}, \eqref{eqn:pnsmceta 2}, \eqref{eqn:pnsmckappa 2},  \eqref{eqn:esmceta 2} and \eqref{eqn:esmckappa 2} converges quickly with a high level of accuracy just after fewer terms. Hence we assume the first 25 terms for each infinite series.} Moreover, with a view to obtaining a clear comparison in terms of secrecy performance between the two considered scenarios, we present a set of curves corresponding to each scenario in each figure.  

The SOPM is depicted as a function of the average SNR of the multicast channels in Figures \ref{fig:2}a and \ref{fig:2}b in order to observe the impacts due to variation in target secrecy rate.  In each scenario, we assume two cases with $\phi_{t1}=\phi_{t2}=$ 5 dB and 10 dB. We can observe as seen from the figure that the SOPM increases with $\sigma_{s1}$ and $\sigma_{s2}$ which is also shown in \cite{zeng2018physical}. Note that we assume a passive eavesdropping scenario in this work, hence the channel state information of eavesdropper networks is not known. For that reason, we assume a target secrecy rate and send information at this particular capacity. But if the instantaneous secrecy capacity falls below this target secure transmission capacity, then there will be a secrecy outage. Moreover, since the wireless channels are random in nature, an increase in target secrecy rate increases the probability of instantaneous secrecy rate falling below the target rate, hence the SOPM, as can be seen from the figure, will increase.

\begin{figure*}[!ht]
\vspace{0mm}
    \centerline{\includegraphics[width=1\textwidth,angle =0]{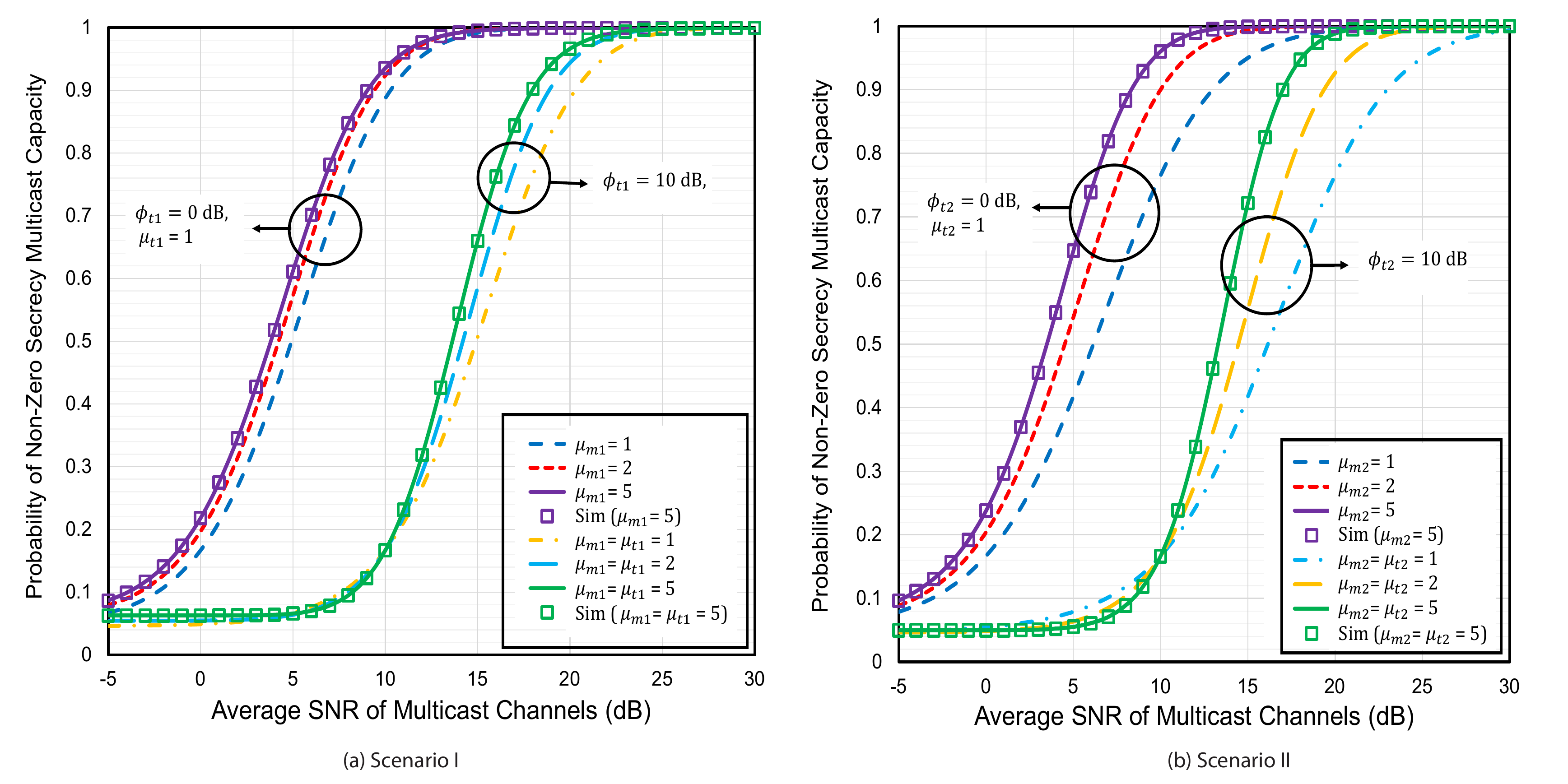}}
        \vspace{0mm }
    \caption{
  The PNSMC versus average SNR of multicast channels varying $\mu_{m1}$, $\mu_{t1}$, $\mu_{m2}$, $\mu_{t2}$  for selected values of  $\phi_{t1}$= 0 dB, 5 dB, and $\phi_{t2}$= 0 dB, 5 dB with $\mathcal{M}=\mathcal{N}=$ 2, $\mathcal{K}=$ 2, (a) $m_{m1}=m_{t1}=5$, $\mu_{p1}=1$ and $\eta_{p}=\eta_{m}=\eta_{t}=$ 0.5, and (b) $m_{m2}=m_{t2}=5$, $\mu_{p2}=2$, and $\kappa_{p}=\kappa_{m}=\kappa_{t}=0.5$.}
    \label{fig:4}
\end{figure*}

In Figures \ref{fig:3}a and \ref{fig:3}b, we illustrate the SOPM versus average SNR of the multicast channels showing the impacts of $\phi_{t1}$ and $\phi_{t2}$. It can be clearly seen from both the scenarios that the outage performance is better for lower values of the eavesdropper channel's average SNR. It is obvious that the eavesdropper links becomes better due to an increased value of $\phi_{t1}$ and $\phi_{t2}$ which lets the eavesdroppers sneak more information from the multicast channels, and accordingly, the SOPM also increases. Same results also can be seen in \cite{badrudduza2020enhancing,lei2015performance} which legibly justify our observations.

\begin{figure*}[!ht]
\vspace{0mm}
    \centerline{\includegraphics[width=1\textwidth,angle =0]{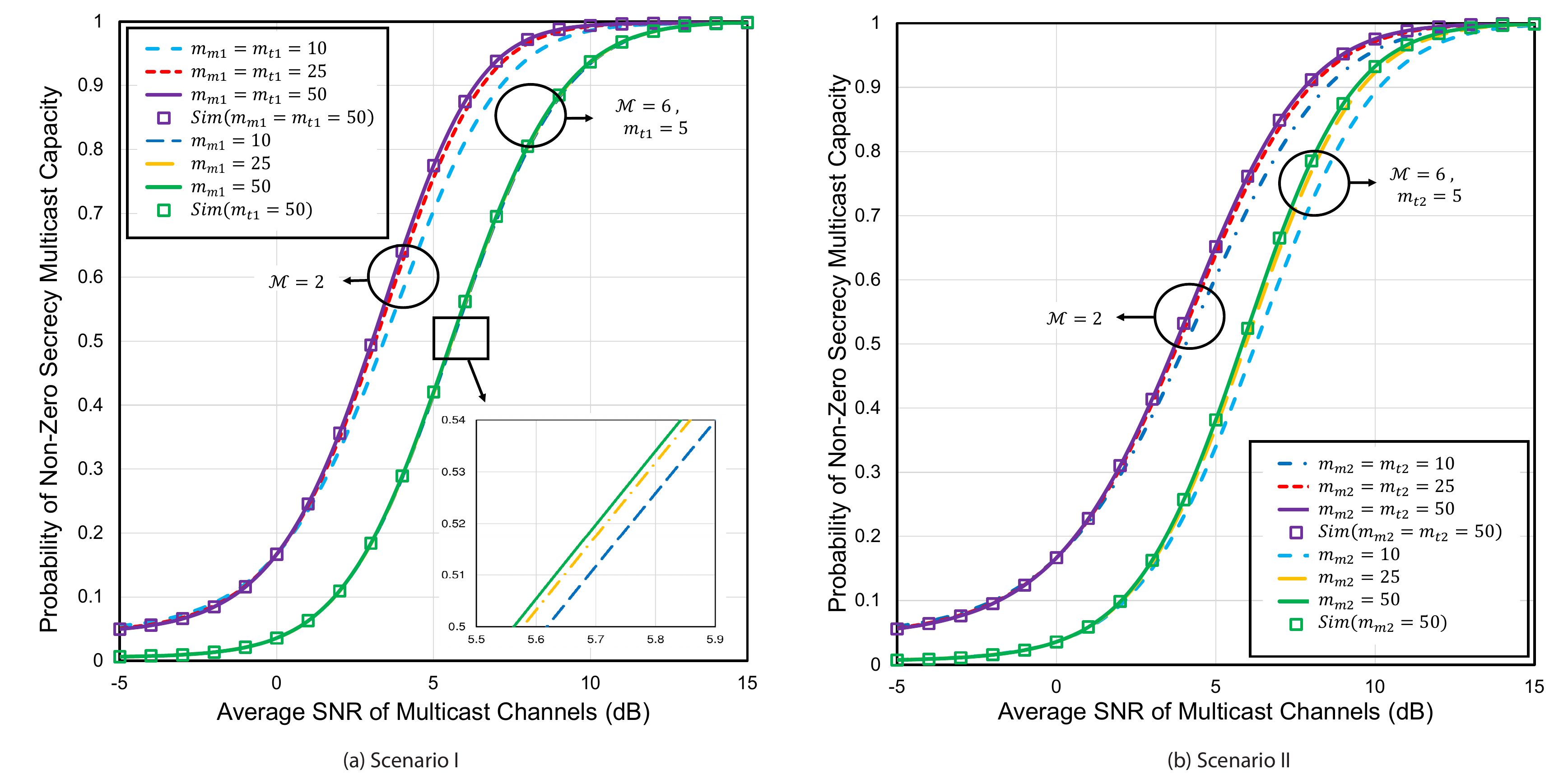}}
        \vspace{0mm }
    \caption{
  The PNSMC versus average SNR of multicast channels varying $m_{m1}$, $m_{t1}$, $m_{m2}$, and $m_{t2}$ for selected values of $\mathcal{M}$ with $\mathcal{N}= 2$, $\mathcal{K}= 2$, (a) $\phi_{t1}=$ 0 dB, $\mu_{p1}=\mu_{m1}=\mu_{t1}=$ 2, and $\eta_{p}=\eta_{m}=\eta_{t}=$ 0.5, and (b) $\phi_{t2}=$ 0 dB, $\mu_{p2}=\mu_{m2}=\mu_{t2}=$ 2, and $\kappa_{p}=\kappa_{m}=\kappa_{t}=$ 0.5.}
    \label{fig:5}
\end{figure*}
The impacts of the number of multipath clusters of the multicast and the eavesdropper channels are depicted in Figures \ref{fig:4}a and \ref{fig:4}b in terms of PNSMC. It is observed that the PNSMC performance becomes better for the higher values of $\mu_{m1}$ and $\mu_{m2}$. The reason is that with the increase in the number of multipath clusters in the $\mathcal{K}\rightarrow\mathcal{M}$ link, the fading of the corresponding link is reduced. On the contrary, the lower amount of fading assists in improving the secrecy capacity, and thus the PNSMC performance also enhances as testified in \cite{peppas2013performance}. Similarly, an increase in $\mu_{t_1}$ and $\mu_{t_2}$ reduce the fading of the $\mathcal{K}\rightarrow\mathcal{N}$ which will be beneficial for the eavesdroppers for wiretapping more confidential data. But if we simultaneously vary the shadowing severity of both the multicast and the eavesdropper links with equal values of $\mu_{m_1}$, $\mu_{m_2}$, $\mu_{t_1}$, and $\mu_{t_2}$, the PNSMC performance becomes better.

\begin{figure*}[!ht]
\vspace{0mm}
    \centerline{\includegraphics[width=1\textwidth,angle =0]{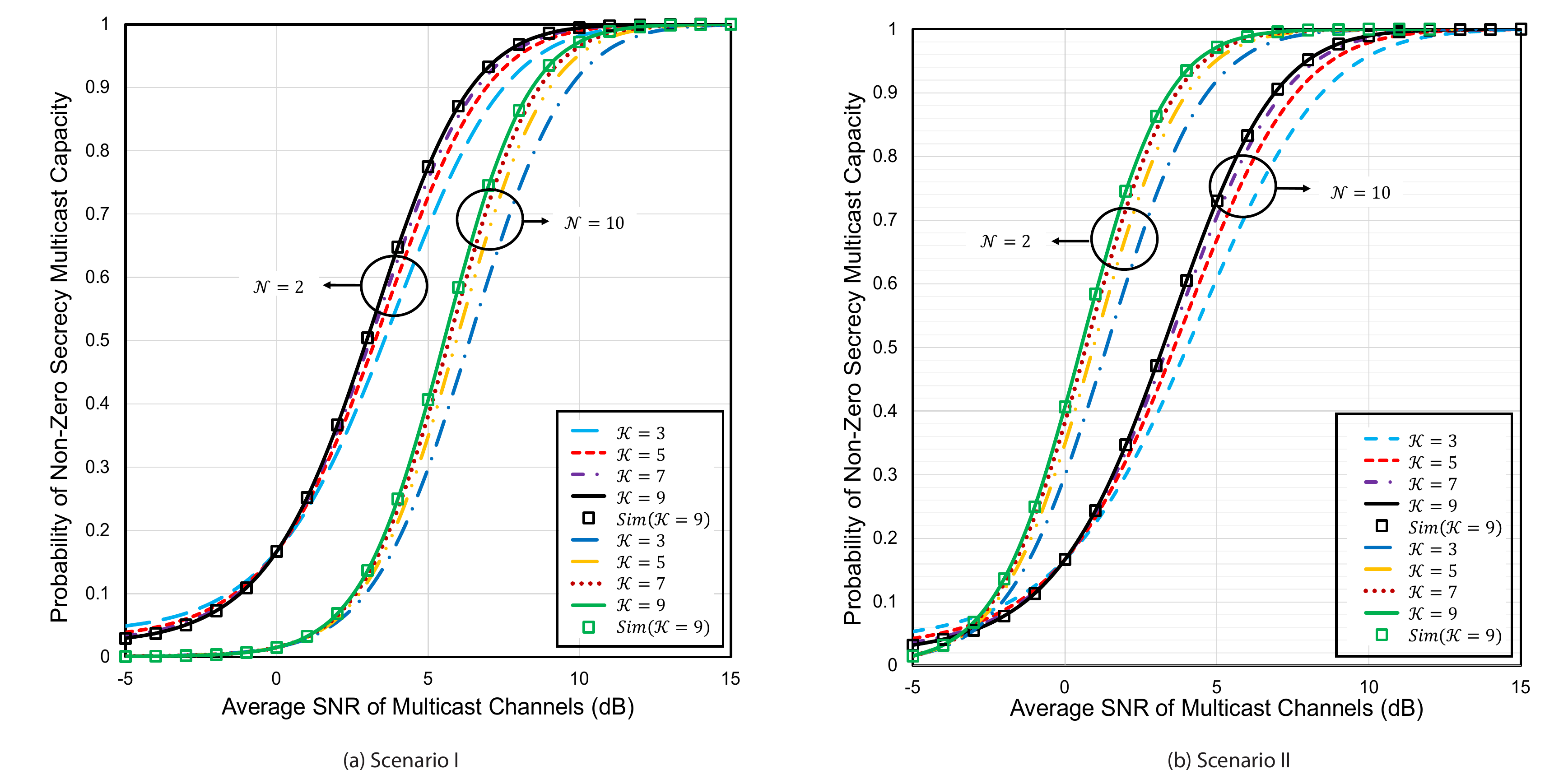}}
        \vspace{0mm }
    \caption{
  The PNSMC versus average SNR of multicast channels varying $\mathcal{K}$ for selected values of $\mathcal{N}$ with $\mathcal{M}=$ 2, (a) $\phi_{t1}=$ 0 dB, $m_{m1} = m_{t1}=$ 5, $\mu_{p1}=\mu_{m1}=\mu_{t1}=$ 2, and $\eta_{p}=\eta_{m}=\eta_{t}=$ 0.5, and (b) $\phi_{t2}=$ 0 dB,  $m_{m2} = m_{t2}=$ 5, $\mu_{p2}=\mu_{m2}=\mu_{t2}=$ 2, and $\kappa_{p}=\kappa_{m}=\kappa_{t}=$ 0.5.}
    \label{fig:6}
\end{figure*}
\begin{figure*}[!ht]
\vspace{0mm}
    \centerline{\includegraphics[width=1\textwidth,angle =0]{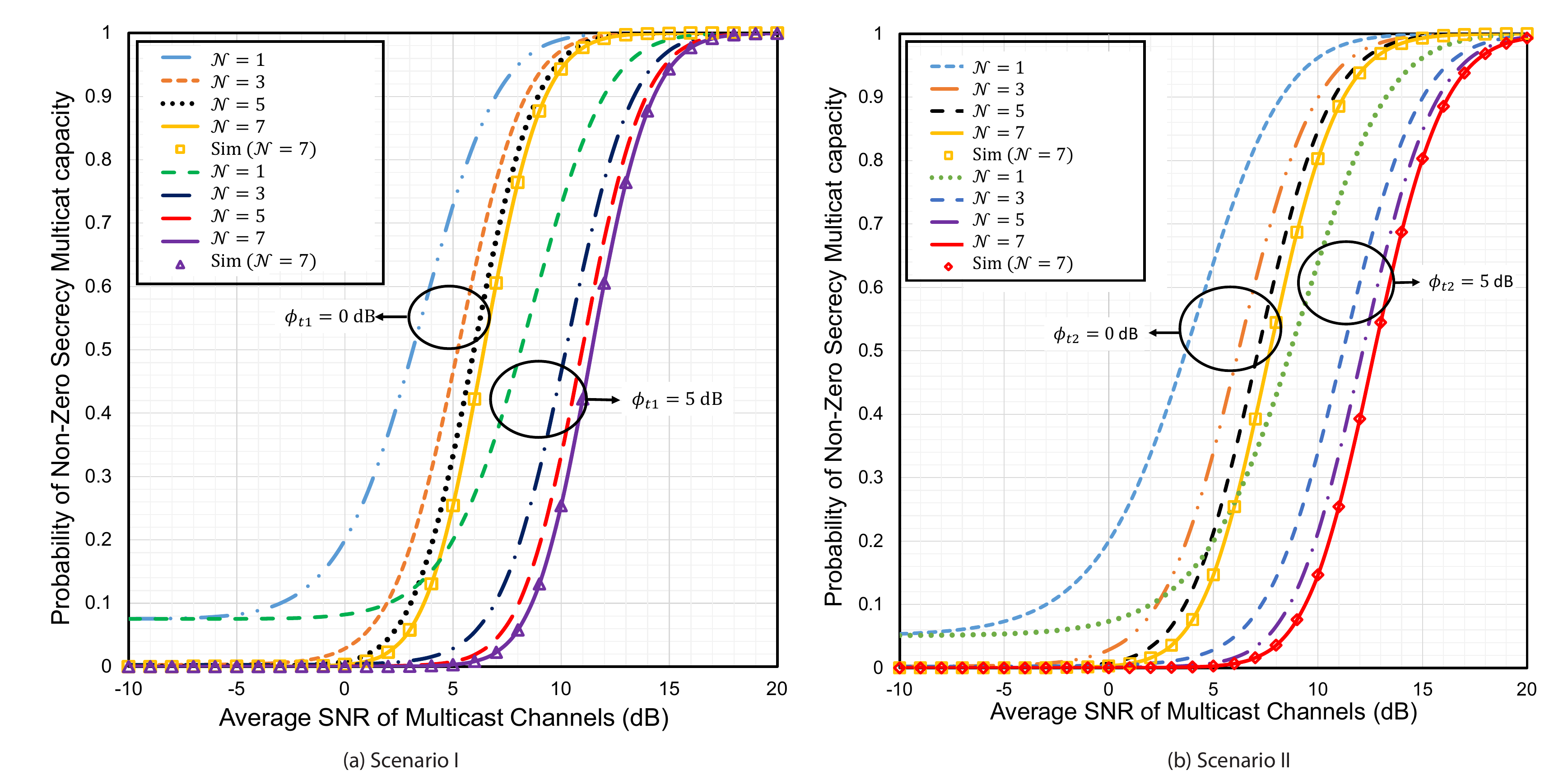}}
        \vspace{0mm }
    \caption{
  The PNSMC versus average SNR of multicast channels varying $\mathcal{N}$ for selected values of $\phi_{t1}$ and $\phi_{t2}$ with $\mathcal{M}=$ 4, $\mathcal{K}=$ 2, (a) $m_{m1} = m_{t1}=$ 5, $\mu_{p1}=\mu_{m1}=\mu_{t1}=$ 2, and $\eta_{p}=$$\eta_{m}=$$\eta_{t}=$ 0.5, and (b) $m_{m2} = m_{t2}=$ 5, $\mu_{p2}=\mu_{m2}=\mu_{t2}=$ 2, and $\kappa_{p}=\kappa_{m}=\kappa_{t}=$ 0.5.}
    \label{fig:7}
\end{figure*}
\begin{figure*}[!ht]
\vspace{0mm}
    \centerline{\includegraphics[width=1\textwidth,angle =0]{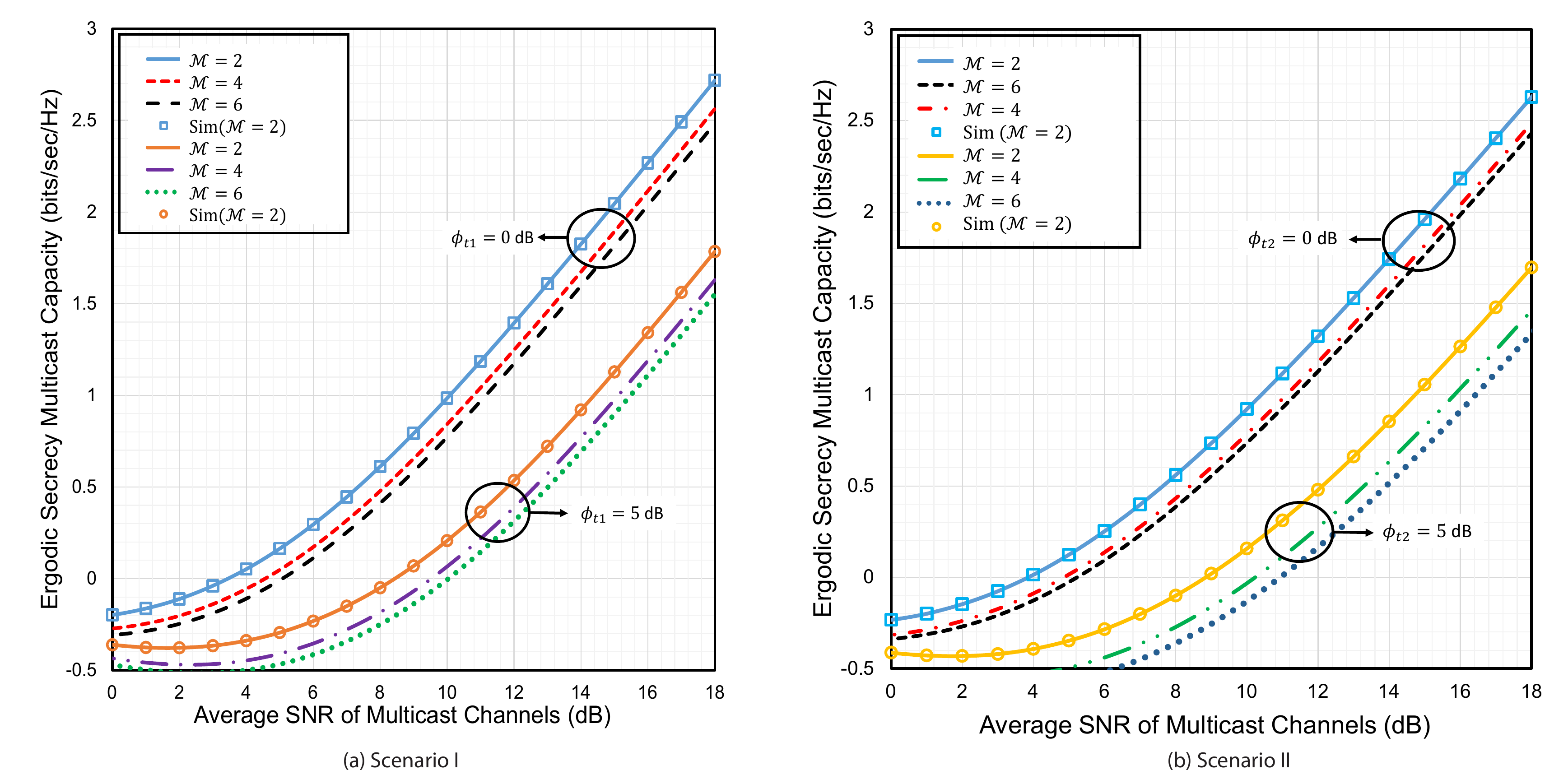}}
        \vspace{0mm }
    \caption{
  The ESMC versus average SNR of  multicast channels varying $\mathcal{M}$ for selected values of $\phi_{t1}$ and $\phi_{t2}$ with $\mathcal{N}=$ 2, $\mathcal{K}=$ 4, (a) $m_{m1} = m_{t1}=$ 5, $\mu_{p1}=\mu_{m1}=\mu_{t1}=$ 2, $\eta_{p}=\eta_{m}=\eta_{t}=$ 0.5, $\sigma_{s1}$ = 0.5 bits/s/Hz, and (b) $m_{m2} = m_{t2}=$ 5, $\mu_{p2}=\mu_{m2}=\mu_{t2}=$ 2, $\kappa_{p}=\kappa_{m}=\kappa_{t}=$ 0.5, and $\sigma_{s2}$ = 0.5 bits/s/Hz.}
    \label{fig:8}
\end{figure*}

\begin{figure*}[!ht]
\vspace{0mm}
    \centerline{\includegraphics[width=1\textwidth,angle =0]{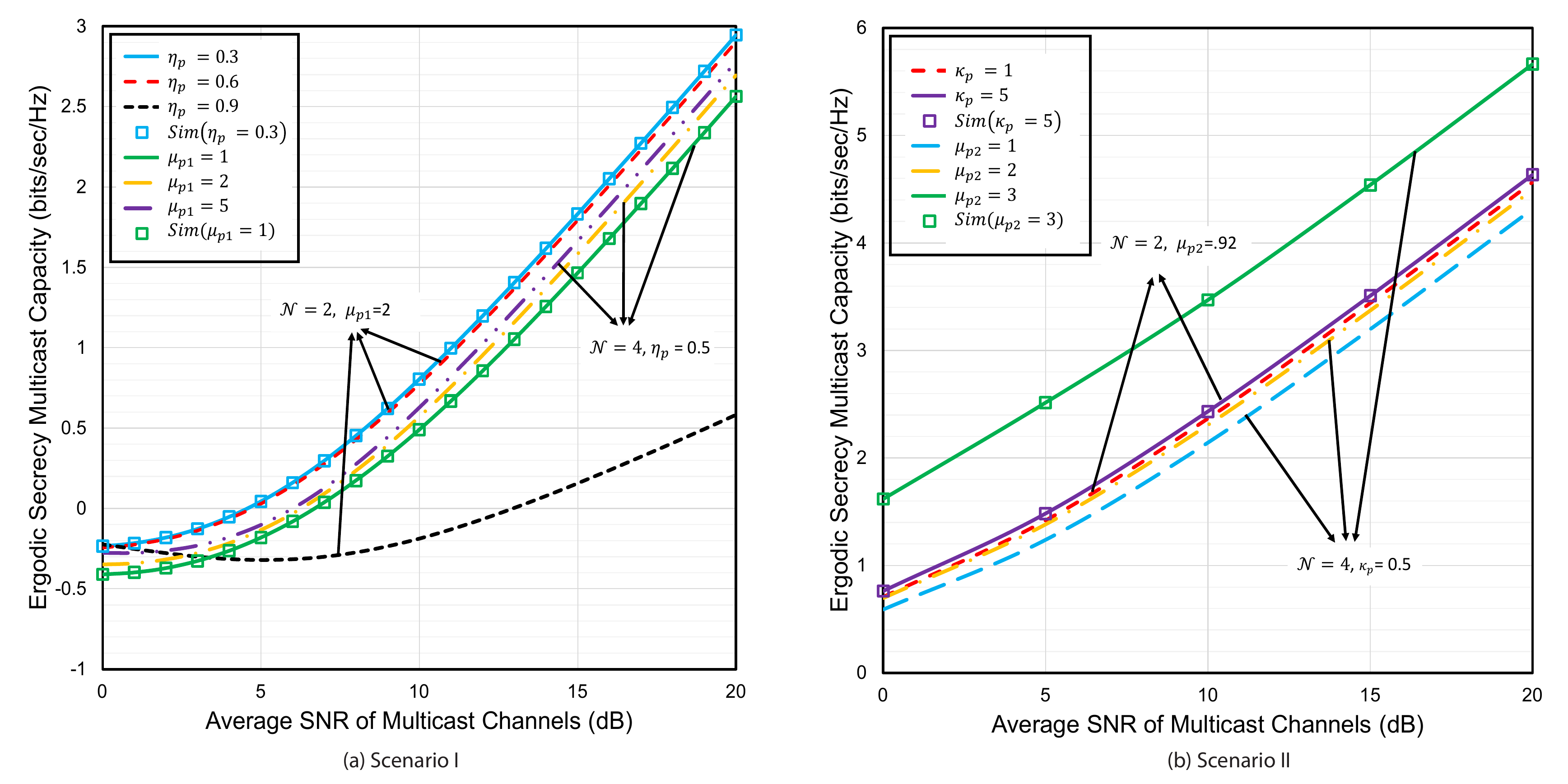}}
        \vspace{0mm }
    \caption{
  The ESMC versus average SNR of multicast channel varying $\eta_{p}$, $\mu_{p1}$, $\kappa_{p}$, and $\mu_{p2}$ for selected values of $\mathcal{N}$ with $\mathcal{M}$=2, $\mathcal{K}$=4, (a) $\phi_{t1}=$ 0 dB,  $m_{m1}=m_{t1}=5$, $\mu_{m1}=\mu_{t1}=2$ , $\eta_{m}=\eta_{t}=0.5$ and $\sigma_{s1}=0.5$ bits/s/Hz, and (b) $\phi_{t2}=$ 0 dB, $m_{m2}=m_{t2}=5$, $\mu_{m2}=\mu_{t2}=2$, $\kappa_{m}=\kappa_{t}=0.5$, and $\sigma_{s2}=0.5$ bits/s/Hz.}
    \label{fig:9}
\end{figure*}


\begin{figure*}[!ht]
\vspace{0mm}
    \centerline{\includegraphics[width=1\textwidth,angle =0]{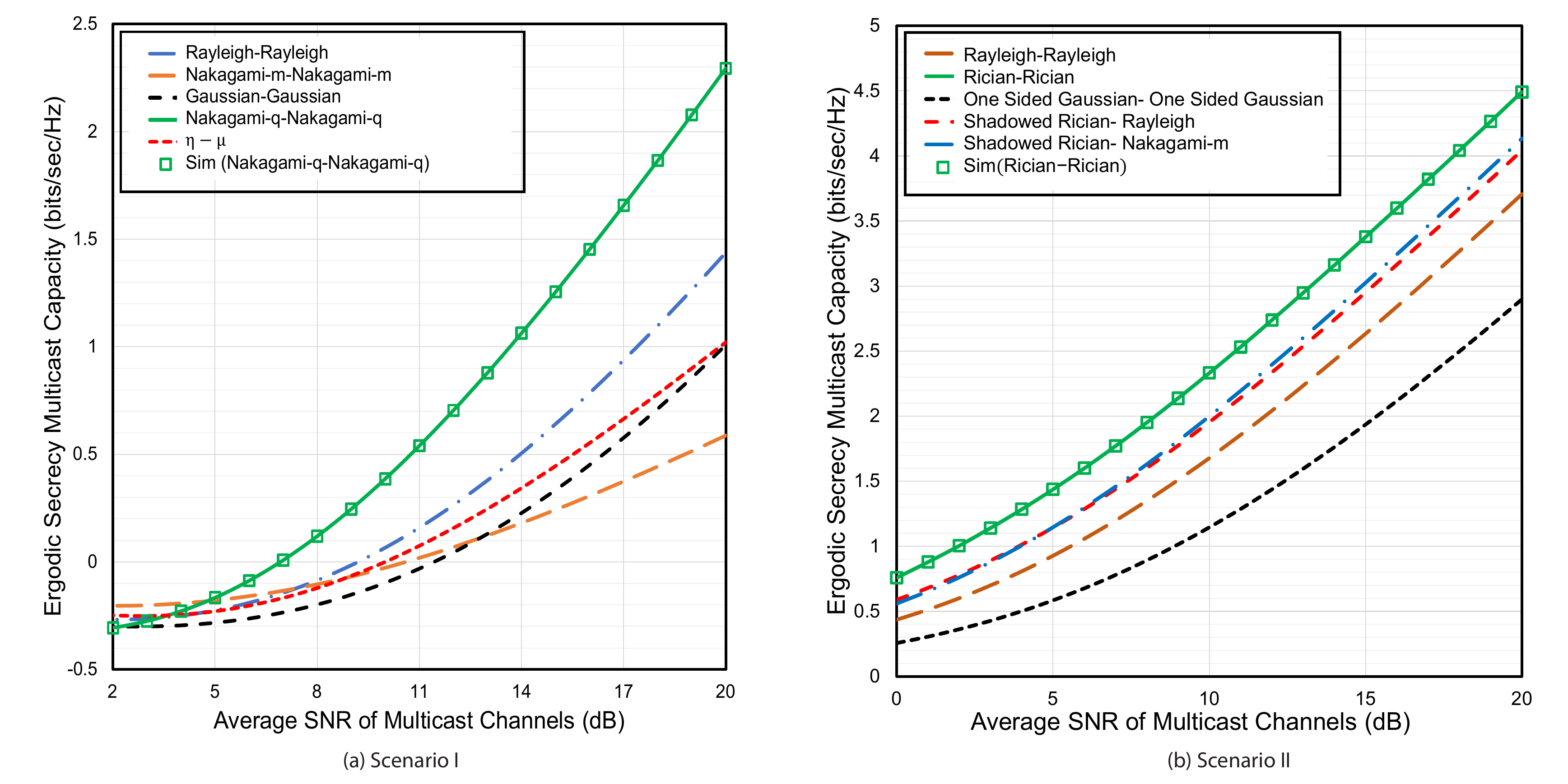}}
        \vspace{0mm }
    \caption{
  The ESMC versus average SNR of multicast channels for selected values of $\mu_{p1}$, $\mu_{m1}$,  $\mu_{t1}$, $\mu_{p2}$, $\mu_{m2}$, $\mu_{t2}$,  $\eta_{p}$, $\eta_{m}$, $\eta_{t}$, $\kappa_{p}$, $\kappa_{m}$, $\kappa_{t}$ with $\mathcal{M}= \mathcal{N}=\mathcal{K}=2$, (a) $m_{m1}=m_{t1}=35$, $\phi_{t1}=$ 0 dB, and $\sigma_{s1}=0.5$ bits/s/Hz, (b) $m_{m2}=m_{t2}=$ 35, $\phi_{t2}=$ 0 dB, and $\sigma_{s2}=0.5$ bits/s/Hz.}
    \label{fig:10}
\end{figure*}



Figures \ref{fig:5}a and \ref{fig:5}b show the impacts of shadowing severity of multicast and eavesdropper channels on the physical layer security performance. We consider two groups with $\mathcal{M}$= 2, and 6. In the first group, we vary the shadowing severity of the $\mathcal{K} \rightarrow \mathcal{M}$ and $\mathcal{K} \rightarrow \mathcal{N}$ links simultaneously while in the second group we vary the severity of shadowing of the $\mathcal{K} \rightarrow \mathcal{M}$ links. It can clearly be observed that with the equally increasing values of $m_{m1}$ and $m_{t1}$, and $m_{m2}$ and $m_{t2}$, the secrecy performance is enhanced, and  for increasing values of $m_{m1}$ and $m_{m2}$ individually, the performance also becomes better. This is because increasing $m_{m1}$ and $m_{m2}$ from 0 to $\infty$ indicates the shadowing severity changing from a stronger to weaker conditions while an increasing $m_{t1}$ and $m_{t2}$ reduces the shadowing severity of the eavesdropper links. The authors in \cite{9331252} also observed some similar results as ours which is a clear indication that the results corresponding to our proposed dual-hop model are valid.

Figures \ref{fig:6}a and \ref{fig:6}b are depicted to explain the impacts of the number of relays on the PNSMC performance. We consider two cases with $\mathcal{N}=2$ and $\mathcal{N}=10$. The numerical results reveal that for both cases the proposed system exhibits an enhanced level of secrecy with the increasing number of relays $\mathcal{K}$. This result is also supported by \cite{6480923}. An increased $\mathcal{K}$ represents an improved cooperative diversity and at the same time, the selection of the opportunistic channels in terms of best relaying ensures maximum capacity at the receive terminals thereby increasing the secrecy capacity. 

In Figures \ref{fig:7}a and \ref{fig:7}b, the PNSMC is shown against the average SNR of multicast channels. The effects of the number of eavesdroppers $\mathcal{N}$ are shown assuming two groups with $\phi_{t1}=\phi_{t2}=$ 0 dB and 5 dB. As we consider maximum impacts of the eavesdroppers (maximum SNR at the eavesdropper) among $\mathcal{N}$ eavesdroppers (i.e. the worst case), it is evident that an increase in $\mathcal{N}$ will increase the probability of the existence of a strong eavesdropper channel in the system and hence the system's security will be degraded as testified in \cite{badrudduza2020enhancing}.

To demonstrate the effects of the number of multicast receivers in the proposed model, we depict Figures \ref{fig:8}a and \ref{fig:8}b by showing ESMC against the average SNR of multicast channels. It can be observed that the secrecy capacity deteriorates with the number of receivers. Note that in wireless multicasting, a specific bandwidth is allocated for a group of multicast receivers which is divided among them equally according to the multicast scheme. With the increase in numbers of multicast receivers within the same allocated bandwidth, the bandwidth per receiver is reduced. Hence the SNR and capacity at the receiver's terminals are also degraded which leads to a deteriorated secrecy performance as shown in \cite{9331252}.

In Figures \ref{fig:9}a and \ref{fig:9}b, the impacts of the shape parameters of the first-hop are depicted by showing ESMC against the average SNR of the multicast channels. We can observe that the ESMC decreases with $\eta_{p}$, but increases with $\mu_{p_{1}}$, $\mu_{p_{2}}$, and $\kappa_{p}$. The authors of \cite{9122938, sofotasios2018error} also obtained similar results that clearly verify our analysis.

\quad 
\noindent
\textbf{Generalization of the Existing Works: }

It is noteworthy that the $\kappa-\mu$/IG and $\eta-\mu$/IG models exhibits extreme versatility as generalized $\kappa-\mu$ and $\eta-\mu$ models are the special cases of those IG models. Moreover, $\kappa-\mu$ and $\eta-\mu$ models also unify the performance evaluation of some classical multipath models. In the IG models, only the shadowing parameters (i.e. $m_{m1}$, $m_{t1}$, $m_{m2}$, and $m_{t2}$) determine the amount of shadowing of the mean signal power. If 
$m_{m1}$, $m_{t1}$, $m_{m2}$, and $m_{t2}$ $\rightarrow \infty$, then we can assume that both the dominant and scattered components suffer from severe shadowing. On the other hand, if $m_{m1}$, $m_{t1}$, $m_{m2}$, and $m_{t2}$ $\rightarrow 0$, then no shadowing is observed in the channels and hence the mean signal power can be considered to be the deterministic one. For that particular case, $\kappa-\mu$/IG and $\eta-\mu$/IG models takes the form of $\kappa-\mu$ and $\eta-\mu$ fading models, respectively.


In Figure \ref{fig:10}a and \ref{fig:10}b the generic nature of the proposed scenarios has been shown. For $\eta-\mu$ fading channel, setting $\eta_{p}$ = $\eta_{m}$ = $\eta_{t}$ = 0.9 and $\mu_{p1}$ = $\mu_{m1}$ = $\mu_{t1}$ = 1 with $m_{m1}$, $m_{t1}$, $m_{m2}$, and $m_{t2}$ $\rightarrow \infty$, we can obtain Nakagami-$m$ channel where $\mu_{m1}$, $\mu_{m2}$, $\mu_{t1}$, and $\mu_{t2}$ becomes equivalent to the Nakagami fading parameter $m$. Based on this, letting $\mu_{m1}$, $\mu_{m2}$, $\mu_{t1}$, and $\mu_{t2}$ equal to 1, we get Rayleigh distribution. Similarly, Gaussian channels can be obtained with $\eta_{p}$ = $\eta_{m}$ = $\eta_{t}$ = 0.9 and $\mu_{p1}$ = $\mu_{m1}$ = $\mu_{t1}$ = 0.25, and Nakagami-$q$ fading channel can be obtained with $\eta_{p}$ = $\eta_{m}$ = $\eta_{t}$ = 0.25 and $\mu_{p1}$ = $\mu_{m1}$ = $\mu_{t1}$ = 0.5. In the case of $\kappa-\mu$ model, we can generate Rician fading channel letting $\kappa_{p}$ = $\kappa_{m}$ = $\kappa_{t}$ = 5.4, $\mu_{p2}$ = $\mu_{m2}$ = $\mu_{t2}$ = 0.9, where $\kappa_{p}$, $\kappa_{m}$, and $\kappa_{t}$ become equivalent to the Rician-$K$ factor. Following those conditions, Rayleigh fading is obtained setting $\kappa_{p}$ = $\kappa_{m}$ = $\kappa_{t}$ = 0. Nakagami-$m$ is generated considering $\kappa_{p}$ = $\kappa_{m}$ = $\kappa_{t}$ = 0.001 and $\mu_{p2}$ = $\mu_{m2}$ = $\mu_{t2}$ = 1.5, where $\mu_{p2}$, $\mu_{m2}$ and $\mu_{t2}$ is equivalent to Nakagami fading parameter $m$. Likewise, we can obtain one-sided Gaussian channel setting $\kappa_{p}$ = $\kappa_{m}$ = $\kappa_{t}$ = 0 and $\mu_{p2}$ = $\mu_{m2}$ = $\mu_{t2}$ = 0.5, and shadowed Rician with $\kappa_{p}$ = $\kappa_{m}$ = $\kappa_{t}$ = 6 , $\mu_{p2}$ = $\mu_{m2}$ = $\mu_{t2}$ = 1. Besides the proposed scenarios also provide a good match to the secure shadowed Rician- Rayleigh \cite{huang2018secrecy, 8506345}, and shadowed Rician- Nakagami-$m$ \cite{bankey2017secrecy, bankey2019physical} models. Hence, it can clearly be seen from the figures that the proposed scenarios (I and II) exhibits enormous versatility than any dual-hop multicast networks of the literature since a wide range of classical dual-hop multicast and broadcast models can be shown as special cases of the proposed work.

\section{CONCLUSION}
This work focuses on the secrecy analysis of dual-hop multicast communication systems in the presence of multiple eavesdroppers exploiting the best relay selection scheme. With a view to observing the impacts of each system parameter, we derive expressions for SOPM, PNSMC, and ESMC in closed-form which are further authenticated via Monte-Carlo simulations. It can clearly be seen from the numerical results that although fading, shadowing (both LOS and multiplicative), number of multicast receivers and eavesdroppers impose detrimental impacts on the secrecy performance, an acceptable secrecy level can till be maintained by first increasing the number of relays and then exploiting the best relay selection strategy. Moreover, the proposed model provides immense versatility since it can approximate a wide range of composite/multipath fading models of the literature. {\color{black} In future, our plan is to analyse the secrecy performance of multi-hop-multicast networks over IG composite fading models incorporating asymptotic analysis at high SNR regime.}

\appendices


\section{Proof of Dual-hop SNRs (Scenario I)}

\subsection{$\mathcal{S} \rightarrow \mathcal{k} \rightarrow \mathcal{M}$ link }
\label{dual1}

The PDF of $\Upsilon_{sm,1}$ and $\Upsilon_{st,1}$ is defined as \cite{shahriyer2021opportunistic}
\setcounter{eqnback}{\value{equation}} \setcounter{equation}{40}
\begin{align}\label{eqn:smp}
f_{sm,1}(\Upsilon)&=\frac{dF_{sm,1}(\Upsilon)}{d\Upsilon},
\\\label{eqn:sep}
f_{st,1}(\Upsilon)&=\frac{dF_{st,1}(\Upsilon)}{d\Upsilon},
\end{align}
where
$F_{sm,1}(\Upsilon)$ and $F_{st,1}(\Upsilon)$ denote the CDFs of $\Upsilon_{sm,1}$ and $\Upsilon_{st,1}$.
Now, the CDF of $\Upsilon_{sm,1}$ is defined as \cite[eq.~12]{kumar2015performance} 
\begin{align}\label{eqn:smc}
F_{sm,1}(\Upsilon)&=1-\Pr(\Upsilon_{sp,1}>\Upsilon)\Pr(\Upsilon_{pm,1}>\Upsilon),
\end{align} 
where
$\Pr(\Upsilon_{sp,1}>\Upsilon_{sm,1})$ and $\Pr(\Upsilon_{pm,1}>\Upsilon_{sm,1})$ are the complementary cumulative distribution functions (CCDFs) of $\Upsilon_{sp,1}$ and $\Upsilon_{pm,1}$, and the CCDFs are, respectively defined by \cite{5089994}
\begin{align}\label{eqn:smcc1}
\Pr(\Upsilon_{sp,1}>\Upsilon_{sm,1})&=\int_{\Upsilon_{sm,1}}^{\infty}f_{sp,1}(\Upsilon)d\Upsilon,
\\\label{eqn:smcc2}
\Pr(\Upsilon_{pm,1}>\Upsilon_{sm,1})&=\int_{\Upsilon_{sm,1}}^{\infty}f_{pm,1}(\Upsilon)d\Upsilon.
\end{align} 
Substituting \eqref{eqn:msp1} into \eqref{eqn:smcc1} and performing integration using \cite[eq~ 3.351.2]{GR:07:Book}, we have
\begin{align}\label{eqn:CDF1}
\nonumber
\Pr(\Upsilon_{sp,1}>\Upsilon_{sm,1})&=\sum _{n_{1}=0}^{\infty}\frac{\alpha_{2}}{\beta_{s}^{2\mu_{p1}+2n_{1}}}
\\
&\times \Gamma(2\mu_{p1}+2n_{1},\beta_{s}\Upsilon_{sm,1}).
\end{align}
Again substituting \eqref{eqn:mmp1} into \eqref{eqn:smcc2} and executing integration using \cite[eq~3.194.2]{GR:07:Book}, we obtain
\begin{align}\label{eqn:CDF2}
\nonumber
&\Pr(\Upsilon_{pm,1}>\Upsilon_{sm,1})=\sum _{r_{1}=0}^{\infty}\frac{\lambda_{2}\Upsilon_{sm,1}^{-m_{m1}}}{\beta_{m}^{m_{m1}+2\mu_{m1}+2r_{1}}m_{m1}}
\\
\nonumber
&\times\,_2F_{1}\left(m_{m1}+2\mu_{m1}+2r_{1},m_{m1};m_{m1}+1;\frac{-1}{\beta_{m}\Upsilon_{sm,1}}\right)
\\
\nonumber
&= \sum _{r_{2}=0}^{\infty}\sum _{r_{1}=0}^{\infty}\frac{\lambda_{2}\Upsilon_{sm,1}^{-m_{m1}}}{\beta_{m}^{m_{m1}+2\mu_{m1}+2r_{1}}m_{m1}}
\\
&\times \frac{\left(m_{m1}\right)_{r_{2}}\left(m_{m1}+2\mu_{m1}+2r_{1}\right)_{r_{2}}} {r_{2}! \left(m_{m1}+1\right)_{r_{2}}\left(-\beta_{m}\right)^{r_{2}}}.
\end{align}
After a few terms, the infinite series here quickly converges. \cite{634675}. Now, substituting \eqref{eqn:CDF1} and \eqref{eqn:CDF2} into \eqref{eqn:smc}, the CDF of $\Upsilon_{sm,1}$ is obtained as
\begin{align}\label{eqn:msmc}
\nonumber 
F_{sm,1}(\Upsilon)&=1-\sum _{r_{2}=0}^{\infty}\sum_{r_{1}=0}^{\infty}\sum _{n_{1}=0}^{\infty}\Lambda_{2}\Upsilon^{-m_{m1}-r_{2}}
\\
&\times \Gamma(2\mu_{p1}+2n_{1},\beta_{s}\Upsilon),
\end{align}
where
$\Lambda_{2}=\frac{\Lambda_{1}\left(m_{m1}\right)_{r_{2}}\left(m_{m1}+2\mu_{m1}+2r_{1}\right)_{r_{2}}} {r_{2}! \left(m_{m1}+1\right)_{r_{2}}\left(-\beta_{m}\right)^{r_{2}}}$, and
\\
$\Lambda_{1}=\frac{\alpha_{2}\lambda_{2}\beta_{s}^{-2\mu_{p1}-2n_{1}}}{\beta_{m}^{m_{m1}+2\mu_{m1}+2r_{1}}m_{m1}}$.
Substituting \eqref{eqn:msmc} into \eqref{eqn:smp} and performing differentiation with respect to $\Upsilon_{sm,1}$, the PDF of $\Upsilon_{sm,1}$ is obtained in \eqref{eqn:theorem 1}.

\subsection{$\mathcal{S} \rightarrow \mathcal{k} \rightarrow \mathcal{N}$ link }
\label{dual2}

Similar to \eqref{eqn:msmc}, the CDF of $\Upsilon_{st,1}$ is given by
\begin{align}\label{eqn:sec}
\nonumber 
F_{st,1}(\Upsilon)&=1-\sum _{\nu_{2}=0}^{\infty}\sum_{\nu_{1}=0}^{\infty}\sum _{n_{1}=0}^{\infty}\chi_{2}\Upsilon^{-m_{t1}-\nu_{2}}
\\
&\times \Gamma(2\mu_{p1}+2n_{1},\beta_{s}\Upsilon),
\end{align}
where $\chi_{2}=\frac{\chi_{1}\left(m_{t1}\right)_{\nu_{2}}\left(m_{t1}+2\mu_{t1}+2\nu_{1}\right)_{\nu_{2}}}
{\nu_{2}!\left(m_{t1}+1\right)_{\nu_{2}}\left(-\beta_{e}\right)^{\nu_{2}}}$, and $\chi_{1}=\frac{\xi_{2}\beta_{s}^{-2\mu_{p1}-2n_{1}}}{\beta_{e}^{m_{t1}+2\mu_{t1}+2\nu_{1}}m_{t1}}$,
Substituting \eqref{eqn:sec} into \eqref{eqn:sep} and performing differentiation with respect to $\Upsilon_{st,1}$, the PDF of $\Upsilon_{st,1}$ is obtained as shown in \eqref{eqn:theorem 2}.

\section{Proof of Best Relay Selection (Scenario I)}
\subsection{$\mathcal{S} \rightarrow \mathcal{k} \rightarrow \mathcal{M}$ link }
\label{best1}

Mathematically, the CDF of $\Upsilon_{bm,1}$ is defined as
\begin{align}\label{eqn:rCDF1}
F_{bm,1}(\Upsilon)=[F_{sm,1}(\Upsilon)]^{\mathcal{K}}.
\end{align}
Substituting \eqref{eqn:msmc} into \eqref{eqn:rCDF1}, the CDF of $\Upsilon_{bm,1}$ is obtained as
\begin{align}\label{eqn:mrCDF1}
\nonumber 
F_{bm,1}(\Upsilon)=&\biggl[1-\sum _{r_{2}=0}^{\infty}\sum _{r_{1}=0}^{\infty}\sum _{n_{1}=0}^{\infty} \Lambda_{2}\Upsilon^{-m_{m1}-r_{2}}
\\
&\times \Gamma(2\mu_{p1}+2n_{1},\beta_{s}\Upsilon)\biggl]^{\mathcal{K}}.
\end{align}
Differentiating \eqref{eqn:rCDF1} with respect to $\Upsilon_{sm,1}$, we get the PDF of $\Upsilon_{bm,1}$ as
\begin{align}\label{eqn:rPDF11}
f_{bm,1}(\Upsilon)&=\mathcal{K}f_{sm,1}(\Upsilon)[F_{sm,1}(\Upsilon)]^{\mathcal{K}-1}.
\end{align}
Substituting \eqref{eqn:msmc} and \eqref{eqn:theorem 1} into \eqref{eqn:rPDF11}, we obtain \eqref{eqn:theorem 3}.

\subsection{$\mathcal{S} \rightarrow \mathcal{k} \rightarrow \mathcal{N}$ link }
\label{best2}
Similar to \eqref{eqn:mrCDF1}, the CDF of $\Upsilon_{bt,1}$ is given by
\begin{align}\label{eqn:rCDF2}
\nonumber
F_{bt,1}(\Upsilon)=&\biggl[1-\sum _{\nu_{2}=0}^{\infty}\sum_{\nu_{1}=0}^{\infty}\sum _{n_{1}=0}^{\infty}\chi_{2}\Upsilon^{-m_{t1}-\nu_{2}}
\\
&\times\Gamma(2\mu_{p1}+2n_{1},\beta_{s}\Upsilon)\biggl]^{\mathcal{K}}.
\end{align}
Differentiating \eqref{eqn:rCDF2} with respect to $\Upsilon_{st,1}$, the PDF of $\Upsilon_{bt,1}$ can be obtained that is given by \eqref{eqn:theorem 4}.

\section{Proof of Dual-hop SNRs (Scenario II)}

\subsection{$\mathcal{S} \rightarrow \mathcal{k} \rightarrow \mathcal{M}$ link }
\label{dual.s2.5}
Denoting $\Upsilon_{sm,2}$, and $\Upsilon_{st,2}$ as the SNRs of $\mathcal{S}\rightarrow \mathcal{K} \rightarrow \mathcal{M}$, and $\mathcal{S}\rightarrow \mathcal{K}\rightarrow \mathcal{N}$ links, respectively, the PDFs of $\Upsilon_{sm,2}$ and $\Upsilon_{st,2}$ are given by
\begin{align}\label{eqn:pPDF1}
 f_{sm,2}(\Upsilon)=\frac{dF_{sm,2}(\Upsilon)}{d\Upsilon},
 \\
 \label{eqn:pPDF2}
 f_{st,2}(\Upsilon)=\frac{dF_{st,2}(\Upsilon)}{d\Upsilon}, 
\end{align}
where $F_{sm,2}(\Upsilon)$ and $F_{st,2}(\Upsilon)$ denote the CDFs of $\Upsilon_{sm,2}$ and $\Upsilon_{st,2}$.
Similar to \eqref{eqn:smc}, the CDF of $\Upsilon_{sm,2}$ is defined by
\begin{align}\label{eqn:cCDF}
F_{sm,2}(\Upsilon)=1-\Pr(\Upsilon_{sp,2}>\Upsilon)\Pr(\Upsilon_{pm,2}>\Upsilon),
\end{align}
where $\Pr(\Upsilon_{sp,2}>\Upsilon_{sm,2})$ and $\Pr(\Upsilon_{pm,2}>\Upsilon_{sm,2})$ are the complementary cumulative distribution functions (CCDFs) of $\Upsilon_{sp,2}$ and $\Upsilon_{pm,2}$ which are, respectively defined by
\begin{align}\label{eqn:cCDF1}
\Pr(\Upsilon_{sp,2}>\Upsilon_{sm,2})=\int_{\Upsilon_{sm,2}}^{\infty}f_{sp,2}(\Upsilon)d\Upsilon,
 \\
 \label{eqn:cCDF2}
\Pr(\Upsilon_{pm,2}>\Upsilon_{sm,2})=\int_{\Upsilon_{sm,2}}^{\infty}f_{pm,2}(\Upsilon)d\Upsilon.
\end{align}
Substituting \eqref{eqn:msp2} into \eqref{eqn:cCDF1} and performing integration using \cite[eq.~ 3.351.2]{GR:07:Book}, we get
\begin{align}\label{eqn:cCDF11}
\nonumber
 \Pr(\Upsilon_{sp,2}>\Upsilon_{sm,2})&=\sum_{\rho_{1}=0}^{\infty}\delta_{2}\delta_{b_{1}}^{-\mu_{p2}-\rho_{1}}
 \\
 &\times \Gamma(\mu_{p2}+\rho_{1},\delta_{b_{1}}\Upsilon_{sm,2}),
\end{align}
where $\delta_{b_{1}}=\frac{\mu_{p2}(1+\kappa_{p})}{\phi_{p2}}$.
Again substituting \eqref{eqn:mmp2} into \eqref{eqn:cCDF2} and performing integration, we get
\begin{align}\label{eqn:cCDF22}
\nonumber 
&\Pr(\Upsilon_{pm,2}>\Upsilon_{sm,2})=\sum _{\rho _2=0}^\infty \frac{\delta_{5}\Upsilon_{sm,2}^{-m_{m2}}}{\delta_{4}^{m_{m2}+\mu_m2+\rho_{2}}m_{m2}}
\\
\nonumber
&\times\,_2F_1\left(m_{m2}+\mu_{m2}+\rho_{2},m_{m2};m_{m2}+1;\frac{-1}{\delta_{4}\Upsilon_{sm,2}}\right)
\\
\nonumber
&=\sum _{\rho_{3}=0}^\infty\sum_{\rho_2=0}^\infty \frac{\delta_{5}\Upsilon_{sm,2}^{-m_{m2}}}{\delta_{4}^{m_{m2}+\mu_m2+\rho_{2}}m_{m2}}
\\
&\times\frac{\left(m_{m2}+\mu_{m2}+\rho _2\right)_{\rho _3}\left(m_{m2}\right)_{\rho_3}}{(-\delta_4)^{\rho_3}\rho_3! \left(m_{m2}+1\right)_{\rho_3}}. 
\end{align}
Where the infinite series converges rapidly as described in \cite{sofotasios2013eta}. Substituting \eqref{eqn:cCDF11}, and \eqref{eqn:cCDF22} into \eqref{eqn:cCDF}, the CDF of $\Upsilon_{sm,2}$ is obtained as 
\begin{align}\label{eqn:mcCDF}
\nonumber
F_{sm,2}(\Upsilon)&=1-\sum_{\rho_3=0}^\infty\sum_{\rho _2=0}^\infty\sum_{\rho_{1}=0}^{\infty}\delta_{7}\Upsilon^{-m_{m2}-\rho_{3}}
\\
&\times \Gamma(\mu_{p2}+\rho_{1},\delta_{b_{1}}\Upsilon),
\end{align}
where $\delta_7=\delta_6\frac{\left(m_{m2}+\mu_{m2}+\rho _2\right)_{\rho_3}\left(m_{m2}\right)_{\rho_3}}{(-\delta_4)^{\rho_3}\rho_3!\left(m_{m2}+1\right)_{\rho_3}}$, and $\delta_{6}=\frac{\delta_{2}\delta_{5}\delta_{b_{1}}^{-\mu_{p2}-\rho_{1}}}{\delta_{4}^{m_{m2}+\mu_{m2}+\rho_{2}}m_{m2}}$. Substituting \eqref{eqn:mcCDF} into \eqref{eqn:pPDF1} and performing differentiation with respect to $\Upsilon_{sm,2}$, the PDF of $\Upsilon_{sm,2}$ is obtained in \eqref{eqn:theorem 5}.

\subsection{$\mathcal{S} \rightarrow \mathcal{k} \rightarrow \mathcal{N}$ link}
\label{dual.s2.6}
Similar to \eqref{eqn:mcCDF}, the CDF of $\Upsilon_{st,2}$ is given by
\begin{align}\label{eqn:ecCDF} 
\nonumber
F_{st,2}(\Upsilon)&=1-\sum _{\sigma _2=0}^\infty\sum_{\rho_{1}=0}^{\infty}\sum _{\sigma _3=0}^\infty\beta_{7}\Upsilon^{-m_{t2}-\sigma_{3}}\\
&\times \Gamma(\mu_{p2}+\rho_{1},\delta_{b_{1}}\Upsilon),
\end{align}
where $\beta_{7}=\beta_6\frac{ \left(m_{t2}+\mu_{t2}+\sigma _2\right)_{\sigma _3}\left(m_{t2}\right)_{\sigma_3}}{(-\beta_2)^{\sigma _3}\sigma _3! \left(m_{t2}+1\right)_{\sigma _3}}$, and
$\beta_{6}=\frac{\delta_{2}\beta_{3}\delta_{b_{1}}^{-\mu_{p2}-\sigma_{1}}}{\beta_{2}^{m_{t2}+\mu_{t2}+\sigma_{2}}m_{t2}}$.
Now, substituting \eqref{eqn:ecCDF} into \eqref{eqn:pPDF2} and performing differentiation with respect to $\Upsilon_{st,2}$, we obtain \eqref{eqn:theorem 6}.

\section{Proof of Best Relay Selection (Scenario II)}
\subsection{$\mathcal{S} \rightarrow \mathcal{k} \rightarrow \mathcal{M}$ link }
\label{dual5}

The CDF of $\Upsilon_{bm,2}$ is defined as
\begin{align}\label{eqn:cdcCDF}
    F_{bm,2}(\Upsilon)=[F_{sm,2}(\Upsilon)]^{\mathcal{K}}.
\end{align}
Now, substituting \eqref{eqn:mcCDF} into \eqref{eqn:cdcCDF}, the CDF of $\Upsilon_{bm,2}$ is obtained as
\begin{align}\label{eqn:fcdcCDF}
\nonumber 
F_{bm,2}(\Upsilon)=&[1-\sum_{\rho_3=0}^\infty\sum_{\rho _2=0}^\infty\sum_{\rho_{1}=0}^{\infty}\delta_{7}\Upsilon_{sm,2}^{-m_{m2}-\rho_{3}}
\\
&\times \Gamma(\mu_{p2}+\rho_{1},\delta_{b_{1}}\Upsilon_{sm,2})]^{\mathcal{K}}.
\end{align}   
Differentiating \eqref{eqn:cdcCDF} with respect to $\Upsilon_{sm,2}$, we get the PDF of $\Upsilon_{bm,2}$ as
\begin{align}\label{eqn:dcdcCDF}
f_{bm,2}(\Upsilon)=\mathcal{K} f_{sm,2}(\Upsilon)[F_{sm,2}(\Upsilon)]^{\mathcal{K}-1}.
\end{align}
 Now substituing \eqref{eqn:mcCDF} , and \eqref{eqn:theorem 5} into \eqref{eqn:dcdcCDF}, we get $f_{bm,2}(\Upsilon)$ as shown in \eqref{eqn:theorem 7}.

\subsection{$\mathcal{S} \rightarrow \mathcal{k} \rightarrow \mathcal{N}$ link }
\label{dual6}

Similar to \eqref{eqn:fcdcCDF}, the CDF of $\Upsilon_{bt,2}$ is given by
\begin{align}\label{eqn:cecCDF}
\nonumber
F_{bt,2}(\Upsilon)&=[1-\sum_{\sigma _3=0}^\infty\sum_{\sigma_2=0}^\infty\sum_{\rho_{1}=0}^{\infty}\beta_{7}\Upsilon^{-m_{t2}-\sigma_{3}}
 \\
 &\times \Gamma(\mu_{p2}+\rho_{1},\delta_{b_{1}}\Upsilon)]^{\mathcal{K}}.
\end{align}
Differentiating \eqref{eqn:cecCDF} with respect to $\Upsilon_{st,2}$, the PDF of $\Upsilon_{bt,2}$ is shown  in \eqref{eqn:theorem 8}.

\section{Proof of Multicast Channel Models}

\subsection{Scenario I }
\label{min1}

In the case of mixed $\eta-\mu$ and $\eta-\mu$ /IG composite fading channel, the PDF of $\sigma _{min,1}$ is given by
\begin{align}\label{eqn:min1}
f_{\sigma _{min,1}}(\Upsilon)=\mathcal{M}f_{bm,1}(\Upsilon)[1-F_{bm,1}(\Upsilon)]^{\mathcal{M}-1}.
\end{align}
Plugging \eqref{eqn:mrCDF1} and \eqref{eqn:theorem 3} into \eqref{eqn:min1} leads to
\begin{align}\label{eqn:min2}
\nonumber
&f_{\sigma_{min,1}}(\Upsilon)=\sum_{r_{2}=0}^{\infty}\sum_{r_{1}=0}^{\infty}\sum_{n_{1}=0}^{\infty}\Lambda_{2}\mathcal{M}\mathcal{K}\biggl[ e^{-\beta_{s}\Upsilon}
\\
\nonumber 
&\times \Upsilon^{-n_{2}+2\mu_{p1}+2n_{1}}+\Upsilon^{-n_{2}}\Gamma(2\mu_{p1}+2n_{1},\beta_{s}\Upsilon)\biggl]
\\
\nonumber 
&\times\biggl[1-\sum_{r_{2}=0}^{\infty}\sum _{r_{1}=0}^{\infty}\sum_{n_{1}=0}^{\infty}\Lambda_{2}\Upsilon^{-m_{m1}-r_{2}}\Gamma(2\mu_{p1}+2n_{1},\beta_{s}\Upsilon)\biggl]^{\mathcal{K}-1}
\\
\nonumber
&\times\biggl[1-\biggl(1-\sum_{r_{2}=0}^{\infty}\sum _{r_{1}=0}^{\infty}\sum _{n_{1}=0}^{\infty} \Lambda_{2}\Upsilon^{-m_{m1}-r_{2}}
\\
& \times \Gamma(2\mu_{p1}+2n_{1},\beta_{s}\Upsilon)\biggl)^{\mathcal{K}}\biggl]^{\mathcal{M}-1}.
\end{align}
Using binomial expansion of \cite[eq~ 1.111]{GR:07:Book}, \eqref{eqn:min2} can be expressed as
\begin{align}\label{eqn:min3}
\nonumber
&f_{\sigma_{min,1}}(\Upsilon)=\sum_{r_{5}=0}^{\mathcal{M}-1}\sum_{r_{6}=0}^{\mathcal{K}+\mathcal{K}r_{5}-1}\sum _{r_{2}=0}^{\infty}\sum _{r_{1}=0}^{\infty}\sum _{n_{1}=0}^{\infty}\mathcal{M}\mathcal{K}\Lambda_{6}\Lambda_{2}
\\
\nonumber
&\times \biggl[\Upsilon^{-n_{2}+2\mu_{p1}+2n_{1}}e^{-\beta_{s}\Upsilon}+\Upsilon^{-n_{2}}\Gamma(2\mu_{p1}+2n_{1},\beta_{s}\Upsilon)\biggl]
\\
&\times\biggl[\sum _{r_{1}=0}^{\infty}\sum _{n_{1}=0}^{\infty}\sum _{r_{2}=0}^{\infty} \Lambda_{2}\Upsilon^{-m_{m1}-r_{2}}\Gamma(2\mu_{p1}+2n_{1},\beta_{s}\Upsilon)\biggl]^{r_{6}},
\end{align}
where
$\Lambda_{6}={(-1)^{r_{5}+r_{6}}}\binom{\mathcal{M}-1}{r_{5}}\binom{\mathcal{K}+\mathcal{K}r_{5}-1}{r_{6}}$. Making use of \cite[eq~ 8.352.7]{GR:07:Book}, \eqref{eqn:min3} can be further simplified as
\begin{align}\label{eqn:min4}
\nonumber 
&f_{\sigma_{min,1}}(\Upsilon)=\sum_{r_{5}=0}^{\mathcal{M}-1}\sum_{r_{6}=0}^{\mathcal{K}+\mathcal{K}r_{5}-1}\sum _{r_{2}=0}^{\infty}\sum _{r_{1}=0}^{\infty}\sum _{n_{1}=0}^{\infty}\mathcal{M}\mathcal{K}\Lambda_{6}\Lambda_{2}
\\
\nonumber
&\times \biggl[\Upsilon^{-n_{2}+2\mu_{p1}+2n_{1}}+\sum_{r_{7}=0}^{2\mu_{p1}+2n_{1}-1}\Lambda_{7}\Upsilon^{-n_{2}+r_{7}}\biggl]
\\
&\times e^{-\beta_{s}\Upsilon}[\Theta_{1}(\Upsilon)]^{r_{6}},
\end{align}
where $\Lambda_{7}= \frac{\Gamma(2\mu_{p1}+2n_{1})}{r_{7}!\beta_{s}^{-r_{7}}}$. Here,
\begin{align}\label{eqn:mult1}
\Theta_{1}(\Upsilon)=\sum_{r_{8}=0}^{2\mu_{p1}+2n_{1}-1}\sum _{r_{2}=0}^{\infty}\sum _{r_{1}=0}^{\infty}\sum _{n_{1}=0}^{\infty}\Lambda_{8}\Upsilon^{-m_{m1}-r_{2}+r_{8}}e^{-\beta_{s}\Upsilon},
\end{align}
where
$\Lambda_{8}=\frac{\Lambda_{2}\Gamma(2\mu_{p1}+2n_{1})}
{r_{8}!\beta_{s}^{-r_{8}}}$. Using the multinomial theorem of \cite[eq.~ 7]{pena2014performance}, we get
\begin{align}\label{eqn:mult2}
\nonumber 
&[\Theta_{1}(\Upsilon)]^{r_{6}}=\Xi_{\psi_{r_{6}}}e^{-\phi_{\psi_{r_{6}}}\Upsilon}\Upsilon^{\varphi_{\psi_{r_{6}}}}
\\
&\times \sum_{\psi_{r_{6}}}\binom{r_{6}}{p_{0,0,0,0,\ldots,}p_{r_{8},r_{2},r_{1},n_{1},\ldots,}n_{2\mu_{p1}+2n_{1}-1,\infty,\infty,\infty}},
\end{align}
where
$\binom{b}{b_{1},b_{2},...,b_{m}}=\frac{b!}{b_{1}!,b_{2}!,\ldots,b_{m}!}$ denotes the multinomial coefficients, $\Xi_{\psi_{r_{6}}}=\Pi_{r_{8},r_{2},r_{1},n_{1}}\Lambda_{8}^{p_{r_{8},r_{2},r_{1},n_{1}}}$,  $\phi_{\psi_{r_{6}}}=\sum_{r_{8}}\sum_{r_{2}}\sum_{r_{1}}\sum_{n_{1}}\beta_{s}p_{r_{8},r_{2},r_{1},n_{1}}$,
and $\varphi_{\psi_{r_{6}}}=\sum_{r_{8}}\sum_{r_{2}}\sum_{r_{1}}\sum_{n_{1}}(-m_{m1}-r_{2}+r_{8})p_{r_{8},r_{2},r_{1},n_{1}}$. The sum of \eqref{eqn:mult2} is to be carried out for each element of $\psi_{r_{6}}$, which is defined by
\begin{align}\label{eqn:mult3}
\nonumber
\psi_{r_{6}}&=\{(p_{0,0,0,0,\ldots,}p_{r_{8},r_{2},r_{1},n_{1},\ldots,}n_{2\mu_{p1}+2n_{1}-1,\infty,\infty,\infty}):
\\\nonumber 
&p_{r_{8},r_{2},r_{1},n_{1}}\in N  0\leqslant r_{8}\leqslant 2\mu_{p1}+2n_{1}-1, 0\leqslant r_{2}\leqslant \infty, 
\\
&0\leqslant r_{1}\leqslant \infty, 0\leqslant n_{1}\leqslant \infty; \sum_{r_{8},r_{2},r_{1},n_{1}}p_{r_{8},r_{2},r_{1},n_{1}}=r_{6}\}.
\end{align}
After modifying \eqref{eqn:mult2}, we get
\begin{align}\label{eqn:mult4}
[\Theta_{1}(\Upsilon)]^{r_{6}}=\sum_{\psi_{r_{6}}}\Lambda_{9}e^{-\phi_{\psi_{r_{6}}}\Upsilon}\Upsilon^{\varphi_{\psi_{r_{6}}}},
\end{align}
where
$\Lambda_{9}=\binom{r_{6}}{p_{0,0,0,0,\ldots,}p_{r_{8},r_{2},r_{1},n_{1},\ldots,}n_{2\mu_{p1}+2n_{1}-1,\infty,\infty,\infty}}\Xi_{\psi_{r_{6}}}$.
Substituting \eqref{eqn:mult4} into \eqref{eqn:min4}, we get the final expression of $f_{\sigma_{min,1}}(\Upsilon)$ in \eqref{eqn:theorem 9}.

\subsection{Scenario II } 
\label{min2}

For the mixed $\kappa-\mu$ and $\kappa-\mu$ /IG composite fading channel, the PDF of $\sigma _{min,2}$ is given by 

\begin{align}\label{eqn:mullPDF}
f_{\sigma _{min,2}}(\Upsilon)=\mathcal{M}f_{bm,2}(\Upsilon)[1-F_{bm,2}(\Upsilon)]^{\mathcal{M}-1}.
\end{align}
Substituting \eqref{eqn:fcdcCDF} and \eqref{eqn:theorem 7} into \eqref{eqn:mullPDF}, we get 
\begin{align}\label{eqn:mullPDF1}
\nonumber
&f_{\sigma _{min,2}}(\Upsilon)=\sum_{\rho _3=0}^\infty\sum_{\rho_2=0}^\infty\sum_{\rho_{1}=0}^{\infty}\delta_{7}\mathcal{M}\mathcal{K}\biggl[e^{-\Upsilon\delta_{b_{1}}}
\\
\nonumber
&\times \Upsilon^{-\rho_{3}-m_{m2}-1+\mu_{p2}+\rho_{1}}
+\Upsilon^{-1-\rho_{3}-m_{m2}}
\\
\nonumber
&\times \Gamma[\mu_{p2}+\rho_{1},\Upsilon\delta_{b_{1}}]\biggl]
\\
\nonumber 
&\times\biggl[1-\sum _{\rho _2=0}^\infty\sum_{\rho_{1}=0}^{\infty}\sum _{\rho _3=0}^\infty\delta_{7}\Upsilon^{-m_{m2}-\rho_{3}}
\Gamma(\mu_{p2}+\rho_{1},\delta_{b_{1}}\Upsilon)\biggl]^{\mathcal{K}-1}
\\
\nonumber
&\times\biggl[1-\biggl(1-\sum _{\rho _2=0}^\infty\sum_{\rho_{1}=0}^{\infty}\sum _{\rho _3=0}^\infty\delta_{7}\Upsilon^{-m_{m2}-\rho_{3}}
\\
&\times \Gamma(\mu_{p2}+\rho_{1},\delta_{b_{1}}\Upsilon)\biggl)^{\mathcal{K}}\biggl]^{\mathcal{M}-1}. 
\end{align} 
Using \cite[eq.~ 1.111]{GR:07:Book}, \eqref{eqn:mullPDF1} can be simplified as  
\begin{align}\label{eqn:fmmulPDF1}
\nonumber
&f_{\sigma _{min,2}}(\Upsilon)=\sum_{\rho_{6}=0}^{\mathcal{M}-1}\sum_{\rho_{7}=0}^{\mathcal{K}+\mathcal{K}\rho_{6}-1}\sum_{\rho_2=0}^\infty\sum_{\rho_{1}=0}^{\infty}\sum _{\rho _3=0}^\infty\mathcal{M}\mathcal{K}\delta_{10}\delta_{7}
\\
\nonumber 
&\times \biggl[e^{-\Upsilon\delta_{b_{1}}}\Upsilon^{-\rho_{3}-m_{m2}-1+\mu_{p2}+\rho_{1}}
\\
\nonumber
&+\Upsilon^{-1-\rho_{3}-m_{m2}}\Gamma[\mu_{p2}+\rho_{1},\Upsilon\delta_{b_{1}}]\biggl]\times\biggl[\sum_{\rho _2=0}^\infty\sum_{\rho_{1}=0}^{\infty}   
\\
&\times\sum_{\rho _3=0}^\infty\delta_{7}\Upsilon^{-m_{m2}-\rho_{3}}
\Gamma(\mu_{p2}+\rho_{1},\delta_{b_{1}}\Upsilon)\biggl]^{\rho_{7}},
\end{align} 
where $\delta_{10}=(-1)^{\rho_{6}+\rho_{7}}\binom{\mathcal{M}-1}{\rho_{6}}\binom{\mathcal{K}+\mathcal{K}\rho_{6}-1}{\rho_{7}}$. Again using \cite[eq.~ 8.352.7]{GR:07:Book}, \eqref{eqn:fmmulPDF1} can be further simplified as
\begin{align}\label{eqn:fmulPDF2}
\nonumber 
&f_{\sigma _{min,2}}(\Upsilon)=\sum_{\rho_{6}=0}^{\mathcal{M}-1}\sum_{\rho_{7}=0}^{\mathcal{K}+\mathcal{K}\rho_{6}-1}\sum _{\rho _3=0}^\infty\sum_{\rho_2=0}^\infty\sum_{\rho_{1}=0}^{\infty}\mathcal{M}\mathcal{K}\delta_{10}\delta_{7} \\  
\nonumber
&\times e^{-\Upsilon\delta_{b_{1}}}\biggl[\Upsilon^{\delta_{11}+\mu_{p2}+\rho_{1}}+\sum_{\rho_{8}=0}^{\mu_{p2}+\rho_{1}-1}\delta_{12}\Upsilon^{\delta_{11}+\rho_{8}}\biggl]
\\
&\times\biggl[\Theta_{2}(\Upsilon)\biggl]^{\rho_{7}},
\end{align} 
where $\delta_{11}=-m_{m2}-\rho_{3}-1$ , $\delta_{11}+1=-m_{m2}-\rho_{3}$, and $\delta_{12}=\frac{\Gamma(\mu_{p2}+\rho_{1})}{(\rho_{8})!\delta_{b_{1}}^{-\rho_{8}}}$.     
Here  
\begin{align}\label{eqn:fmulPDF3}
\Theta_{2}(\Upsilon)= \sum_{\rho_{9}=0}^{\mu_{p2}+\rho_{1}-1}
\sum_{\rho_{3}=0}^{\infty}\sum_{\rho _2=0}^{\infty}\sum_{\rho_{1}=0}^{\infty}
\delta_{13}\Upsilon^{\delta_{11}+\rho_{9}+1}e^{-\Upsilon\delta_{b_{1}}},   
\end{align}
where
$\delta_{13}=\frac{\delta_{7}\Gamma(\mu_{p2}+\rho_{1})}{\rho_9! (\delta_{b_{1}})^{-\rho_{9}}}$. Using the multinomial theorem of \cite[eq.~ 7]{pena2014performance}, we get   
\begin{align}\label{eqn:bisum}
\nonumber 
&[\Theta_{2}(\Upsilon)]^{\rho_{7}}=\chi_{\psi_{\rho_{7}}}e^{-\Theta_{\psi_{\rho_{7}}}\Upsilon}\Upsilon^{\eta_{\psi_{\rho_{7}}}}
\\
&\times \sum_{\psi_{\rho_{7}}}^{}\binom{\rho_{7}}{g_{0,0,0,0},\ldots,g_{\rho_{9},\rho_{3}\rho_{2},\rho_{1}},\ldots,n_{\mu_{p2}+\rho_{1}-1,\infty,\infty,\infty}},
\end{align}  
where $\binom{a}{a_{1},a_{2},\ldots,a_{m}}=\frac{a!}{a_{1}!a_{2}!\ldots a_{m}!}$ denotes the multinomial coefficients, $\chi_{\psi_{\rho_{7}}}=\Pi_{\rho_{9},\rho_{3},\rho_{2},\rho_{1}}\delta_{13}^{g_{\rho_{9},\rho_{3},\rho_{2},\rho_{1}}}$, $\eta_{\psi_{\rho_{7}}}=\sum_{\rho_{9}}\sum_{\rho_{3}}\sum_{\rho_{2}}\sum_{\rho_{1}}^{(\delta_{11}+\rho_{9}+1){g_{\rho_{9},\rho_{3},\rho_{2},\rho_{1}}}}$ and $\Theta_{\psi_{\rho_{7}}}=\sum_{\rho_{9}}\sum_{\rho_{3}}\sum_{\rho_{2}}\sum_{\rho_{1}}\delta_{b_{1}}{g_{\rho_{9},\rho_{3},\rho_{2},\rho_{1}}}$. For each element of $\psi_{\rho_{7}}$, the sum in \eqref{eqn:bisum} is to be performed, which can be defined as
\begin{align}\label{eqn:bisum2}
\nonumber
\psi_{\rho_{7}}=&(g_{0,0,0,0},\ldots,g_{\rho_{9},\rho_{3}\rho_{2},\rho_{1}},\ldots,n_{\mu_{p2}+\rho_{1}-1,\infty,\infty,\infty}):
\\\nonumber 
&g_{\rho_{9},\rho_{3},\rho_{2},\rho_{1}}\in \mathbb{N}  0\leq\rho_{9}\leq\mu_{p2}+\rho_{1}-1, 0\leq\rho_{3}\leq \infty, 
\\
& 0\leq\rho_{2}\leq \infty, 0\leq\rho_{1}\leq \infty; \sum_{\rho_{9},\rho_{3},\rho_{2},\rho_{1}}^{}g_{\rho_{9},\rho_{3},\rho_{2},\rho_{1}}=\rho_{7}.
\end{align}
Substituting \eqref{eqn:bisum} into \eqref{eqn:fmulPDF2}, we get the final expression of  $f_{\sigma_{min,2}}(\Upsilon_{bm,2})$ in \eqref{eqn:theorem 10}.

\section{Proof of Eavesdropper Channel Models}
 
\subsection{Scenario I }
\label{max1}
The PDF of $\sigma _{max,1}$ in the case of mixed $\eta-\mu$ and $\eta-\mu$ /IG composite fading channel is given by 
\begin{align}\label{eqn:max1}
f_{\sigma _{max,1}}(\Upsilon)=\mathcal{N}f_{bt,1}(\Upsilon)[F_{bt,1}(\Upsilon)]^{\mathcal{N}-1}.
\end{align}
Substituting \eqref{eqn:rCDF2}, and \eqref{eqn:theorem 4} into \eqref{eqn:max1}, we get
\begin{align}\label{eqn:max2}
\nonumber
&f_{\sigma_{max,1}}(\Upsilon)=\sum_{\nu_{1}=0}^{\infty}\sum _{n_{1}=0}^{\infty}\sum _{\nu_{2}=0}^{\infty}\chi_{2}\mathcal{N}\mathcal{K}\biggl[e^{-\beta_{s}\Upsilon}
\\
\nonumber 
&\times \Upsilon^{-n_{3}+2\mu_{p1}+2n_{1}}+\Upsilon^{-n_{3}}\Gamma(2\mu_{p1}+2n_{1},\beta_{s}\Upsilon)
\biggl]
\\
\nonumber 
&\times \biggl[1-\sum_{\nu_{2}=0}^{\infty}\sum_{\nu_{1}=0}^{\infty}\sum _{n_{1}=0}^{\infty}\chi_{2}\Upsilon^{-m_{t1}-\nu_{2}}\Gamma(2\mu_{p1}+2n_{1},\beta_{s}\Upsilon)\biggl]^{\mathcal{K}-1}
\\
\nonumber
&\times\biggl[[1-\sum_{\nu_{2}=0}^{\infty}\sum_{\nu_{1}=0}^{\infty}\sum _{n_{1}=0}^{\infty}\chi_{2}\Upsilon^{-m_{t1}-\nu_{2}}
\\
&\times \Gamma(2\mu_{p1}+2n_{1},\beta_{s}\Upsilon)]^{\mathcal{K}}\biggl]^{\mathcal{N}-1}.
\end{align}
Simplifying \eqref{eqn:max2}, we get
\begin{align}\label{eqn:max3}
\nonumber 
&f_{\sigma_{max,1}}(\Upsilon)=\sum_{\nu_{6}=0}^{\mathcal{K}\mathcal{N}-1}\sum_{\nu_{2}=0}^{\infty}\sum_{\nu_{1}=0}^{\infty}\sum _{n_{1}=0}^{\infty}\mathcal{N}\mathcal{K}\chi_{6}\chi_{2}e^{-\beta_{s}\Upsilon}
\\
\nonumber
&\times\biggl[\Upsilon^{-n_{3}+2\mu_{p1}+2n_{1}}+\sum_{\nu_{7}=0}^{2\mu_{p1}+2n_{1}-1}\chi_{7}\Upsilon^{-n_{3}+\nu_{7}}\biggl]
\\
&\times\biggl[\Theta_{3}(\Upsilon)\biggl]^{\nu_{6}},
\end{align}
where
$\chi_{6}=\binom{\mathcal{K}\mathcal{N}-1}{\nu_{6}}(-1)^{\nu_{6}}$, and $\chi_{7}=\frac{\Gamma(2\mu_{p1}+2n_{1})}{\nu_{7}!\beta_{s}^{-\nu_{7}}}$.
Here
\begin{align}\label{eqn:emult1}
\Theta_{3}(\Upsilon)=\sum_{\nu_{8}=0}^{2\mu_{p1}+2n_{1}-1}\sum_{\nu_{2}=0}^{\infty}\sum_{\nu_{1}=0}^{\infty}\sum_{n_{1}=0}^{\infty}\chi_{8}\Upsilon^{-m_{t1}-\nu_{2}+\nu_{8}}e^{-\beta_{s}\Upsilon},    
\end{align}                                     
where  
$\chi_{8}=\frac{\chi_{2}\Gamma(2\mu_{p1}+2n_{1})}
{\nu_{8}!\beta_{s}^{-\nu_{8}}}$. Applying multinomial theorem of \cite[eq.~ 7]{pena2014performance}, we obtain
\begin{align}\label{eqn:emult2}
\nonumber 
&[\Theta_{3}(\Upsilon)]^{\nu_{6}}=\Xi_{\nu_{6}}e^{-\phi_{\nu_{6}}\Upsilon}\Upsilon^{\varphi_{\nu_{6}}}
\\
&\times \sum_{\psi_{\nu_{6}}}\binom{\nu_{6}}{q_{0,0,0,0,\ldots,}q_{\nu_{8},\nu_{2},\nu_{1},n_{1}.\ldots,}n_{2\mu_{p1}+2n_{1}-1,\infty,\infty,\infty}},
\end{align}
where $\Xi_{\psi_{\nu_{6}}}=\Pi_{\nu_{8},\nu_{2},\nu_{1},n_{1}}\chi_{8}^{q_{\nu_{8},\nu_{2},\nu_{1},n_{1}}}$,  $\phi_{\psi_{\nu_{6}}}=\sum_{\nu_{8}}\sum_{\nu_{2}}\sum_{\nu_{1}}\sum_{n_{1}}\beta_{s}q_{\nu_{8},\nu_{2},\nu_{1},n_{1}}$,and 
\\
$\varphi_{\psi_{\nu_{6}}}=\sum_{\nu_{8}}\sum_{\nu_{2}}\sum_{\nu_{1}}\sum_{n_{1}}(-m_{t1}-\nu_{2}+\nu_{8})q_{\nu_{8},\nu_{2},\nu_{1},n_{1}}$. Substituting \eqref{eqn:emult2} into \eqref{eqn:max3}, we get
 we get the final expression of $f_{\sigma_{max,1}}(\Upsilon)$ in \eqref{eqn:theorem 11}.

\subsection{Scenario II }
\label{max2}

In the case of mixed $\kappa-\mu$ and $\kappa-\mu$ /IG composite fading channel, the PDF of $\sigma _{max,2}$ is given by 
\begin{align}\label{eqn:eavPDF}
f_{\sigma _{max}}(\Upsilon)=\mathcal{N}f_{bt,2}(\Upsilon)[F_{bt,2}(\Upsilon)]^{\mathcal{N}-1}.
\end{align}
Substituting \eqref{eqn:cecCDF} and \eqref{eqn:theorem 8} into \eqref{eqn:eavPDF}, we get
\begin{align}\label{eqn:eavPDF1}
\nonumber
&f_{\sigma _{max,2}}(\Upsilon)=\sum_{\sigma _3=0}^\infty\sum_{\sigma_2=0}^\infty\sum_{\rho_{1}=0}^{\infty}\beta_{7}\mathcal{N}\mathcal{K}\biggl[e^{-\Upsilon\delta_{b_{1}}}
\\ 
\nonumber
&\times \Upsilon^{-\sigma_{3}-m_{t2}-1+\mu_{p2}+\sigma_{1}}+\Upsilon^{-1-\sigma_{3}-m_{t2}}
\\
\nonumber
&\times \Gamma[\mu_{p2}+\rho_{1},\Upsilon\delta_{b_{1}}]\biggl]
\\
\nonumber 
&\times \biggl[1-\sum_{\sigma _3=0}^\infty\sum_{\sigma _2=0}^\infty\sum_{\rho_{1}=0}^{\infty}\beta_{7}\Upsilon^{-m_{t2}-\sigma_{3}}\Gamma(\mu_{p2}+\rho_{1},\delta_{b_{1}}\Upsilon)\biggl]^{\mathcal{K}-1}
\\
\nonumber
&\times\biggl[[1-\sum_{\sigma _3=0}^\infty\sum_{\sigma _2=0}^\infty\sum_{\rho_{1}=0}^{\infty}\beta_{7}\Upsilon^{-m_{t2}-\sigma_{3}}
\\
&\times \Gamma(\mu_{p2}+\rho_{1},\delta_{b_{1}}\Upsilon)]^{\mathcal{K}}\biggl]^{\mathcal{N}-1}.
\end{align}
Simplifying \eqref{eqn:eavPDF1}, we get
\begin{align}\label{eqn:bieavmax}
\nonumber 
&f_{\sigma _{max,2}}(\Upsilon)=\sum_{\sigma_{10}=0}^{\mathcal{K}\mathcal{N}-1}\sum_{\sigma _3=0}^\infty\sum_{\sigma _2=0}^\infty\sum_{\rho_{1}=0}^{\infty}\mathcal{N}\mathcal{K}\beta_{15}\beta_{7}e^{-\Upsilon\delta_{b_{1}}}
\\\nonumber 
&\times\biggl[\Upsilon^{\beta_{10}+\mu_{p2}+\rho_{1}}+\sum_{\sigma_{11}=0}^{\mu_{p2}+\rho_{1}-1}\beta_{16}\Upsilon^{\beta_{10}+\sigma_{11}}\biggl]
\\
&\times \biggl[\Theta_{4}(\Upsilon)\biggl]^{\sigma_{10}},
\end{align}
where $\beta_{10}=-m_{t2}-\sigma_{3}-1$, $\beta_{10}+1=-m_{t2}-\sigma_{3}$, 
$\beta_{15}=\binom{\mathcal{K}\mathcal{N}-1}{\sigma_{10}}(-1)^{\sigma_{10}}$, and $\beta_{16}=\frac{\Gamma(\mu_{p2}+\rho_{1})}{(\sigma_{11})!\delta_{b_{1}}^{-\sigma_{11}}}$. Here

\begin{align}\label{eqn:bieavmax2}
\theta_{4}(\Upsilon)= \sum_{\sigma_{12}=0}^{\mu_{p2}+\rho_{1}-1} \sum_{\sigma_{3}=0}^{\infty}\sum_{\sigma_{2}=0}^{\infty}\sum_{\rho_{1}=0}^{\infty}\beta_{17}\Upsilon^{\beta_{10}+\sigma_{12}+1}e^{-\Upsilon\delta_{b_{1}}},
\end{align}
where
$\beta_{17}=\frac{\beta_{7}\Gamma(\mu_{p2}+\rho_{1})}{\sigma_{12}!(\delta_{b_{1}})^{-\sigma_{12}}}$. Using the multinomial theorem of \cite[eq.~ 7]{pena2014performance}, we get
\begin{align}\label{eqn:bieav}
\nonumber 
&[\Theta_{4}(\Upsilon)]^{\sigma_{10}}=\chi_{\psi_{\sigma_{10}}}e^{-\Theta_{\psi_{\sigma_{10}}}\Upsilon}\Upsilon^{\eta_{\psi_{\sigma_{10}}}}
\\
&\times \sum_{\psi_{\sigma_{10}}}\binom{\sigma_{10}}{f_{0,0,0,0},\ldots,f_{\sigma_{12},\sigma_{3},\sigma_{2},\rho_{1}},\ldots,n_{\mu+\rho_{1}-1,\infty,\infty,\infty}},
\end{align}
where 
$\chi_{\psi_{\sigma_{10}}}=\Pi_{\sigma_{12},\sigma_{3},\sigma_{2},\rho_{1}}\beta_{17}^{f_{\sigma_{12},\sigma_{3},\sigma_{2},\rho_{1}}}$, $\eta_{\psi_{\sigma_{10}}}=\sum_{\sigma_{12}}\sum_{\sigma_{3}}\sum_{\sigma_{1}}\sum_{\rho_{1}}(\beta_{10}+\sigma_{12}+1){f_{\sigma_{12},\sigma_{3},\sigma_{2},\rho_{1}}}$, and $\Theta_{\psi_{\sigma_{10}}}=\sum_{\sigma_{12}}\sum_{\sigma_{3}}\sum_{\sigma_{2}}\sum_{\rho_{1}}\delta_{b_{1}}{f_{\sigma_{12},\sigma_{3},\sigma_{2},\rho_{1}}}$.
Substituting \eqref{eqn:bieav} into \eqref{eqn:bieavmax}, we get the final expression of $f_{\sigma _{max,2}}(\Upsilon)$ in \eqref{eqn:theorem 12}. 

\bibliographystyle{IEEEtran}
\bibliography{IEEEabrv,main.bib}

\begin{thebibliography}{10}
\providecommand{\url}[1]{#1}
\csname url@samestyle\endcsname
\providecommand{\newblock}{\relax}
\providecommand{\bibinfo}[2]{#2}
\providecommand{\BIBentrySTDinterwordspacing}{\spaceskip=0pt\relax}
\providecommand{\BIBentryALTinterwordstretchfactor}{4}
\providecommand{\BIBentryALTinterwordspacing}{\spaceskip=\fontdimen2\font plus
\BIBentryALTinterwordstretchfactor\fontdimen3\font minus
  \fontdimen4\font\relax}
\providecommand{\BIBforeignlanguage}[2]{{%
\expandafter\ifx\csname l@#1\endcsname\relax
\typeout{** WARNING: IEEEtran.bst: No hyphenation pattern has been}%
\typeout{** loaded for the language `#1'. Using the pattern for}%
\typeout{** the default language instead.}%
\else
\language=\csname l@#1\endcsname
\fi
#2}}
\providecommand{\BIBdecl}{\relax}
\BIBdecl

\bibitem{lv2019secure}
L.~Lv, F.~Zhou, J.~Chen, and N.~Al-Dhahir, ``{S}ecure {C}ooperative
  {C}ommunications {W}ith an {U}ntrusted {R}elay: {A} {NOMA}-{I}nspired
  {J}amming and {R}elaying {A}pproach,'' \emph{IEEE Transactions on Information
  Forensics and Security}, vol.~14, no.~12, pp. 3191--3205, 2019.

\bibitem{moualeu2018physical}
J.~M. Moualeu, D.~B. da~Costa, W.~Hamouda, U.~S. Dias, and R.~A. de~Souza,
  ``Physical {L}ayer {S}ecurity {O}ver $\alpha $-$\kappa $-$\mu $ and $\alpha
  $-$\eta $-$\mu $ {F}ading {C}hannels,'' \emph{IEEE Transactions on Vehicular
  Technology}, vol.~68, no.~1, pp. 1025--1029, 2018.

\bibitem{diamanti2016practical}
E.~Diamanti, H.-K. Lo, B.~Qi, and Z.~Yuan, ``Practical challenges in quantum
  key distribution,'' \emph{npj Quantum Information}, vol.~2, no.~1, pp. 1--12,
  2016.

\bibitem{yoo2017kappa}
S.~K. Yoo, N.~Bhargav, S.~L. Cotton, P.~C. Sofotasios, M.~Matthaiou,
  M.~Valkama, and G.~K. Karagiannidis, ``The $\kappa$-$\mu$/inverse gamma and
  $\eta$-$\mu $/inverse gamma composite fading models: {F}undamental statistics
  and empirical validation,'' \emph{IEEE Trans. Commun.}, vol.~69, no.~8, pp.
  5514--5530, 2017.

\bibitem{bhatt2018asep}
M.~Bhatt and S.~K. Soni, ``\textsc{ASEP} {A}nalysis {O}ver {U}nified
  {L}ognormal {S}hadowed $\alpha$--$\eta$--$\mu$ and $\alpha$--$\kappa$--$\mu$
  {C}omposite {F}ading {C}hannels,'' in \emph{2018 Second International
  Conference on Intelligent Computing and Control Systems (ICICCS)}.\hskip 1em
  plus 0.5em minus 0.4em\relax IEEE, 2018, pp. 1126--1129.

\bibitem{yoo2015k}
S.~K. Yoo, S.~L. Cotton, P.~C. Sofotasios, M.~Matthaiou, M.~Valkama, and G.~K.
  Karagiannidis, ``The $\kappa$—$\mu$/{I}nverse gamma fading model,'' in
  \emph{2015 IEEE 26th annual international symposium on personal, indoor, and
  mobile radio communications (PIMRC)}.\hskip 1em plus 0.5em minus 0.4em\relax
  IEEE, 2015, pp. 425--429.

\bibitem{yoo2015eta}
S.~K. Yoo, P.~C. Sofotasios, S.~L. Cotton, M.~Matthaiou, M.~Valkama, and G.~K.
  Karagiannidis, ``The $\eta$—$\mu$/{I}nverse gamma composite fading model,''
  in \emph{2015 IEEE 26th annual international symposium on personal, indoor,
  and mobile radio communications (PIMRC)}.\hskip 1em plus 0.5em minus
  0.4em\relax IEEE, 2015, pp. 166--170.

\bibitem{yacoub2007kappa}
M.~D. Yacoub, ``The $\kappa$-$\mu$ distribution and the $\eta$-$\mu$
  distribution,'' \emph{IEEE Antennas and Propagation Magazine}, vol.~49,
  no.~1, pp. 68--81, 2007.

\bibitem{yacoub2007alpha}
------, ``The $\alpha-\mu$ distribution: {A} {P}hysical {F}ading {M}odel for
  the {S}tacy {D}istribution,'' \emph{IEEE Transactions on Vehicular
  Technology}, vol.~56, no.~1, pp. 27--34, 2007.

\bibitem{fraidenraich2003spl}
G.~Fraidenraich and M.~D. Yacoub, ``The $\lambda-\mu$ general fading
  distribution,'' in \emph{Proceedings of the 2003 SBMO/IEEE MTT-S
  International Microwave and Optoelectronics Conference-IMOC 2003.(Cat. No.
  03TH8678)}, vol.~1.\hskip 1em plus 0.5em minus 0.4em\relax IEEE, 2003, pp.
  49--54.

\bibitem{salahat2014performance}
E.~Salahat and A.~Hakam, ``Performance analysis of $\alpha$-$\eta$-$\mu$ and
  $\alpha$-$\kappa$-$\mu$ generalized mobile fading channels,'' in
  \emph{European Wireless 2014; 20th European Wireless Conference}.\hskip 1em
  plus 0.5em minus 0.4em\relax VDE, 2014, pp. 1--6.

\bibitem{yacoub2016alpha}
M.~D. Yacoub, ``The $\alpha-\eta-\kappa-\mu$ {F}ading {M}odel,'' \emph{IEEE
  Transactions on Antennas and Propagation}, vol.~64, no.~8, pp. 3597--3610,
  2016.

\bibitem{papazafeiropoulos2009eta}
A.~K. Papazafeiropoulos and S.~A. Kotsopoulos, ``The $\eta$-$\lambda$-$\mu$:
  {A} general fading distribution,'' in \emph{GLOBECOM 2009-2009 IEEE Global
  Telecommunications Conference}.\hskip 1em plus 0.5em minus 0.4em\relax IEEE,
  2009, pp. 1--5.

\bibitem{issaid2017fast}
C.~B. Issaid, M.-S. Alouini, and R.~Tempone, ``On the {F}ast and {P}recise
  {E}valuation of the {O}utage {P}robability of {D}iversity {R}eceivers {O}ver
  $\alpha-\mu$, $\kappa-\mu$, and $\eta-\mu$ {F}ading {C}hannels,'' \emph{IEEE
  Transactions on Wireless Communications}, vol.~17, no.~2, pp. 1255--1268,
  2017.

\bibitem{9230592}
A.~Hanif, A.~S.~M. Badrudduza, M.~S. Hossen, M.~K. Kundu, and M.~Z.~I. Sarkar,
  ``Secrecy {P}erformance {A}nalysis over \textsc{I}nverse \textsc{G}amma
  {C}omposite {M}ulticast {F}ading {C}hannels,'' in \emph{2020 IEEE Region 10
  Symposium (TENSYMP)}, 2020, pp. 949--952.

\bibitem{yoo2020effective}
S.~K. Yoo, S.~L. Cotton, P.~C. Sofotasios, S.~Muhaidat, and G.~K.
  Karagiannidis, ``Effective {C}apacity {A}nalysis over \textsc{G}eneralized
  {C}omposite {F}ading {C}hannels,'' \emph{IEEE Access}, 2020.

\bibitem{pena2014performance}
J.~P. Pena-Martin, J.~M. Romero-Jerez, and C.~Tellez-Labao, ``Performance of
  \textsc{S}election \textsc{C}ombining {D}iversity in $\eta-\mu$ {F}ading
  {C}hannels {W}ith {I}nteger {V}alues of $\mu$,'' \emph{IEEE Transactions on
  Vehicular Technology}, vol.~64, no.~2, pp. 834--839, 2014.

\bibitem{9330523}
A.~S.~M. {Badrudduza}, M.~{Ibrahim}, S.~M.~R. {Islam}, M.~S. {Hossen}, M.~K.
  {Kundu}, I.~S. {Ansari}, and H.~{Yu}, ``Security at the {P}hysical {L}ayer
  {O}ver \textsc{GG} {F}ading and \textsc{mEGG} {T}urbulence {I}nduced
  \textsc{RF-UOWC} {M}ixed {S}ystem,'' \emph{IEEE Access}, vol.~9, pp.
  18\,123--18\,136, 2021.

\bibitem{xu2021performance}
G.~Xu and Z.~Song, ``Performance analysis of a \textsc{UAV}-{A}ssisted
  \textsc{RF/FSO} {R}elaying {S}ystems for {I}nternet of {V}ehicles,''
  \emph{IEEE Internet of Things Journal}, 2021.

\bibitem{gupta2018performance}
J.~Gupta, V.~K. Dwivedi, and V.~Karwal, ``On the performance of \textsc{RF-FSO}
  system over \textsc{R}ayleigh and \textsc{K}appa-\textsc{M}u/inverse
  {G}aussian fading environment,'' \emph{IEEE Access}, vol.~6, pp. 4186--4198,
  2018.

\bibitem{nafis2021}
A.~Nafis, A.~Badrudduza, Z.~Borshon, M.~Kundu, and M.~Sarkar, ``Secrecy
  \textsc{T}rade-off at the \textsc{P}hysical \textsc{L}ayer over
  \textsc{M}ixed \textsc{F}ading \textsc{M}ulticast \textsc{C}hannels
  \textsc{E}mploying \textsc{A}ntenna \textsc{D}iversity,'' \emph{Wireless
  Personal Communications}, pp. 1--18, 2021.

\bibitem{ibrahim2021enhancing}
M.~Ibrahim, A.~Badrudduza, M.~Hossen, M.~K. Kundu, I.~S. Ansari \emph{et~al.},
  ``Enhancing security of \textsc{TAS/MRC} based mixed \textsc{RF-UOWC} system
  with induced underwater turbulence effect,'' \emph{arXiv preprint
  arXiv:2105.09088}, 2021.

\bibitem{yadav2021comprehensive}
P.~Yadav, S.~Kumar, and R.~Kumar, ``A comprehensive survey of physical layer
  security over fading channels: {C}lassifications, applications, and
  challenges,'' \emph{Transactions on Emerging Telecommunications
  Technologies}, p. e4270, 2021.

\bibitem{a.s.m.physical2021}
A.~Badrudduza, S.~Islam, M.~Kundu, and I.~Ansari, ``Secrecy {P}erformance of
  $\alpha-\kappa-\mu$ {S}hadowed {F}ading {C}hannel,'' \emph{ICT Express},
  2021.

\bibitem{sumona2021security}
A.~S. Sumona, M.~K. Kundu, and A.~Badrudduza, ``Security {A}nalysis in
  {M}ulticasting over {S}hadowed {R}ician and $\alpha-\mu$ {F}ading {C}hannels:
  {A} \textsc{D}ual-hop {H}ybrid {S}atellite {T}errestrial {R}elaying
  {N}etwork,'' \emph{IET Communication}, in press.

\bibitem{sofotasios2018error}
P.~C. Sofotasios, S.~K. Yoo, S.~Muhaidat, S.~L. Cotton, M.~Matthaiou,
  M.~Valkama, and G.~K. Karagiannidis, ``Error analysis of wireless
  transmission over generalized multipath/shadowing channels,'' in \emph{2018
  IEEE Wireless Communications and Networking Conference (WCNC)}.\hskip 1em
  plus 0.5em minus 0.4em\relax IEEE, 2018, pp. 1--6.

\bibitem{rana2017novel}
V.~Rana, R.~Joshi, and S.~Soni, ``A novel closed-form of \textsc{ASEP} and
  channel capacity with \textsc{MRC} over $\eta$-$\mu$/\textsc{IG}
  distribution,'' in \emph{2017 IEEE 38th Sarnoff Symposium}.\hskip 1em plus
  0.5em minus 0.4em\relax IEEE, 2017, pp. 1--5.

\bibitem{sofotasios2013eta}
P.~C. Sofotasios, T.~A. Tsiftsis, M.~Ghogho, L.~R. Wilhelmsson, and M.~Valkama,
  ``The $\eta$- $\mu$/\textsc{IG} distribution: {A} novel physical
  multipath/shadowing fading model,'' in \emph{2013 IEEE International
  Conference on Communications (ICC)}.\hskip 1em plus 0.5em minus 0.4em\relax
  IEEE, 2013, pp. 5715--5719.

\bibitem{ramirez2019utility}
P.~Ram{\'\i}rez-Espinosa and F.~J. Lopez-Martinez, ``On the utility of the
  inverse gamma distribution in modeling composite fading channels,'' in
  \emph{2019 IEEE Global Communications Conference (GLOBECOM)}.\hskip 1em plus
  0.5em minus 0.4em\relax IEEE, 2019, pp. 1--6.

\bibitem{sofotasios2018capacity}
P.~C. Sofotasios, S.~K. Yoo, N.~Bhargav, S.~Muhaidat, S.~L. Cotton,
  M.~Matthaiou, M.~Valkama, and G.~K. Karagiannidis, ``Capacity analysis under
  generalized composite fading conditions,'' in \emph{2018 International
  Conference on Advanced Communication Technologies and Networking
  (CommNet)}.\hskip 1em plus 0.5em minus 0.4em\relax IEEE, 2018, pp. 1--10.

\bibitem{pant2019error}
D.~Pant, P.~S. Chauhan, and S.~K. Soni, ``Error probability and channel
  capacity analysis of wireless system over inverse gamma shadowed fading
  channel with selection diversity,'' \emph{International Journal of
  Communication Systems}, vol.~32, no.~16, p. e4083, 2019.

\bibitem{gao2016physical}
Y.~Gao, J.~Ge, and H.~Gao, ``Physical layer security with maximal ratio
  combining over heterogeneous $\kappa-\mu$ and $\eta-\mu$ fading channels,''
  \emph{Wireless Personal Communications}, vol.~86, no.~3, pp. 1387--1400,
  2016.

\bibitem{badrudduza2019enhancement}
A.~Badrudduza, S.~Shahriyer, M.~Kundu, and S.~Shabab, ``Enhancement of secrecy
  multicast capacity over $\kappa$-$\mu$ shadowed fading channel,'' in
  \emph{2019 IEEE International Conference on Telecommunications and Photonics
  (ICTP)}.\hskip 1em plus 0.5em minus 0.4em\relax IEEE, 2019, pp. 1--4.

\bibitem{sarkar2009secure}
M.~Z.~I. Sarkar, T.~Ratnarajah, and M.~Sellathurai, ``Secure wireless
  multicasting through rayleigh fading channels—a secrecy tradeoff,'' in
  \emph{2009 First UK-India International Workshop on Cognitive Wireless
  Systems (UKIWCWS)}.\hskip 1em plus 0.5em minus 0.4em\relax IEEE, 2009, pp.
  1--5.

\bibitem{shabab2019enhancement}
S.~Shabab, A.~Badrudduza, and M.~Kundu, ``Enhancement of physical layer
  security over generalized nakagami-$ m $ fading multicast channel,'' in
  \emph{2019 4th International Conference on Electrical Information and
  Communication Technology (EICT)}.\hskip 1em plus 0.5em minus 0.4em\relax
  IEEE, 2019, pp. 1--5.

\bibitem{badrudduza2019performance}
A.~Badrudduza, M.~Sarkar, M.~Kundu, and D.~Sarker, ``Performance analysis of
  multicasting over rician-k fading channels: A secrecy tradeoff,'' in
  \emph{2019 International Conference on Computer, Communication, Chemical,
  Materials and Electronic Engineering (IC4ME2)}.\hskip 1em plus 0.5em minus
  0.4em\relax IEEE, 2019, pp. 1--4.

\bibitem{peppas2013performance}
K.~P. Peppas, G.~C. Alexandropoulos, and P.~T. Mathiopoulos, ``Performance
  {A}nalysis of \textsc{D}ual-\textsc{H}op \textsc{AF} {R}elaying {S}ystems
  over {M}ixed $\eta$-$\mu$ and $\kappa$-$\mu$ {F}ading channels,'' \emph{IEEE
  Transactions on Vehicular Technology}, vol.~62, no.~7, pp. 3149--3163, 2013.

\bibitem{almaeeni2016error}
S.~Al~Maeeni, P.~C. Sofotasios, S.~Muhaidat, and M.~Valkama, ``Error analysis
  of differentially modulated cooperative systems under generalized fading,''
  in \emph{2016 23rd International Conference on Telecommunications
  (ICT)}.\hskip 1em plus 0.5em minus 0.4em\relax IEEE, 2016, pp. 1--5.

\bibitem{GR:07:Book}
I.~S. Gradshteyn and I.~M. Ryzhik, \emph{Table of Integrals, Series, and
  Products}, 7th~ed.\hskip 1em plus 0.5em minus 0.4em\relax San Diego, CA:
  Academic, 2007.

\bibitem{bhargav2016secrecy}
N.~Bhargav, S.~L. Cotton, and D.~E. Simmons, ``Secrecy {C}apacity {A}nalysis
  {O}ver $\kappa$-$\mu$ {F}ading {C}hannels: {T}heory and {A}pplications,''
  \emph{IEEE Transactions on Communications}, vol.~64, no.~7, pp. 3011--3024,
  2016.

\bibitem{wyner1975wire}
A.~D. Wyner, ``The wire-tap channel,'' \emph{Bell system technical journal},
  vol.~54, no.~8, pp. 1355--1387, 1975.

\bibitem{zeng2018physical}
W.~Zeng, J.~Zhang, S.~Chen, K.~P. Peppas, and B.~Ai, ``Physical layer security
  over fluctuating two-ray fading channels,'' \emph{IEEE Transactions on
  Vehicular Technology}, vol.~67, no.~9, pp. 8949--8953, 2018.

\bibitem{badrudduza2020enhancing}
A.~Badrudduza, M.~Sarkar, and M.~Kundu, ``Enhancing security in multicasting
  through correlated {N}akagami-m fading channels with opportunistic
  relaying,'' \emph{Physical Communication}, vol.~43, p. 101177, 2020.

\bibitem{lei2015performance}
H.~Lei, H.~Zhang, I.~S. Ansari, C.~Gao, Y.~Guo, G.~Pan, and K.~A. Qaraqe,
  ``Performance analysis of physical layer security over generalized-$ k $
  fading channels using a mixture {G}amma distribution,'' \emph{IEEE
  Communications Letters}, vol.~20, no.~2, pp. 408--411, 2015.

\bibitem{9331252}
M.~S. {Hossen}, A.~S.~M. {Badrudduza}, A.~{Hanif}, M.~K. {Kundu}, M.~F.
  {Mahmud}, and M.~Z.~I. {Sarkar}, ``On the {E}nhancement of {S}ecrecy
  {M}ulticast {C}apacity over $\kappa-\mu/\textsc{IG}$ {C}omposite {F}ading
  {C}hannel,'' in \emph{2021 2nd International Conference on Robotics,
  Electrical and Signal Processing Techniques (ICREST)}, 2021, pp. 618--623.

\bibitem{6480923}
K.~T. {Hemachandra} and N.~C. {Beaulieu}, ``Outage {A}nalysis of
  {O}pportunistic {S}cheduling in \textsc{D}ual-\textsc{H}op {M}ultiuser
  {R}elay {N}etworks in the {P}resence of {I}nterference,'' \emph{IEEE
  Transactions on Communications}, vol.~61, no.~5, pp. 1786--1796, 2013.

\bibitem{9122938}
D.~{Pant}, P.~S. {Chauhan}, S.~K. {Soni}, and S.~{Naithani}, ``Channel
  {C}apacity {A}nalysis of {W}ireless {S}ystem under \textsc{ORA} scheme over
  $\kappa-\mu/$ \textsc{I}nverse \textsc{G}amma and $\eta-\mu/$
  \textsc{I}nverse \textsc{G}amma {C}omposite {F}ading {M}odels,'' in
  \emph{2020 International Conference on Electrical and Electronics Engineering
  (ICE3)}, 2020, pp. 425--430.

\bibitem{huang2018secrecy}
Q.~Huang, M.~Lin, K.~An, J.~Ouyang, and W.-P. Zhu, ``Secrecy performance of
  hybrid satellite-terrestrial relay networks in the presence of multiple
  eavesdroppers,'' \emph{IET Communications}, vol.~12, no.~1, pp. 26--34, 2018.

\bibitem{8506345}
W.~Cao, Y.~Zou, Z.~Yang, and J.~Zhu, ``Relay {S}election for {I}mproving
  {P}hysical-{L}ayer {S}ecurity in {H}ybrid {S}atellite-{T}errestrial {R}elay
  {N}etworks,'' \emph{IEEE Access}, vol.~6, pp. 65\,275--65\,285, 2018.

\bibitem{bankey2017secrecy}
V.~Bankey and P.~K. Upadhyay, ``Secrecy outage analysis of hybrid
  satellite-terrestrial relay networks with opportunistic relaying schemes,''
  in \emph{2017 IEEE 85th Vehicular Technology Conference (VTC Spring)}.\hskip
  1em plus 0.5em minus 0.4em\relax IEEE, 2017, pp. 1--5.

\bibitem{bankey2019physical}
------, ``Physical layer security of hybrid satellite-terrestrial relay
  networks with multiple colluding eavesdroppers over non-identically
  distributed {N}akagami-m fading channels,'' \emph{IET Communications},
  vol.~13, no.~14, pp. 2115--2123, 2019.

\bibitem{shahriyer2021opportunistic}
S.~M.~S. Shahriyer, A.~S.~M. Badrudduza, S.~Shabab, M.~K. Kundu, and H.~Yu,
  ``Opportunistic relay in multicast channels with generalized shadowed fading
  effects: A physical layer security perspective,'' \emph{IEEE Access}, in
  pres.

\bibitem{kumar2015performance}
N.~Kumar and V.~Bhatia, ``Performance analysis of amplify-and-forward
  cooperative networks with best-relay selection over {W}eibull fading
  channels,'' \emph{Wireless Personal Communications}, vol.~85, no.~3, pp.
  641--653, 2015.

\bibitem{5089994}
R.~H.~Y. {Louie}, Y.~{Li}, H.~A. {Suraweera}, and B.~{Vucetic}, ``Performance
  analysis of beamforming in two hop amplify and forward relay networks with
  antenna correlation,'' \emph{IEEE Transactions on Wireless Communications},
  vol.~8, no.~6, pp. 3132--3141, 2009.

\bibitem{634675}
C.~Tan and N.~Beaulieu, ``Infinite series representations of the bivariate
  {R}ayleigh and {N}akagami-m distributions,'' \emph{IEEE Transactions on
  Communications}, vol.~45, no.~10, pp. 1159--1161, 1997.

\end{thebibliography}

\end{document}